\documentclass[mnsc,nonblindrev]{informs_modified}
\OneAndAHalfSpacedXI % current default line spacing
\usepackage{endnotes}
\let\enotesize=\normalsize
\def\notesname{Endnotes}%
\def\makeenmark{$^{\theenmark}$}
\def\enoteformat{\rightskip0pt\leftskip0pt\parindent=1.75em
	\leavevmode\llap{\theenmark.\enskip}}

\usepackage{natbib}
\bibpunct[, ]{(}{)}{,}{a}{}{,}%
\def\bibfont{\small}%
\TheoremsNumberedThrough     % Preferred (Theorem 1, Lemma 1, Theorem 2)
\ECRepeatTheorems

\EquationsNumberedThrough    % Default: (1), (2), ...
\usepackage{setspace}
\usepackage{amsmath}
\usepackage{amssymb}  
\usepackage{amsfonts}
\usepackage{xcolor}
\usepackage{array}
\usepackage{booktabs}
\usepackage{comment}
\usepackage{graphicx}
\usepackage{multirow} 
\usepackage{mathtools}
\usepackage{subfigure}
\usepackage{tabularx}
\definecolor{DarkBlue}{rgb}{0.0,0.0,0.55}
\usepackage[backref = false, bookmarks, colorlinks = true, plainpages = false, citecolor = DarkBlue , urlcolor = DarkBlue, filecolor = DarkBlue, linkcolor = DarkBlue]{hyperref}
\usepackage[hyphenbreaks]{breakurl}  
\usepackage{natbib}
\usepackage{pifont}
\usepackage{bm}
\usepackage{threeparttable}

\usepackage{enumitem}

\usepackage{algorithmic}
\usepackage[ruled,linesnumbered]{algorithm2e}
\newcommand{\E}{\mathbb{E}}
\newcommand{\R}{\mathbb{R}}

\bibpunct[, ]{(}{)}{,}{a}{}{,}%
\def\bibfont{\small}%

\newif\ifshownotes
\shownotestrue

\ifshownotes
  \newcommand{\Authornote}[3]{{\small\textcolor{#2}{\sf[{#1: #3}]}}}
  \newcommand{\chnote}[1]{\Authornote{Chonghuan}{red}{#1}}
  \newcommand{\sznote}[1]{\Authornote{Shuze}{blue}{#1}}
\else
  \newcommand{\Authornote}[3]{}
  \newcommand{\chnote}[1]{}
  \newcommand{\sznote}[1]{}
\fi

\usepackage{amsmath,amsfonts,bm}

\def\eqref#1{equation~\ref{#1}}
\def\Eqref#1{Eq.~(\ref{#1})}
\def\1{\bm{1}}

\def\vzero{{\bm{0}}}

\def\vmu{{\bm{\mu}}}
\def\vtheta{{\bm{\theta}}}
\def\vrho{{\bm{\rho}}}
\def\vdelta{{\bm{\delta}}}

\def\vDelta{{\bm{\Delta}}}
\def\va{{\bm{a}}}

\def\ve{{\bm{e}}}

\def\vh{{\bm{h}}}

\def\vl{{\bm{l}}}

\def\vp{{\bm{p}}}

\def\vr{{\bm{r}}}

\def\vu{{\bm{u}}}

\def\vR{{\bm{R}}}

\def\mA{{\bm{A}}}

\def\mE{{\bm{E}}}

\def\mI{{\bm{I}}}

\def\mK{{\bm{K}}}
\def\mL{{\bm{L}}}

\def\mP{{\bm{P}}}
\def\mQ{{\bm{Q}}}
\def\mR{{\bm{R}}}

\def\mU{{\bm{U}}}
\def\mV{{\bm{V}}}

\def\mY{{\bm{Y}}}

\DeclareMathAlphabet{\mathsfit}{\encodingdefault}{\sfdefault}{m}{sl}
\SetMathAlphabet{\mathsfit}{bold}{\encodingdefault}{\sfdefault}{bx}{n}

\def\gA{{\mathcal{A}}}

\def\gD{{\mathcal{D}}}

\def\gF{{\mathcal{F}}}
\def\gG{{\mathcal{G}}}

\def\gM{{\mathcal{M}}}
\def\gN{{\mathcal{N}}}
\def\gO{{\mathcal{O}}}

\def\gS{{\mathcal{S}}}
\def\gT{{\mathcal{T}}}

\def\gX{{\mathcal{X}}}

\def\fc{{\mathfrak{c}}}
\def\ft{{\mathfrak{t}}}

\newcommand{\AB}{\mathrm{AB}}
\newcommand{\IS}{\mathrm{IS}}

\newcommand{\PI}{\mathrm{PI}}

\newcommand{\Var}{\mathrm{Var}}

\newcommand{\Cov}{\mathrm{Cov}}
\DeclareMathOperator{\vect}{vec}
\DeclareMathOperator{\Diag}{Diag}

\newcommand{\pa}[1]{\left({#1}\right)}
\newcommand{\br}[1]{\left[{#1}\right]}
\newcommand{\brk}[1]{\left\{{#1}\right\}}
\newcommand{\norm}[1]{\left\|{#1}\right\|}

\newif\ifRMP
\RMPfalse % \RMPfalse

\begin{document}
	%%%%%%%%%%%%%%%%
	
	%\setcounter{page}{1}
	%\DOI{10.1287/xxxx.0000.0000}%
	
        \ifRMP
            \RUNAUTHOR{Anonymous Authors}
        \else
    	\RUNAUTHOR{Chen, Simchi-Levi, and Wang}
        \fi
	
	\RUNTITLE{Experimenting with Treatment Locality}
	
	% \TITLE{Bundling Information Goods of Decreasing Value}
	%\TITLE{Experimenting on Markov Decision Processes with Local Treatments}
	\TITLE{Improving the Estimation of Lifetime Effects in A/B Testing via Treatment Locality}
    \ifRMP
        \ARTICLEAUTHORS{%
  \AUTHOR{Anonymous Authors}}
    \else
	\ARTICLEAUTHORS{%
  \AUTHOR{Shuze Chen}
	\AFF{Graduate School of Business, Columbia University, New York, NY 10027, \EMAIL{shuze.chen@columbia.edu}
 }
	\AUTHOR{David Simchi-Levi}
	\AFF{Institute for Data, Systems, and Society, Department of Civil and Environmental Engineering, and Operations Research
Center, Massachusetts Institute of Technology, Cambridge, MA 02139, \EMAIL{dslevi@mit.edu}}
	\AUTHOR{Chonghuan Wang}
	\AFF{ Naveen Jindal School of Management, University of Texas at Dallas, Richardson, TX 75080, \EMAIL{chonghuan.wang@utdallas.edu}}
	%Enter all authors
	} % end of the block

    \fi
	\ABSTRACT{%
Utilizing randomized experiments to evaluate the effect of short-term treatments on the short-term outcomes has been well understood and become the golden standard in industrial practice. However, as service systems become increasingly dynamical and personalized, much focus is shifting toward maximizing long-term outcomes, such as \emph{customer lifetime value}, through \emph{lifetime exposure to interventions}. Our goal is to assess the impact of treatment and control policies on long-term outcomes from relatively short-term observations, such as those generated by A/B testing. A key managerial observation is that many practical treatments are \emph{local}, affecting only targeted states while leaving other parts of the policy unchanged. This paper rigorously investigates whether and how such locality can be exploited to improve estimation of long-term effects in Markov Decision Processes (MDPs), a fundamental model of dynamic systems. We first develop optimal inference techniques for general A/B testing in MDPs and establish corresponding efficiency bounds. We then propose methods to harness the \emph{localized} structure by sharing information on the non-targeted states. Our new estimator can achieve a \emph{linear} reduction with the number of test arms for a major part of the variance without sacrificing unbiasedness. It also matches a tighter variance lower bound that accounts for locality. Furthermore, we extend our framework to a broad class of \emph{differentiable estimators}, which encompasses many widely used approaches in practice. We show that all such estimators can benefit from variance reduction through information sharing without increasing their bias. Together, these results provide both theoretical foundations and practical tools for conducting efficient experiments in dynamic service systems with local treatments.

	}%
	
	%\KEYWORDS{deterministic inventory theory; infinite linear programming duality;
	%  existence of optimal policies; semi-Markov decision process; cyclic schedule}
	
	%\HISTORY{XXXXXX}
	
	\maketitle
	%%%%%%%%%%%%%%%%%%%%%%%%%%%%%%%%%%%%%%%%%%%%%%%%%%%%%%%%%%%%%%%%%%%%%%

% lifetime > local treatments > MDP
\section{Introduction}
Randomized experiments, commonly known as A/B testing, have emerged as an indispensable tool, enabling organizations across various sectors—including service, retail, manufacturing, and healthcare—to empirically test and refine their strategies for enhancing customer engagement and satisfaction (\citealt{imbens2015causal,kohavi17the,Kohavi_Tang_Xu_2020, koning22exp}). In particular, A/B testing is well-established and widely used to compare short-term or immediate outcomes (e.g., whether a customer clicks on an advertisement) resulting from short-term treatments (e.g., displaying a new version of the advertisement).
%Particularly, empirically comparing short-term or instant outcomes (e.g., whether the customer click-through the advertisement) from short-term treatments (e.g., showing a new version of the advertisement) by A/B testing has been well understood and widely applied.

However, as service systems grow more complex and dynamic, the focus is increasingly shifting from \textit{immediate or short-term outcomes} to \textit{long-term effects} on customers or users who are \textit{continually exposed} to new interventions. For example, online retailers may employ strategies such as regular coupon distribution or cash back, aiming to trade-off short-term profit reductions for potential life-time revenue gains (\citealt{simester06dynamic}). Similarly, user interface updates on web platforms tend to have continual exposure to customers after their launches, thus impacting user experience and engagement over extended periods (\citealt{huang2023estimating,maystre2023optimizing}). These practical scenarios require estimating the impact of \emph{long-term treatments} on \emph{long-term outcomes}, such as \emph{customer lifetime value}, using relatively short-term observations from randomized experiments.

In this work, we investigate the randomized experiments on Markov Decision Processes (MDPs), which are widely used to model dynamic service systems over extended time horizons including customer relationships (\citealt{pfeifer2000modeling}), inventory control (\citealt{gong23bandits}), two-sided platforms (\citealt{ramesh22tsr}), mobile health (\citealt{zhang2023statistical}) and many others. Additionally, 
as modern systems become increasingly complex with numerous states, the practical challenges of implementing entirely new global policies intensify. Such policies could be overly aggressive, in terms of posing significant risks and requiring substantial infrastructure changes. Consequently, many treatments in practice are becoming more fine-grained and localized. This means that the new treatments may only apply on a small number of targeted states locally.
%as the states of the modern systems could be numerous and complex and the practical challenges of applying a complicated globally completely new policies, more and more treatments become more fine-grained and localized. Also, the completely new policies might be too aggressive in terms of risk and complete infrastructure change. This means that the new treatments may only apply on a small number of targeted states locally.
%A typical local treatment policy intervenes only in specific customer interactions, basing its actions on recent customer behaviors or other contextual factors (i.e., the customer's state).
%One of the examples is that it is increasingly common for treatments or policies to be localized, targeting specific customer interactions based on recent behaviors or other contextual factors (i.e., state).
For instance, in the customer relationships, the customer's states are referring to her recent behaviors or other contextual factors. An online marketplace or service platform might offer a ``welcome-back" coupon to the customer only when she has not engaged with the service for an extended period (e.g., Sephora’s Welcome Back Event \citealt{danielle20sephora}). If she is a daily active user, she might not receive it.  Similarly, physical retail locations like pharmacies and markets may provide incentives like cash back to the customer only when she makes in-store purchases (e.g., Bed Bath \& Beyond \citealt{rafi15bed}). Likewise, a typical intervention for recommender systems may involve \emph{precise exposure}, only taking effect when user click into certain pages such as banners or author profiles, without altering other system components throughout the iterative recommendation process (e.g., Spotify \citealt{maystre2023optimizing} and Douyin). These localized treatments are strategic, usually aiming to optimize the lifetime value of customers by tailoring experiences and offers to individual needs and histories. 
% Moreover, the well-known classical $(s,S)$-policy (\citealt{scarf1960optimality}) in inventory management is another typical example of local treatment. In this policy, the inventory ordering decision only happens when the inventory level falls below a specific threshold $s$. 
Given the prevalence of local treatments in practice, our research goal is to explore the potential of leveraging treatment locality to optimize data utilization for efficient A/B testing on long-term effects within an MDP environment.

\subsection{An Illustrative Example: Customer Lifetime Value}
% To further illustrate our modeling of SST, we provide an application example on improving customer lifetime value, 
% borrowed from the existing literature in marketing like \cite{berger1998customer, pfeifer2000modeling, simester06dynamic, zhang22dynamic}. Particularly, we adopt the following model in \cite{pfeifer2000modeling} and this example will be revisited throughout the paper.
To be more concrete about our research question, we further introduce an application example on improving customer lifetime value, adapted from a seminal work in the marketing literature by \cite{pfeifer2000modeling} and closely related to customer retention and churn management (\citealt{hamilton2023churning,lemmens20managing, kan24managing}). This example will be revisited throughout the paper.

\begin{example}[Customer Lifetime Value]\label{exp: CLV}
   Drawing from the classic Recency, Frequency, Monetary value (RFM) framework, it is often posited by companies that the probability a customer will purchase at the end of any period is solely a function of her \emph{recency}---the number of periods since her last purchase. \cite{pfeifer2000modeling} model recency as the state space of an MDP and characterize customer lifetime value via the value function, which is the discounted sum of revenue collected in the long-run. 
   
   Currently, in order to encourage repeat business and build customer loyalty, the company is exploring the impact of issuing coupons to customers with high recency (i.e., inactive customers at state $s^5$) to attract them back. The coupon treatment is considered \emph{local}, affecting only the state transitions and immediate reward at state $s^5$, while leaving all other states unchanged before and after the treatment. Our study can then enable the company to assess how such a coupon policy might affect customer lifetime value in the long run. 
\end{example}

\subsection{Overview of Main Results}
To demonstrate our results, we start with what is arguably the most fundamental and illustrative scenario: the new local treatment available in only one specific state. We refer to such a local treatment as \emph{Single-State Treatment (SST)}. In this setup, the treatment policy of interest applies the new SST whenever possible and adheres to the control policy in all other states. Our objective is to estimate and infer the Average Treatment Effect (ATE), which quantifies the difference in long-term cumulative rewards between the treatment policy and the control policy. We summarize and highlight our main results in detail below.

First, we design an efficient estimator of ATE in general A/B testing on MDPs, where the treatment does not need to be local. Specifically, we show that the two principal approaches from the reinforcement learning (RL) literature, the model-based estimation and temporal difference (TD) learning, remain effective for estimating ATE if the data are split correctly. 
%A significant difference from classical RL setting is that we need to estimate two values functions simultaneously from data collected by A/B testing, which mixes the treatment policy and the control policy. 
Our proposed estimator is asymptotically normal and unbiased. Moreover, we show that no unbiased estimator can achieve a smaller variance without considering treatment locality. This result builds on a new efficiency proof framework via the \textit{constrained} Cramér-Rao bound. 

We then introduce an estimator that incorporates the SST structure by sharing data collected from states where the new treatment is not applied. While data sharing naturally increases the sample size for both the treatment and control policies, it also introduces additional \textit{correlations} that complicate the overall variance. Fortunately, we are able to show that this estimator can always achieve a smaller variance compared to the original estimator that does not share data. Our analysis further reveals that the variance reduction arises specifically from the component of the variance associated with the states that share data. Importantly, when extending to multiple test arms, we illustrate that a major part of the asymptotic variance can be reduced \emph{linearly} with the number of test arms compared to the estimator that does not use the SST structure. Additionally, we derive another variance lower bound for all unbiased estimators even with the knowledge of the SST, which confirms the optimality of our proposed estimator.

While we have focused on the unbiased estimator for a specific ATE, decision-makers can have many other design choices for A/B testing in dynamic systems. For example, in practice, a slight bias may be acceptable if it substantially reduces variance. Motivated by this, we define a broad class of estimators, called \textit{differentiable estimators}, which includes many existing estimators in the literature (e.g., \citealt{farias2022markovian,farias2023correcting,huang2023estimating}). We show that information sharing \emph{universally} reduces variance without increasing bias for all differentiable estimators. Consequently, any proposed differentiable estimators, regardless of its specific design objective, could immediately benefit from the variance reduction through information sharing if the treatment is local. When there are multiple testing arms, a major component of the variance can also get a linear reduction in the number of arms. These results further illustrate the power of treatment locality.

Having demonstrated the variance reduction achieved through SST, we also expand our framework and generalize the results to cover more general local treatments, which are applicable to multiple states, broadening the scope and applicability of our results. 
%In Section~\ref{sec: exp} we verify the effectiveness of our proposed estimators via numerical simulations. 
We also test the effectiveness of our design on a synthetic environment as well as a Sepsis simulator based on real-world data. Finally, we reveal that under SST, the treatment effect remains consistently positive (or negative) for all states before and after the intervention. This indicates that if the new treatment benefits one state, it benefits all others as well.
% Also, we discuss that our information sharing concept is orthogonal to and 
%compatible with the bias-variance trade-offs (e.g., \citealt{farias2022markovian}) and function approximation approaches (e.g.,  \citealt{shi2021statistical, syrgkanis2023postepisodic}) developed for the global treatments. This means that many developments under the global treatments can be further improved with the presence of the local treatments by our information sharing idea. 

\subsection{Related Works}
\noindent\textbf{Structure-aware Experiment Design. } Experimental design originated in economics with basically model-free methods. In recent years, researchers in operations research and management science instead realize that there are many useful structures in operations scenarios to understand, analyze and improve experimentation. %the experimental design ideas with the elaborating models in operations scenarios. 
\cite{bojinov23design,hu2022switchback,ni2023design,jia2023clustered,xiong2024data} consider the switchback design under various scenarios, from temporal interference to spatial interference, from stationary environment to non-stationary environment, and from theory driven perspective to data driven perspective. \cite{glynn2020adaptive,munro2024treatmenteffectsmarketequilibrium,wagner21experimenting,bajari2021multiplerandomizationdesigns,ramesh22tsr, zhu2024seller,weng24exp} study the experiments happening on two-sided markets and a common practice is to model the system as a continuous-time Markov chain. \cite{chen2025b} investigate the A/B testing in e-commerce where the product stock-out can introduce extra bias. \cite{chen2025bias} analyze the bias of experiments for inventory control policies.  \cite{li15thevalue} leverage the sparsity structure of cross-product demand elasticities to greatly reduce the number of experiments required. \cite{li2023experimenting} carefully take the queuing structure into account and propose efficient estimators under the stochastic congestion.  \cite{xiong2023optimal} investigate multi-unit settings with varying treatment starting times. \cite{zhan2021off, zhan23policy} study the post-experiment inference for contextual bandits, which does not capture the state transition as our MDPs. \cite{simester20efficiently} pioneer at considering treatment structures and suggest comparing outcomes only for customers exposed to different treatment and control actions to improve efficiency empirically. Compared to theirs, we further incorporate complex transition dynamics and provide theoretical guarantees of estimations. The long-term effect with the long-term treatment under the short-term measurement is recently studied by \cite{huang2023estimating}. This paradigm of structure-aware experiment design has witnessed great success in theory and practice in above works with either significant reduction of bias or substantial improvement of efficiency.

%To name a few, \cite{zhan2021off, zhan23policy} studies the post-experiment inference for contextual bandit; \cite{glynn2020adaptive,wagner21experimenting,ramesh22tsr} studies experiments on two-sided market that can be modeled as a continuous-time Markov chain; \cite{li2023experimenting} studies the stochastic congestion that forms a queuing system; \cite{bojinov23design,hu2022switchback,huang2023estimating} considers different level of carryover effect; then most generally, \cite{farias2022markovian,farias2023correcting, shi2021statistical,shi2023dynamic,syrgkanis2023postepisodic,zhang2023statistical,hu2022switchback, tran2023inferring} studies experiments on MDPs. This paradigm of structure-aware experiment design has witnessed great success in theory and practice in above works with either significant reduction of bias or substantial improvement of efficiency .  

% Experimental design (+MDP) \cite{glynn2020adaptive} \cite{shi2023dynamic} \cite{shi2021statistical}, \cite{syrgkanis2023postepisodic}
% \cite{zhang2023statistical}
% \cite{ramesh22tsr}
% \cite{li2023experimenting}
% OPE (Tianyi)\cite{farias2022markovian,farias2023correcting}

\noindent\textbf{Experimenting on MDPs with General Treatments.} Experimenting on MDPs is more relevant to our work. Many works adopt MDPs to capture the dynamical system and investigate the long-term outcomes 
(\citealt{pfeifer2000modeling,Murphy_2003,simester06dynamic,tran2023inferring, shi2021statistical,shi2023dynamic, farias2022markovian, farias2023correcting}), where most previous works consider general treatment, i.e. a treatment will be posed to every state of MDP. The seminal work by \cite{farias2022markovian} studies the average reward setting. They prove the asymptotic lower bound for unbiased inference on MDPs and show that this lower bound can sometimes grow exponentially with the number of states. They also construct an LSTD(0)-type OPE estimator that is efficient and achieves this lower bound. In our work, we show that similar phenomena can also be observed in the discounted reward settings. Moreover, \cite{farias2022markovian} study the bias of a Naive estimator and propose an alternative estimator that achieves a notable bias-variance trade-off with an upper bound of asymptotic variance. In the sequel, we demonstrate all three estimators studied in \cite{farias2022markovian} fall into a general class of \emph{differentiable estimator} and can thus all benefit from the variance reduction leveraging treatment locality. Finally, some recent work like \cite{shi2023dynamic,zhang2023statistical} considers asymptotic results with linear function approximation. In contrast, this paper focuses on a fundamental split of the experimental data and is therefore orthogonal to previous literature on estimator design.

In addition, there is another line of work including \cite{syrgkanis2023postepisodic, cheikhi23on, shi2021statistical, khamaru21istemp, tran2023inferring} that considers asymptotic inference on MDPs but under \emph{episodic} setting or with the assumption of a \emph{generative model}, which means the MDP can be independently simulated many times. In contrast, we consider estimations that can only sample one trajectory from a non-episodic MDP, which restricts our access to the data and thus imposes additional challenges. %As we will see later, our setting can incur technical challenges like correlations compared to theirs.

\subsection{Structures and Notations}
In Section~\ref{sec: model}, we formally introduce the model and the formulation, introducing the MDP, the treatment policy and the control policy, and the average treatment effect that we care about. Section~\ref{sec: inference-fixed-policy} presents the estimation methods for the long-term cumulative rewards of a single, specific policy. In Section~\ref{sec: general-ab-testing}, we discuss the idea of classical A/B testing in our setting and how to estimate the long-term cumulative rewards for the treatment and control policies from A/B testing. Section~\ref{sec: IS} provides a new estimator based on the idea of information sharing for the local treatment. Section~\ref{sec: local} generalizes the analysis to the general local treatments. Section~\ref{sec: general-estimator} introduces the differentiable estimators, which can always benefit from the information sharing. In Section~\ref{sec: exp}, we provide some numerical results verifying the idea of information sharing including the real-world sepsis simulation results. Section~\ref{sec: discussion} covers the discussion on some useful properties of the local treatment as well as future work. We conclude our paper in  Section~\ref{sec: conclusion}.

We denote for any $N\in\mathbb{Z}^+, [N]=\{1,2,\cdots,N\}$. For any vector $x$ and matrix $\mA$, we define the semi-norm induced by the matrix $A$ as $\|x\|_\mA\coloneqq \sqrt{x^\top \mA x}$. Additionally, for any function on the state space $f\colon \gS \to \R$, we denote $\Diag_{s\in\gS}(f(s))\coloneqq \Diag(f(s^1),\cdots,f(s^K))\in\R^{K\times K}$. Moreover, we define $a\vee b\coloneqq\max\{a,b\}$ and $a\land b\coloneqq\min\{a,b\}$.

\section{Model}\label{sec: model}
% Drawing on the existing literature concerning the lifetime value of customers, such as works like \citealt{berger1998customer, pfeifer2000modeling, simester06dynamic}, we frame our exploration and inference challenges within the framework of MDPs. 

We start with the general definition of experimenting on MDPs and then specify the \textit{Single-State Treatment (SST)}, which serves as a simple yet illustrative example of local treatment. In Section~\ref{sec: local}, we generalize this concept to broader local treatment scenarios.

Consider a (non-episodic, finite-state, infinite-horizon) MDP with state space $\gS:=\{s^{i}\}_{i\in [K]}$ with cardinality $K$. At time $t\in\mathbb{N}$, we denote the state of the MDP by $s_t\in\gS$. Associated with each state is a set of available actions, $\gA$, which influence the state transitions through the transition probability function $P: \gS\times \gA\times\gS \rightarrow [0,1]$. Here, the next state $s_{t+1}$ follows the distribution $s_{t+1} \sim P(\cdot\mid s_t,a_t)$.
%Each state is associated with a set of available actions $\gA$, which govern the transition probabilities between states via the unknown function $P: \gS\times \gA\times\gS \rightarrow [0,1]$. This is saying that $s_{t+1} \sim P(\cdot\mid s_t,a_t)$. %For brevity, we assume that $\gA=\{c,t\}$ regardless of states, where the action $t$ is the prospective treatment that we want to test and thus $c$ stands for the control action i.e., without employing the intervention. 
For each state-action pair $(s,a)$, the instant reward $r$ earned is independently generated by a Gaussian distribution\footnotemark{}, i.e., $r\sim \gN(r(s,a), \sigma_r^2(s,a))$, where we assume $\sigma^2_r(s,a)<\infty$ for all $(s,a)$ pairs.
\footnotetext{With standard arguments, our results can be extended to the setting where the second moment of the noise is finite.} A policy $\pi: \gS \rightarrow \gA$ maps states to (random) actions. The value function is defined as the cumulative discounted reward starting from state $s\in\gS$ under a policy $\pi$,
\begin{equation}\label{def: value_function}
    V^\pi(s)\coloneqq \E\br{\sum_{i=0}^\infty\gamma^ir(s_i,a_i)\mid \pi, s_0=s},
\end{equation}
where $0\le \gamma<1$ is a discount factor that reflects the trade-off between instant and future rewards. The value function has been widely applied to quantify long-term outcomes in various fields including marketing (\citealt{pfeifer2000modeling, simester06dynamic}), ride-hailing platforms (\citealt{shi2023dynamic}), and video streaming service (\citealt{tran2023inferring, maystre2023optimizing}). We further use $\mV^\pi \in \mathbb{R}^K \coloneqq  (V^\pi(s^1),\cdots,V^\pi(s^K))^\top$ to denote value functions of all states under policy $\pi$. Moreover, associated with each $\pi$, $\vr^\pi\in\R^K$ is the expected reward under each state, defined as $\vr^\pi(s) \coloneqq \sum_{a\in\gA}r(s,a)\pi(a\mid s)$ for $s\in\gS$. $\mP^\pi\in \R^{K\times K}$ is the transition matrix under $\pi$ with $\mP^\pi(s,s^\prime)\coloneqq \sum_{a\in\gA}P(s^\prime\mid s,a)\pi(a\mid s)$ for $s,s^\prime\in\gS$.

We now introduce a special class of treatment policies that we call the \emph{Single-State Treatment (SST)}. Specifically, there exists a pre-determined ``crucial" state, which without loss of generality, we assume to be $s^1$. The treatment of interest is applied exclusively when the system's state, $s_t$, coincides with $s^1$. 
%To illustrate, in the previous motivating example, the new treatment is to offer a coupon and the crucial state is that the customer recently visit the store. 
Formally, we assume that $\gA=\{\fc,\ft\}$ regardless of states, while the action $\ft$, representing the new treatment, is only applicable in the crucial state $s^1$, while $\fc$ denotes the control or default action. This restriction ensures that the treatment is targeted and its effects are isolated to the crucial state.

Building on the definition of SST, we next define two policies and the corresponding treatment effect we aim to assess.
\begin{itemize}[wide]
    \item \textbf{The Control Policy $\pi^\fc$:} This policy does not implement the new intervention, maintaining $\pi^\fc(s)\equiv \fc$ across all states $s\in\gS$. This is typically the default or existing policy used by the organization. Denote the associated transition matrix as $\mP^\fc$, where the entries are exactly $P(\cdot\mid \cdot,\fc)$. The vector $\vr^\fc \in \mathbb{R}^K$ is the unknown mean reward for each state under the control whose entries are just $r(\cdot,\fc)$. 
    \item \textbf{The Treatment Policy $\pi^\ft$:} This policy implements the new intervention $\ft$ exclusively whenever the state is $s^1$, while defaulting to $\fc$ in all the other states. Formally, $\pi^\ft(s^1)\equiv \ft$ and  $\pi^\ft(s)\equiv \fc$ for all $s\in \gS\setminus s^1$. Accordingly, we denote the associated transition matrix and the mean reward vector as $\mP^\ft$ and $\vr^\ft$, respectively.
\end{itemize}
The treatment effect that we are interested in is the difference of the value function under the treatment policy and the control policy. For simplicity, we specially abbreviate the value functions under $\pi^\ft$ and $\pi^\fc$, $V^{\pi^\ft}(s)$ and $V^{\pi^\fc}(s)$, as   $V^{\ft}(s)$ and $V^{\fc}(s)$, respectively. Specifically, for a given initial state $s$, the estimand of our interest is the average treatment effect (ATE),
\begin{equation}\notag
    \Delta(s)=V^{\ft}(s)-V^{\fc}(s).
\end{equation}
For all the states, we write the treatment effect in a more compact way as $\vDelta \coloneqq \mV^{\ft}-\mV^{\fc}$, where $\mV^{\ft}$ and $\mV^{\fc}$ are abbreviations of $\mV^{\pi^\ft}$ and $\mV^{\pi^\fc}$, respectively. The central research question of our paper is to estimate the ATE, and especially, how the SST structure can be taken into account and utilized.

% \section{A Warm-up: Inference under General Policies}
%\section{Essence of Temporal Difference Learning}
\subsection{Preliminaries: Inference under a Fixed Policy} \label{sec: inference-fixed-policy}
In this section, we briefly introduce how to perform inference on the value function under a fixed policy $\pi$ which is not a new task in the literature, as an important cornerstone. There are two streams of methods, both of which are reliant on the well-known Bellman Equation, i.e.,  
\begin{equation}\label{eq:bellman-eq}
    \mV^\pi = \vr^\pi + \gamma \mP^\pi \mV^\pi.
\end{equation}

The first stream of methods, known as \textit{model-based estimation}, is detailed in Algorithm~\ref{alg: model-based}. Model-based estimation $\widehat\mV^{\textnormal{MB}}$ utilizes the reformulation of the Bellman Equation as $\mV^\pi=(\mI-\gamma\mP^\pi)^{-1} \vr^\pi$, where $\mI$ is the identity matrix. This method estimates $\mV^\pi$ by plugging in the estimations of the MDP parameters including $\vr^\pi$ and $\mP^\pi$. Such an estimation is based on the comprehensive understanding of all model parameters, hence it is termed model-based. 

The second methodology, referred to as \textit{model-free estimation}, is outlined in Algorithm~\ref{alg: model-free}. It directly estimates the value function without estimating the model's parameters. Specifically, for a given tuple $(s_i,a_i,s_{i+1},r_i)$, $r_i+\gamma V(s_{i+1})$ is an unbiased estimation of $V(s_i)$ according to the Bellman Equation. Thus, we obtain the estimator $\widehat\mV^{\textnormal{TD}}$ following the well-known least square temporal difference (TD) algorithm, which solves
\begin{equation}\label{eq: LSTD}
    \min_{\widehat \mV}\sum_{s\in\gS} \pa{\sum_{s_i=s} r(s_i,a_i)+\gamma \widehat V(s_{i+1})-\widehat V(s_i) }^2.
\end{equation}

The two inference algorithms are refereed as Algorithm~\ref{alg: model-based} and~\ref{alg: model-free}. Despite of the difference in nature,  $\widehat\mV^{\textnormal{MB}}$ and $\widehat\mV^{\textnormal{TD}}$ are actually providing the same point estimation of the value function $\mV$ under very mild assumptions. 
\begin{proposition}\label{prop: TD_mb}
    Let $\widehat \mV^{\textnormal{MB}}$ and $\widehat \mV^{\textnormal{TD}}$ be the respective output of Algorithm~\ref{alg: model-based} and~\ref{alg: model-free} given the trajectory $\tau = \{(s_i,a_i,s_{i+1}, r_i)\}_{i\in[T]}$ sampled under policy $\pi$. Assume all states are visited at least once in $\tau$, then it holds $\widehat \mV^{\textnormal{MB}} = \widehat \mV^{\textnormal{TD}}$. 
\end{proposition}

%We have now demonstrated the equivalence between the two estimators within finite-space and finite action MDPs without function approximations, indicating that they share identical theoretical properties. 
With such an equivalence, establishing a property for one estimator inherently confirms it for the other. It is important to emphasize that this equivalence does not imply that the model-based approach and TD learning are identical across all scenarios. We are only talking about finite-space and finite-action MDPs without function approximations in this work. There has been a long debate in the RL literature between model-based learning and TD learning, and we refer to the more relevant surveys and textbooks like \cite{moerland2023model} and \cite{Sutton18RL}.

Under mild technical conditions, $\widehat \mV^{\textnormal{MB}}$ and $\widehat \mV^{\textnormal{TD}}$ can be shown to enjoy asymptotic normality. 
\begin{proposition}\label{prop: MB_Normal}
    Assume the Markov chain induced by a fixed policy $\pi$ is uniformly ergodic. Given a trajectory % generated by a fixed policy $\pi$, 
    under any initial distribution, it holds as $T\to\infty$,
    $\sqrt{T}(\widehat \mV^{\textnormal{MB}} - \mV^\pi)\xrightarrow{d} \gN(\vzero, \Sigma_{\mathrm{MB}})$.
\end{proposition}
Here uniform ergodicity is a common technical assumption also used in \cite{farias2022markovian,hu2022switchback,jones04CLT} and will be specified in Appendix~\ref{app: one_policy} along with the detailed form of $\Sigma_{\mathrm{MB}}$. While we expected this result to have been established previously in the RL literature, we were unable to locate an exact reference. For completeness, we provide a full proof in Appendix \ref{app: one_policy}. In addition, we highlight another important result, the asymptotic efficiency of these two classical estimators, which may be of independent interest to the RL literature. Specifically, $\Sigma_{\mathrm{MB}}$ is essentially the smallest variance that can be obtained by any unbiased estimator. The proof is given in Appendix~\ref{app: CRB_TD}.
\begin{proposition}\label{prop: CRB_TD}
     Let the trajectory $\tau=\{(s_i,a_i,s_{i+1},r_i)\}_{i\in [T]}$ be sampled from MDP under policy $\pi$ with the stationary  distribution $\vmu^{\pi}$. Then for any unbiased estimator of value functions with covariance matrix $\Sigma$, it holds that $\Sigma_{\mathrm{MB}}\preceq T\Sigma$.
\end{proposition}

\section{A/B Testing on MDPs}\label{sec: general-ab-testing}
In the previous section, we have established the inference results with the trajectory generated by a fixed policy $\pi$. To infer the ATE, however, it is necessary to simultaneously estimate $\mV^\ft$ and $\mV^\fc$ from a single trajectory. A commonly employed method for this purpose is A/B testing, where, at any given state $s$, a fair coin flip determines whether to apply $\pi^\ft$ or $\pi^\fc$. For simplicity, we introduce the experimental policy $\pi^{1/2}$, defined such that $\pi^{1/2}(a\mid s)\coloneqq {1}/{2}\cdot \pi^\ft(a\mid s)+{1}/{2}\cdot \pi^\fc(a\mid s)$ for any $(s,a)$ pair. Assume that the experimenter executes $\pi^{1/2}$ for $T$ epochs and obtains the trajectory $\tau = \brk{(s_i,a_i,s_{i+1},r_i)}_{i\in[T]}$. An example of a realized trajectory might include sequences such as:
\[\textcolor{blue}{(s^1, \ft, s^3, r_1)}\quad \textcolor{orange}{(s^3, \fc, s^3, r_2)} \quad \textcolor{blue}{(s^3, \fc, s^1, r_3)} \quad \textcolor{orange}{(s^1, \fc, s^2, r_4)} \quad\textcolor{orange}{(s^2, \fc, s^4, r_5)} \quad\textcolor{blue}{(s^4, \fc, s^5, r_6)} \cdots \]
The blue and orange pairs are the observations generated by the treatment policy $\pi^\ft$ and the control policy $\pi^\fc$, respectively. Given that we are considering the SST, action $\ft$ is exclusively available at state $s^1$, while action $\fc$ is the only option in all other states, regardless of the policy applied. Without considering the SST structure—that is, treating $\pi^\ft$ and $\pi^\fc$ as two completely distinct and general policies—a straightforward approach would involve splitting the trajectory $\tau$ into two separate datasets based on the policy implemented at each time step:
%is adopted. Following the definition of SST, we can verify that we took action $\ft$ only at state $s^1$ and took action $\fc$ for all other states. The orange ones are actions taken under control policy $\pi^\fc$. We can verify that $\fc$ is taken for all states.   
%We then divide $\tau$ into two disjoint datasets based on the policy we took, 
\[\textcolor{blue}{\gD^\ft_\AB} = \{\textcolor{blue}{(s_i,a_i,s_{i+1}, r_i)}: \pi_i=\pi^\ft,i\in [T]\}\,,\qquad\textcolor{orange}{\gD^\fc_\AB} = \{\textcolor{orange}{(s_i,a_i,s_{i+1}, r_i)}: \pi_i =\pi^\fc, i\in [T]\}\,, \]
where $\pi_i$ denotes policy used at time $i$. Leveraging the Markov property of the system, unbiased estimations of $\mV^\ft$ and $\mV^\fc$ can be derived from $\gD^\ft$ and $\gD^\fc$ respectively, using either Algorithm~\ref{alg: model-based} or Algorithm~\ref{alg: model-free}. The corresponding algorithm is outlined in Algorithm~\ref{alg: ab_testing}. Although the idea is straightforwards, the estimator can perform in our desired way. More specifically, we can establish the asymptotic normality. 

\begin{center}
\begin{algorithm}
\caption{A/B Testing on Markov Decision Process}
\begin{algorithmic}[1]\label{alg: ab_testing}
\REQUIRE{Policy $\pi^\ft,\pi^\fc$; Experiment horizon $T$; Discount factor $\gamma$.}
\STATE Execute the experiment policy $\pi^{1/2}$ for $T$ epochs and obtain datasets $\gD^\ft_\AB = \{(s_i,a_i,s_{i+1}, r_i): \pi_i=\pi^\ft,i\in [T]\}$ and $\gD^\fc_\AB = \{(s_i,a_i,s_{i+1}, r_i): \pi_i =\pi^\fc, i\in [T]\}$.
\STATE Call Algorithm~\ref{alg: model-based} or Algorithm~\ref{alg: model-free} for $\widehat {\mV^\ft}$ and $\widehat {\mV^\fc}$ with $\gD^\ft_\AB$ and $\gD^\fc_\AB$ respectively. 
\RETURN $\widehat {\mV^\ft} - \widehat {\mV^\fc}\,.$
\end{algorithmic}
\end{algorithm}
\end{center}

\begin{theorem}[Asymptotic Normality of A/B Testing]\label{thm: AB_Testing}
    Assuming the Markov chain induced by the experiment policy $\pi^{1/2}$ is uniformly ergodic, it holds that 
    $\sqrt{T}\pa{\widehat \vDelta ^{\mathrm{AB}} - \vDelta}\xrightarrow{d}\gN(\vzero, \Sigma_{\mathrm{AB}})$ as $T\to\infty$. The detailed form of $\Sigma_{\mathrm{AB}}$ is provided in the appendix (Theorem \ref{thm: app_AB_Testing}).
    
    Specifically, for the crucial state $s^1$, it holds that as $T\to\infty$,
    \[\sqrt{T}\pa{\widehat \Delta ^{\mathrm{AB}} (s^1) - \Delta(s^1)}\xrightarrow{d}\gN(0, \sigma_{\mathrm{AB}}^2(s^1))\,,\]
    where \[\sigma_{\mathrm{AB}}^2(s^1) = 
    2 \sum_{s\in\gS}\frac{\rho^\ft_{s^1}(s)^2}{\mu_{1/2}(s)}\Big(\sigma_\ft^2(s) + \norm{\gamma\mV^\ft}_{\Sigma_{p^\ft(s)}}^2\Big) + 
    2 \sum_{s\in\gS}\frac{\rho^\fc_{s^1}(s)^2}{\mu_{1/2}(s)}\pa{\sigma_\fc^2(s) + \norm{\gamma\mV^\fc}_{\Sigma_{p^\fc(s)}}^2} \,,\]
    where $\rho^\pi_{s^1}(s)$ is defined as $\sum_{i=0}^\infty \gamma^i \mathbb{P}^\pi(s_i=s\mid s_0=s^1)$.
We use $\sigma^2_\pi(s)\coloneqq \sum_{a\in\gA}\mathbb{P}^\pi(a\mid s)\sigma^2(s,a)$ to denote the variance of the instant rewards of state $s$ under policy $\pi$. The term $\mu_{1/2}(s)$ is the stationary probability of state $s$ under the experiment policy $\pi^{1/2}$, and $\Sigma_{p^\pi(s)}\coloneqq \Diag(\vp_s)-\vp_s\vp_s^\top$, where $\vp_s\in\mathbb{R}^K$ is the vector of the transition probability from state $s$ to the other states, $(\mathbb{P}^\pi(s^1\mid s),\cdots,\mathbb{P}^\pi(s^K\mid s))^\top$.
\end{theorem}
%Again, to simplify and provide more insights, we specially focus on the treatment effect of the crucial state $s^1$, which is $\Delta(s^1)$. The following corollary presents the asymptotic normality result of the estimated $\Delta(s^1)$ using A/B testing as Algorithm~\ref{alg: ab_testing}.

The term $\rho^\pi_{s^1}(s)$ is also called the discounted state-visitation measure defined and widely used in the RL literature (\citealt{Agarwal2019ReinforcementLT}). The asymptotic variance $\sigma_{\mathrm{AB}}^2(s^1)$ is influenced not only by the state $s^1$ itself but also by all the other states. This is consistent with our understanding that our objective is the long-term value and the samples are drawn from an MDP. Although the variance depends on all states, we want to specially highlight that $\sigma^2_{\mathrm{AB}}(s^1)$ can be divided into two groups: one associated exclusively with $\pi^\ft$ and the other with $\pi^\fc$. Each group has $|\gS|$ distinct terms, where each term corresponds almost exclusively to one state. Furthermore, each term is inversely proportion to the stationary probability $\mu_{1/2}(s)$. This implies that states which are less frequently visited pose greater challenges in sampling, potentially leading to increased variance. Additionally, each term has two parts, $\sigma_\pi^2(s)$ and $\norm{\gamma\mV^\pi}_{\Sigma_{p^\pi(s)}}^2$, which correspond to the (short-term) variance of the instant rewards and the (long-term) variance related to transitions to the other states.

Finally, we want to remind that Theorem \ref{thm: AB_Testing} is applicable for any general $\pi^\ft$ and $\pi^\fc$.  Our approach segments the trajectory solely based on whether the policy $\pi^\ft$ or $\pi^\fc$ is used at each step, independent of what exactly $\pi^\ft$ and $\pi^\fc$ are. Consequently, this estimator can be readily extended to estimate ATE for a broader range of treatment policies $\pi^\ft$, beyond those confined to SST.

\subsection{Variance Lower Bound for General A/B Testing on MDPs}
%Recall that when we split the trajectory according to A/B testing, we do not consider the SST structure; rather, we split them solely based on whether the policy $\pi^\ft$ or $\pi^\fc$ is used at each step. This idea of data splitting can then be naturally extended to estimating the ATE for general treatment policy $\pi^\ft$ that is not necessarily SST. Formally, the general treatment and control polices are defined as, for all $s\in\gS$, 
% \[\pi^\ft(s)\equiv \ft\,,\qquad \pi^\fc(s)\equiv \fc\,.\]
% And the average treatment effect is defined in the same manner as $\vDelta\coloneqq \mV^\ft-\mV^\fc$. 

In this subsection, we demonstrate that if we ignore the structure of SST, or if the treatment does not adhere to SST principles at all, then the asymptotic variance $\Sigma_{\textrm{AB}}$ given in Theorem~\ref{thm: AB_Testing} is almost optimal for any unbiased estimator of the ATE $\vDelta$. The proof is delayed to Appendix~\ref{app: CRB_AB}.

\begin{theorem}[Asymptotic Efficiency of A/B Testing]\label{thm: CRB_AB}
 If running the mixed policy $\pi^{1/2}$ with the initial state distribution $\vmu_{1/2}$, for any unbiased estimator of ATE $\vDelta$ with covariance matrix $\Sigma$, it holds that $\Sigma_{\mathrm{AB}}\preceq T\Sigma$, where $\Sigma_{\mathrm{AB}}$ is defined in Theorem~\ref{thm: AB_Testing}. 
 %the trajectory $\tau=\{(s_i,a_i,s_{i+1},r_i)\}_{i\in [T]}$ be sampled from MDP under mixed policy $\pi^{1/2}$ with initial distribution $\vmu_{1/2}$. Then for any unbiased estimator of ATE $\vDelta$ with covariance matrix $\Sigma$, it holds that $\Sigma_{\mathrm{AB}}\preceq T\Sigma$. 
   % If the trajectory $\tau=\{(s_i,a_i,s_{i+1},r_i)\}_{i\in [T]}$ be sampled from MDP under mixed policy $\pi^{1/2}$ with initial distribution $\vmu_{1/2}$. Then for any unbiased estimator of ATE $\vDelta$ with covariance matrix $\Sigma$, it holds that $\Sigma_{\mathrm{AB}}\preceq T\Sigma$. 
\end{theorem}
This lower bound effectively implies that asymptotic variance of A/B testing cannot be further reduced under general treatment pattern. In recent literature, \cite{farias2022markovian} derive a similar efficiency result for a similar unbiased A/B testing approach on average-reward MDPs. Their ATE is defined as the difference between average rewards for $\pi^\ft$ and $\pi^\fc$, i.e. $\vmu^{\ft^\top}\vr^\ft - \vmu^{\fc^\top}\vr^\fc$. Compared to theirs, our study further incorporates reward variance. From a technical perspective, we employs a distinct proof framework via the constrained Cramér-Rao bound.

% \subsection{Generalization to A/B/n Testing}
% We conclude the section of A/B testing with a slight generalization of our  results. 

% \section{Inference under Single Point Treatment}
\section{Variance Reduction: Information Sharing between Test Arms} \label{sec: IS}
% According to Corollary~\ref{coro: linear_Variance}, the variance for classical A/B Testing or ABn will grow Linearly with the number of test arms $N$. Such linear growth of variance is indeed optimal if we have no further assumption for the experiment structure.

% To break the barrier of linear variance, we propose the method of information sharing for heterogeneous experiments. For simplicity, we only consider the single-point experiments. Similar results can also hold for multi-point regime.

In this section, we explore how the structure of SST can be leveraged to reduce variance. The underlying principle is that, thanks to the Markov property, the state transitions and reward signals of states $s_t\ne s^1$ can be shared across different test arms. To illustrate this, let us revisit the trajectory in Section \ref{sec: general-ab-testing}. Although $(s^3,\fc,s^3,r_2)$ is generated by $\pi^\fc$, it can also be seen as having been generated by $\pi^\ft$ since $\pi^\ft(s^3)=\pi^\fc(s^3)=\fc$. This observation suggests a potential strategy to use such samples for estimating both $\mV^\ft$ and $\mV^\ft$. Consequently, we recolor the trajectory with additional purple on the samples that can be shared between the two policies.
\[\textcolor{blue}{(s^1, \ft, s^3, r_1)}\quad \textcolor{purple}{(s^3, \fc, s^3, r_2)} \quad \textcolor{purple}{(s^3, \fc, s^1, r_3)} \quad \textcolor{orange}{(s^1, \fc, s^2, r_4)} \quad\textcolor{purple}{(s^2, \fc, s^4, r_5)} \quad\textcolor{purple}{(s^4, \fc, s^5, r_6)} \cdots \]
Accordingly, we can construct two datasets,
\begin{align*}
   &\textcolor{blue}{\gD^\ft_\textrm{IS}} = \{\textcolor{blue}{(s_i,a_i,s_{i+1}, r_i)}: (s_i,a_i)=(s^1,\ft)\}\cup \{\textcolor{purple}{(s_i,a_i,s_{i+1}, r_i)}: s_i\not=s^1\},\\
   &\textcolor{orange}{\gD^\fc_\textrm{IS}} = \{\textcolor{orange}{(s_i,a_i,s_{i+1}, r_i)}: (s_i,a_i)=(s^1,\fc)\}\cup \{\textcolor{purple}{(s_i,a_i,s_{i+1}, r_i)}: s_i\not=s^1\}.
\end{align*}

Applying either Algorithm \ref{alg: model-based} or Algorithm \ref{alg: model-free} on the datasets $\gD^\ft_\textrm{IS}$ and $\gD^\fc_\textrm{IS}$ yields two new estimators $\widehat\mV^\ft_{\textrm{IS}}$ and $\widehat\mV^\fc_{\textrm{IS}}$ respectively. Subsequently, the estimator for ATE, denoted as $\widehat \vDelta^\IS$, is calculated as $\widehat\mV^\ft_{\textrm{IS}}-\widehat\mV^\fc_{\textrm{IS}}$. We will always use ``IS" to denote information sharing.

To assess whether information sharing can effectively reduce the variance, we first derive the central limit theorem with the information sharing. The full proof is delayed to Appendix~\ref{app: CLT_IS}.
\begin{theorem}[Asymptotic Normality of Information Sharing]\label{thm: info_sharing}
    Assuming the Markov chain induced by the experiment policy $\pi^{1/2}$ is uniformly ergodic, with SST, it holds that $\sqrt{T}\pa{\widehat \vDelta^\IS  - \vDelta}\xrightarrow{d}\gN(\vzero, \Sigma_{\mathrm{IS}})$ as $T\to\infty$.
    The detailed form of $\Sigma_{\mathrm{IS}}$ is provided in Appendix~\ref{app: CLT_IS}.
%     \begin{align*}
%         \Sigma&_{\mathrm{IS}}  = \\
%         & +\pa{(\mI-\gamma \mP^\ft)^{-1} -(\mI-\gamma \mP^\fc)^{-1}}
%         \begin{bmatrix}
%             0 &\\
%             & \Diag_{s\in\gS\setminus s^1}\pa{\frac{\sigma_\ft^2(s)}{\mu_{1/2}(s)} + \frac{\norm{\gamma\mV^\ft}_{\Sigma_{p^\ft}(s)}^2}{\mu_{1/2}(s)}}
%         \end{bmatrix} 
%         \pa{(\mI-\gamma \mP^\ft)^{-\top} - (\mI-\gamma \mP^\fc)^{-\top}}\,.
% &\qquad + 2(\mI-\gamma \mP^\fc)^{-1} \Diag_{s\in\gS}\pa{\frac{\sigma_\fc^2(s)}{\mu_{1/2}(s)} +\frac{ \norm{\gamma\mV^\fc}_{\Sigma_{p^\fc}(s)}^2}{\mu_{1/2}(s)}} (\mI-\gamma \mP^\fc)^{-\top}\,. 
%     \end{align*}
    Specifically, for the crucial state $s^1$, it holds that as $T\to\infty$, 
    \[\sqrt{T}\pa{\widehat \Delta^\IS (s^1) - \Delta(s^1)}\xrightarrow{d}\gN(0, \sigma_{\mathrm{IS}}^2(s^1))\,,\]
    where 
    \begin{align*}
        \sigma_{\mathrm{IS}}^2(s^1) = 2 &\pa{\frac{\rho^\ft_{s^1}(s^1)^2}{\mu_{1/2}(s^1)}\Big(\sigma_\ft^2(s^1) + \norm{\gamma\mV^\ft}_{\Sigma_{p^\ft}(s^1)}^2\Big) +  \frac{\rho^\fc_{s^1}(s^1)^2}{\mu_{1/2}(s^1)}\pa{\sigma_\fc^2(s^1) + \norm{\gamma\mV^\fc}_{\Sigma_{p^\fc}(s^1)}^2} } \\
        &\qquad +  \sum_{s\ne s^1} \frac{1}{\mu_{1/2}(s)}\pa{\big(\rho^\ft_{s^1}(s)-\rho^\fc_{s^1}(s)\big)^2 \sigma_\fc^2(s) + \norm{\gamma\rho^\ft_{s^1}(s)\mV^\ft - \gamma\rho^\fc_{s^1}(s)\mV^\fc}_{\Sigma_{p^\fc}(s)}^2}\,.
    \end{align*}
\end{theorem}

%part of the fitted data (the purple ones) are the same for both value functions, potentially introducing correlations between them. 

%\[\textcolor{blue}{\gD^\ft_\textrm{AB}}\subseteq \textcolor{blue}{\gD^\ft_\textrm{IS}}\,,\qquad \textnormal{and} \qquad\textcolor{orange}{\gD^\fc_\textrm{AB}}\subseteq \textcolor{orange}{\gD^\fc_\textrm{IS}}\,.

\subsection{Analysis of Variance Reduction with Two Test Arms}
In this subsection, we dive into the comparison between the asymptotic variance of general A/B testing derived in Theorem \ref{thm: AB_Testing} and that of A/B testing with information sharing in Theorem \ref{thm: info_sharing}. For simplicity, we restrict our focus to the crucial state $s^1$. 
Decomposing the variance based on whether the state is the crucial state or not,  we can reorganize the asymptotic variance of A/B testing in Theorem~\ref{thm: AB_Testing} as
\begin{align*}
    \sigma_{\mathrm{AB}}^2(s^1)     & =\underbrace{2\pa{\frac{\rho^\ft_{s^1}(s^1)^2}{\mu_{1/2}(s^1)}\Big(\sigma_\ft^2(s^1) + \norm{\gamma\mV^\ft}_{\Sigma_{p^\ft}(s^1)}^2\Big) +  \frac{\rho^\fc_{s^1}(s^1)^2}{\mu_{1/2}(s^1)}\pa{\sigma_\fc^2(s^1) + \norm{\gamma\mV^\fc}_{\Sigma_{p^\fc}(s^1)}^2} }}_{\Sigma^{\mathrm{het}}_{\mathrm{AB}} } \\
        &\qquad +  \underbrace{2 \sum_{s\ne s^1}\frac{\rho^\ft_{s^1}(s)^2}{\mu_{1/2}(s)}\Big(\sigma_\ft^2(s) + \norm{\gamma\mV^\ft}_{\Sigma_{p^\ft}(s)}^2\Big) + 2 \sum_{s\ne s^1}\frac{\rho^\fc_{s^1}(s)^2}{\mu_{1/2}(s)}\pa{\sigma_\fc^2(s) + \norm{\gamma\mV^\fc}_{\Sigma_{p^\fc}(s)}^2}}_{\Sigma^{\mathrm{hom}}_{\mathrm{AB}} }\,.
\end{align*}
The $\Sigma^{\mathrm{het}}_{\mathrm{AB}}$ part is the variance directly related to the crucial state $s^1$, and  the superscript \texttt{het} reflects that state $s^1$ is the only state where $\pi^\ft$ and $\pi^\fc$ may be heterogeneous under SST. The $\Sigma^{\mathrm{hom}}_{\mathrm{AB}} $ part represents the variance associated with all other states $s\ne s^1$ where $\pi^\ft$ and $\pi^\fc$ are homogeneous. We want to remind that under SST, the variances $\sigma^2_\ft(s)$ and $\sigma^2_\fc(s)$ are the same for all the other state $s$ except the crucial state $s^1$.

Similarly, we can decompose the asymptotic variance of information sharing, $\sigma_{\mathrm{IS}}^2(s^1)$, in the same fashion,
\begin{align*}
    \sigma_{\mathrm{IS}}^2(s^1) &= \underbrace{2\pa{\frac{\rho^\ft_{s^1}(s^1)^2}{\mu_{1/2}(s^1)}\Big(\sigma_\ft^2(s^1) + \norm{\gamma\mV^\ft}_{\Sigma_{p^\ft}(s^1)}^2\Big) +  \frac{\rho^\fc_{s^1}(s^1)^2}{\mu_{1/2}(s^1)}\pa{\sigma_\fc^2(s^1) + \norm{\gamma\mV^\fc}_{\Sigma_{p^\fc}(s^1)}^2} }}_{\Sigma^{\mathrm{het}}_{\mathrm{IS}} } \\
        &\qquad +  \underbrace{\sum_{s\ne s^1} \frac{1}{\mu_{1/2}(s)}\pa{\pa{\rho^\ft_{s^1}(s)-\rho^\fc_{s^1}(s)}^2 \sigma_\fc^2(s) + \norm{\gamma\rho^\ft_{s^1}(s)\mV^\ft - \gamma\rho^\fc_{s^1}(s)\mV^\fc}_{\Sigma_{p^\fc}(s)}^2}}_{\Sigma^{\mathrm{hom}}_{\mathrm{IS}}}\,.
\end{align*}
The above decomposition demonstrates clear structures of asymptotic variance of both estimators. Formally, the following Theorem \ref{thm: VR_two_arm} compares $\sigma_{\mathrm{AB}}^2(s^1)$ and $\sigma_{\mathrm{IS}}^2(s^1)$.
\begin{theorem}\label{thm: VR_two_arm}
    Consider SST with treatment policy $\pi^\ft$ and control policy $\pi^\fc$, it holds 
    \begin{equation}\label{eq: VR-Compare}
        \Sigma^{\mathrm{het}}_{\mathrm{IS}} = \Sigma^{\mathrm{het}}_{\mathrm{AB}}\quad \textrm{ and } \quad\Sigma^{\mathrm{hom}}_{\mathrm{IS}} \leq \Sigma^{\mathrm{hom}}_{\mathrm{AB}}.
    \end{equation}
\end{theorem}
Theorem~\ref{thm: VR_two_arm} reveals that IS achieves at least an equivalent level of performance to the general A/B testing in terms of asymptotic variance. It aligns with our intuition that the variances associated with $s^1$ remain unchanged, as information sharing is not feasible in state $s^1$ due to the potential differences in the transition matrix and rewards. Consequently, the observed reduction in variance primarily results from the sharing of information across all other states $s\ne s^1$. We would like to highlight that, interestingly, each term of the sum in $\Sigma^{\mathrm{hom}}_{\mathrm{IS}}$ can be directly related to each term of the sum in $\Sigma^{\mathrm{hom}}_{\mathrm{AB}}$ through two Cauthy-Schwarz inequalities. 
More specifically, for any $s\ne s^1$,
    \begin{align*}
        \norm{\gamma\rho^\ft_{s^1}(s)\mV^\ft - \gamma\rho^\fc_{s^1}(s)\mV^\fc}_{\Sigma_{p^\fc}(s)}^2 &\leq 2\norm{\gamma\rho^\ft_{s^1}(s)\mV^t}_{\Sigma_{p^\fc}(s)}^2 + 2\norm{ \gamma\rho^\fc_{s^1}(s)\mV^c}_{\Sigma_{p^\fc}(s)}^2,\\
         (\rho^\ft_{s^1}(s)-\rho^\fc_{s^1}(s))^2\sigma_\fc^2(s) &\le 2\rho^\ft_{s^1}(s)^2\sigma_\ft^2(s)+2\rho^\fc_{s^1}(s)^2\sigma_\fc^2(s),
        % & = 2\norm{\gamma\rho^\ft_{s^1}(s)\mV^t}_{\Sigma_{p^\ft}(s)}^2 + 2\norm{ \gamma\rho^\fc_{s^1}(s)\mV^c}_{\Sigma_{p^\fc}(s)}^2\,,
    \end{align*}
    where the sum of LHS over  $s\ne s^1$ constitutes the $\Sigma^{\mathrm{hom}}_{\mathrm{IS}}$ and the sum of RHS forms $\Sigma^{\mathrm{hom}}_{\mathrm{AB}}$. Such a claim leading to the inequality in \Eqref{eq: VR-Compare}.  
To achieve equality in $\Sigma^{\mathrm{hom}}_{\mathrm{IS}}\le\Sigma^{\mathrm{hom}}_{\mathrm{AB}}$, the two Cauthy-Schwarz inequalities above must be equal for all $s\ne s^1$, which seems to be highly restrictive. In fact, Proposition~\ref{prop: necessary_condition} presents the necessary condition for the equality to hold. 

\begin{samepage}
\begin{proposition} [Necessary Condition of Equality]\label{prop: necessary_condition}
    Equality in \Eqref{eq: VR-Compare} is achieved only if the following two conditions are satisfied simultaneously: 
    \begin{enumerate}
        \item the rewards are fixed constants, i.e, $\sigma^2_\fc(s)=0$ for all $s\ne s^1$;
        \item Either $r(s,a)=0$ for all $s\in\gS,a\in\gA$ or $\exists  \kappa\ne 0$ such that $\rho^\ft_{s^1}(s)=\kappa\rho^\fc_{s^1}(s)$ for all $s\ne s^1$.
    \end{enumerate}
\end{proposition} 
\end{samepage}
The proof is delayed to Appendix~\ref{app: proof_VR_two_arm}. This proposition effectively implies that for the equality to hold: either the MDP is trivial with fixed zero reward at every state, or the discounted state visitation measure has to satisfy linear dependence before and after the treatment.

More importantly, we want to highlight that if the new treatment $\ft$ is identical to the control $\fc$ such that for all states $s$, $\rho^\ft_{s^1}(s)=\rho^\fc_{s^1}(s)$ and $\mV^\ft=\mV^\fc$, the term $\Sigma^{\mathrm{hom}}_{\mathrm{IS}}$ will be zero. In contrast, $\Sigma^{\mathrm{hom}}_{\mathrm{AB}}$ is a value that can be strictly large than zero. This finding is particularly relevant in contexts like A/A testing. Such an observation is somewhat surprising. With information sharing, the amount of data used to estimate the transition probabilities and the rewards can be thought to be doubled, and thus it seems reasonable to conjecture that IS can at most reduce $\Sigma^{\mathrm{hom}}_{\mathrm{AB}}$ to a half, i.e., $\Sigma^{\mathrm{hom}}_{\mathrm{IS}}\ge \frac{1}{2} \Sigma^{\mathrm{hom}}_{\mathrm{AB}}$. However, with the example of A/A testing, the conjecture may not always hold true, suggesting that $\Sigma^{\mathrm{hom}}_{\mathrm{IS}}$ could sometimes be significantly smaller than $\Sigma^{\mathrm{hom}}_{\mathrm{AB}}$.

%Even if the necessary condition in Theorem \ref{thm: VR_two_arm} is satisfied, 

%Moreover, through the decomposition of variance according to state, we explicitly show the variance reduction comes from the information sharing of all other state $s\in\gS\setminus s^1$. In contrast, the variance associated with the crucial state $s^1$, $\Sigma^{\mathrm{het}}_{\mathrm{IS}}$ is the same as $\Sigma^{\mathrm{het}}_{\mathrm{AB}}$. This aligns with the fact that the transition matrix and reward may be different under treatment and control for the crucial state. Hence we do not share information for $s^1$. 

%Here we decompose $\sigma_{\mathrm{AB}}^2(s^1)$ in terms of state to the summation of the heterogeneous part $\Sigma^{\mathrm{het}}_{\mathrm{AB}}$, which represents the variance associated with crucial state $s^1$, and the homogeneous part $\Sigma^{\mathrm{hom}}_{\mathrm{AB}} $, which represents the variance associated with all other state $s\in\gS\setminus s^1$. The superscript \texttt{het} and \texttt{hom} is used to reflect the fact that state $s^1$ is the only state where $\pi^\ft$ and $\pi^\fc$ may differ under SST.   

\subsection{Variance Lower Bound for Experimentation with SST}
In this subsection, we show that IS, though intuitive and simple, is the optimal way to harness treatment locality for consistent A/B testing. Specifically, we present the variance lower bound of any unbiased estimator taking the SST structure into account, which reflects the intrinsic difficulty of A/B testing with SST.
\begin{theorem}[Asymptotic Efficiency of Information-Sharing]\label{thm: CRB_IS}
    Assume the treatment is SST with two test arms. If running mixed policy $\pi^{1/2}$ with initial distribution $\vmu_{1/2}$, for any unbiased estimator of ATE $\vDelta$ with covariance matrix $\Sigma$, it holds $\Sigma_{\mathrm{IS}}\preceq T\Sigma$, where $\Sigma_{\mathrm{IS}}$ is defined in Theorem~\ref{thm: VR_two_arm}.
\end{theorem}
Theorem \ref{thm: CRB_IS} indicates that our estimator $\widehat \vDelta^\IS$ is asymptotically efficient. While Theorem~\ref{thm: VR_two_arm} highlights that IS does not reduce the variance associated with the crucial state $s^1$, Theorem \ref{thm: CRB_IS} affirms that this variance is, in fact, optimal and fundamental.

\subsection{(More) Power of IS with Multiple Test Arms} \label{sec: multiple-arms-IS}
In previous sections, we have shown that IS can do at least as well as A/B testing with two test arms. A practical experiment often contains more than two test arms, such as A/B/n Testing (\citealt{Kohavi_Tang_Xu_2020}, \citealt{russac2021b}) or bandit experiments (\citealt{david23multi}). In this subsection, we generalize our study of the variance reduction of IS to SST with more than two test arms and show that IS becomes even more powerful in reducing the variance comparing with traditional A/B/n testing. For rigor, we start with the definition of SST with $n$ test arms.

%will then generalize our study of the variance reduction of IS to SST with more than two test arms . 

Consider an MDP with state space $\gS\coloneqq\brk{s^i}_{i\in [K]}$, where $s^1$ is still assumed to be the crucial state of treatment. Unlike the previous model in Section~\ref{sec: model} with two test arms, the action set $\gA=\brk{\ft^j}_{j\in[n]}$ now includes $n$ possible actions. Without loss of generality, we let $\ft^1$ denote the control action. Recall Example~\ref{exp: CLV}, a plausible experiment might involve the company testing multiple coupon values to determine which one maximizes customer lifetime value. This implies $n$ SST polices $\Pi\coloneqq\brk{\pi^i}_{i\in[n]}$ with each $\pi^i$ satisfying: $\pi^i(s^1)\equiv \ft^i$ (i.e., $\pi^i(\ft^i \mid s^1)=1$) and $\pi^i(s)\equiv \ft^1$ (i.e., $\pi^i(\ft^1 \mid s)=1$) for all $s\in\gS\setminus s^1$. Additionally, each SST policy is associated with its own transition matrix $\mP^{\ft^i}$ and mean reward $\vr^{\ft^i}$.

% The objective of experimentation in such a setting is to understand the difference impacts on the customer's lifetime value with different test arms. 
The primary objective of experimentation in such a setting is to understand the long-term outcome differences among various test arms. Formally, for any pair of SST polices $(\pi^{i}, \pi^{j})_{i,j\in[n]}$, we define the ATE as $\Delta_{i,j} (s)= V^{i}(s)-V^{j}(s)$, for given state $s\in\gS$. For simplicity, throughout this subsection, we restrict our interest to $\Delta_{i,j} (s^1)$, the ATE under the crucial state $s^1$. For A/B/n testing, the experimental policy becomes $\pi^{1/n}(a\mid s)\coloneqq \frac{1}{n}\sum_{i\in[n]}\pi^{i}(a\mid s)$ for any $(s,a)$ pair, meaning that we apply each policy in $\Pi$ with equal probability.

After collecting data, the traditional A/B/n testing would split the trajectory according to different polices and obtain dataset $\gD^{j}_{\mathrm{ABn}}\coloneqq\{(s_i,a_i,s_{i+1}, r_i):\pi_i=\pi^j, i\in [T]\}$ for each policy $\pi^j$. With information sharing, the observations with $s_i\ne s^1$ among all $n$ test arms can be shared for all SST polices $\pi$. Then, IS yields the corresponding dataset $\gD^{j}_{\mathrm{ISn}}\coloneqq\{(s_i,a_i,s_{i+1}, r_i):(s_i,a_i)\ne (s^1, \ft^k), \forall k\ne j,  i\in [T]\}$ for each policy $\pi^j$. Applying Algorithm~\ref{alg: model-based} or Algorithm~\ref{alg: model-free} to fit $\mV^j$ with $\gD^{j}_{\mathrm{ABn}}$ and $\gD^{j}_{\mathrm{ISn}}$ respectively can produce two estimators $\widehat \mV^{\textnormal{ABn}}_j$ and $\widehat \mV^{\textnormal{ISn}}_j$. The estimators for the treatment effect become $\widehat \vDelta_{i,j}^{\textnormal{ABn}}=\widehat \mV^{\textnormal{ABn}}_i-\widehat \mV^{\textnormal{ABn}}_j$ and $\widehat \vDelta_{i,j}^{\textnormal{ISn}}=\widehat \mV^{\textnormal{ISn}}_i-\widehat \mV^{\textnormal{ISn}}_j$. The following Corollary \ref{coro: linear_Variance} generalizes Theorems~\ref{thm: AB_Testing} and \ref{thm: info_sharing} revealing the asymptotic normality of these two estimators on the crucial state $s^1$.

\begin{corollary}[Asymptotic Normality of A/B/n Testing with(out) IS]\label{coro: linear_Variance}
     Assuming the Markov Chain induced by the mixed policy $\pi^{1/n}$ is uniformly ergodic, with SST, it holds as $T\to\infty$, for both $\Gamma \in \{\mathrm{ABn},\mathrm{ISn}\}$,
    \[\sqrt{T}\pa{\widehat \Delta_{i,j} ^{\Gamma}(s^1) - \Delta_{i,j}(s^1)}\xrightarrow{d}\gN(0, \sigma_{\Gamma,(i,j)}^2(s^1))\,,\]
    where $\sigma_{\Gamma,(i,j)}^2(s^1)$ can be decomposed into two terms, the variance associated with the crucial state $\Sigma_{\Gamma}^{\mathrm{het}}(i,j)$ and the one with all the other states $\Sigma_{\Gamma}^{\mathrm{hom}}(i,j)$.
    % where 
    % \begin{align*}
    %     \sigma_{\mathrm{ABn},(i,j)}^2(s^1) & = \underbrace{n\pa{\frac{\rho^i_{s^1}(s^1)^2}{\mu_{1/n}(s^1)}\pa{\sigma_i^2(s^1) + \norm{\gamma\mV^i}_{\Sigma_{p^i}(s^1)}^2} +  \frac{\rho^j_{s^1}(s^1)^2}{\mu_{1/n}(s^1)}\pa{\sigma_j^2(s^1) + \norm{\gamma\mV^j}_{\Sigma_{p^j}(s^1)}^2} } }_{\Sigma_{\mathrm{ABn}}^{\mathrm{het}}(i,j)}\\
    %     &\qquad + \underbrace{n \sum_{s\ne s^1}\frac{\rho^{i}_{s^1}(s)^2}{\mu_{1/n}(s)}\pa{\sigma_i^2(s) +\norm{\gamma\mV^{i}}_{\Sigma_{p^{i}}(s)}^2} + n \sum_{s\ne s^1}\frac{\rho^{j}_{s^1}(s)^2}{\mu_{1/n}(s)}\pa{\sigma_j^2(s) +\norm{\gamma\mV^{j}}_{\Sigma_{p^{j}}(s)}^2}
    %     }_{\Sigma_{\mathrm{ABn}}^{\mathrm{hom}}(i,j)},
    % \end{align*}
    % and
    % \begin{align*}
    %     \sigma_{\mathrm{ISn},(i,j)}^2(s^1) = &\underbrace{n \pa{\frac{\rho^i_{s^1}(s^1)^2}{\mu_{1/n}(s^1)}\pa{\sigma_i^2(s^1) + \norm{\gamma\mV^i}_{\Sigma_{p^i}(s^1)}^2} +  \frac{\rho^j_{s^1}(s^1)^2}{\mu_{1/n}(s^1)}\pa{\sigma_j^2(s^1) + \norm{\gamma\mV^j}_{\Sigma_{p^j}(s^1)}^2} }}_{\Sigma_{\mathrm{ISn}}^{\mathrm{het}}(i,j)} \\
    %     &\qquad +  \underbrace{\sum_{s\ne s^1} \frac{1}{\mu_{1/n}(s)}\pa{\big(\rho^i_{s^1}(s)-\rho^j_{s^1}(s)\big)^2 \sigma_1^2(s) + \norm{\gamma\rho^i_{s^1}(s)\mV^i - \gamma\rho^j_{s^1}(s)\mV^j}_{\Sigma_{p^1}(s)}^2}}_{\Sigma_{\mathrm{ISn}}^{\mathrm{hom}}(i,j)}\,.
    % \end{align*}
\end{corollary}
We ensure that the both two estimators $\widehat \Delta_{i,j} ^{\mathrm{ABn}}(s^1)$ and $\widehat \Delta_{i,j} ^{\mathrm{ISn}}(s^1)$ enjoy the asymptotic normality, but with different asymptotic variances. Similar as before, the variance can be decomposed according to states. Theorem \ref{thm: VR_n_arm} below characterizes the variance reduction of IS for multiple test arms demonstrating the more power that IS has when facing multiple test arms.
\begin{theorem}\label{thm: VR_n_arm}
    Consider SST with $n$ test arms $\brk{\pi^i}_{i\in[n]}$, it holds for any pair of test arms $(\pi^i,\pi^j)$, 
    \begin{equation}\label{eq: VR_n_arm}
        \Sigma^{\mathrm{het}}_{\mathrm{ISn}}(i,j) = \Sigma^{\mathrm{het}}_{\mathrm{ABn}}(i,j)\,,\qquad \Sigma^{\mathrm{hom}}_{\mathrm{ISn}}(i,j) \leq \frac{2}{n}\Sigma^{\mathrm{hom}}_{\mathrm{ABn}}(i,j),
    \end{equation}
    where the equality of the second part can hold only if the reward of states $s\in\gS\setminus s^1$ are fixed, i.e. $\sigma_{\ft^1}^2(s)=0$ for all $s\ne s^1$.
\end{theorem}
This result underscores the effectiveness of IS in leveraging the structure of SST to significantly reduce the variance compared to traditional A/B/n testing.  Particularly, we demonstrate that the variance associated with all other states $s \ne s^1$, where the control policy is always employed, can be reduced at least linearly with the number of test arms. Thus, the more test arms we have, the greater the reduction in variance achieved by IS. When $n=2$, Theorem \ref{thm: VR_n_arm} replicates the findings of Theorem~\ref{thm: VR_two_arm}. When $n\ge 3$, Theorem \ref{thm: VR_n_arm} consistently ensures a substantially reduced variance. Remarkably, with $n\ge 4$, IS is expected to cut the variance associated with the non-crucial states by at least half. Similar to Theorem~\ref{thm: VR_two_arm}, the variance associated with crucial state $s^1$ is the same as A/B/n testing, as information about this state cannot be shared.  Additionally, the equality condition in the inequality of \Eqref{eq: VR_n_arm} remains as elusive as before, largely because one necessary condition for equality is the absence of noise in the observed rewards.

% \input{Contents/ABn_Testing}
% \section{Inference under Perfect Information of the Control}
%\input{Contents/PI_Estimator}
\section{From SST to Local Treatment}\label{sec: local}
In practice, a local treatment policy may affect more than a single state. Consequently, we  generalize the SST concept to accommodate \emph{Local Treatment}. In this expanded framework, a treatment action is applied to a specific segment of the state space rather than a single state. This approach allows for a more flexible and realistic modeling of interventions in complex MDPs. To illustrate, recall Example~\ref{exp: CLV}. The company now wants to test a new strategy: posting advertisements to customers when their recency is relatively high—for instance, when it exceeds 2. Such a treatment policy is a Local Treatment, as it is applied only to a specific segment ($s^3,s^4,s^5$) of the state space. The remaining states ($s^1,s^2$), representing lower recency, are not subject to this treatment.

% Now that we have demonstrated the power of IS and PI for variance reduction under SST, we are ready to generate SST to general \emph{Local Treatment} on MDPs. 
We then establish the model for Local Treatment. Specifically, there exists a pre-determined ``crucial" subset $\gS^\ft\subseteq \gS$. The treatment policy $\pi^\ft$ and control policy $\pi^\fc$ will only vary at state $s\in\gS^\ft$. Formally, we define
\begin{itemize}[wide]
    \item \textbf{The Control Policy $\pi^\fc$:} This policy maintains $\pi^\fc(s)\equiv \fc$ across all states $s\in\gS$, following the same definition as Section~\ref{sec: model}.  
    \item \textbf{The Treatment Policy $\pi^\ft$:} This policy now implements the new intervention $\ft$ exclusively whenever at the state $s\in\gS^\ft$, while defaulting to $\fc$ in all the other states in $\gS\setminus \gS^\ft$. In other words, $\pi^\ft(s)\equiv \ft$ for all $s\in \gS^\ft$ and  $\pi^\ft(s)\equiv \fc$ for all $s\in \gS\setminus \gS^\ft$. 
\end{itemize}
Notice that when the size of the crucial subset $|\gS^\ft|=1$, we recover the definition of SST. The average treatment effect $\vDelta$ follows the same definition as section~\ref{sec: model}.

Having demonstrated the effectiveness of IS for variance reduction under SST, we can now extend these results to Local Treatment scenarios. It can be expected that we are able to establish similar results to Theorem~\ref{thm: AB_Testing}, \ref{thm: info_sharing}. Compared to SST, the heterogeneous part of the system becomes a subset $\gS^\ft$ instead of the single just state $s^1$. As a result, we can only share information of state $s\in \gS\setminus \gS^\ft$. The following theorem characterizes this extension from SST.

\begin{theorem}[Asymptotic Normality with Local Treatment]
     Assuming the Markov Chain induced by policy $\pi^\ft,\pi^{1/2}$ are both uniformly ergodic, with Local Treatment on state $\gS^\ft \subseteq \gS$, it holds as $T\to\infty$, for $\Gamma \in \{\mathrm{AB},\mathrm{IS}\}$ and any state $s\in \gS$,
    \[\sqrt{T}\pa{\widehat \Delta^{\Gamma}(s) - \Delta(s)}\xrightarrow{d}\gN(0, \sigma_{\Gamma}^2(s))\,,\]
    where $\sigma_{\Gamma}^2(s)$ can be decomposed according to the state as
    \begin{align*}
        \sigma_{\Gamma}^2(s) & = \Sigma_{\Gamma}^{\mathrm{het}} + \Sigma_{\Gamma}^{\mathrm{hom}}\,,
    \end{align*}
    where $\Sigma_{\Gamma}^{\mathrm{het}}$ and $\Sigma_{\Gamma}^{\mathrm{hom}}$ refers to the variance of states $\gS^\ft$ and $\gS\setminus \gS^\ft$ respectively for $\Gamma \in \{\mathrm{AB},\mathrm{IS}\}$. The full expression of $\sigma_{\AB}^2(s), \sigma_{\IS}^2(s)$ can be found in Appendix~\ref{app: local_treament}.
\end{theorem}
Recall in SST, $\Sigma_{\Gamma}^{\mathrm{het}}$ is associated with the single crucial state $s^1$, while now we generalize it to the crucial subset $\gS^\ft$. For the asymptotic relative efficiency of three estimators, we then have the following generalization of Theorem~\ref{thm: VR_two_arm}.

\begin{theorem}\label{thm: local_VAR_compare}
    Consider Local Treatment Effect on $\gS^\ft\subseteq \gS$ with two test arms $\pi^\ft,\pi^\fc$, it holds 
        \[\Sigma^{\mathrm{het}}_{\mathrm{IS}} = \Sigma^{\mathrm{het}}_{\mathrm{AB}} \quad\textrm{and}\quad  \Sigma^{\mathrm{hom}}_{\mathrm{IS}} \leq \Sigma^{\mathrm{hom}}_{\mathrm{AB}}\,.\]
        %\item $\Sigma^{\mathrm{het}}_{\mathrm{PI}}< \brk{1\wedge \frac{1+\delta L}{2}}\Sigma^{\mathrm{het}}_{\mathrm{AB}} \quad\textrm{and}\quad\Sigma^{\mathrm{hom}}_{\mathrm{PI}}=0\,.$
    %     \item $\Sigma^{\mathrm{hom}}_{\mathrm{PI}}=0\,.$
    %  \[\Sigma^{\mathrm{het}}_{\mathrm{PI}}< \min \brk{1, \frac{1+\delta L}{2}}\Sigma^{\mathrm{het}}_{\mathrm{AB}}\quad \textrm{and}\quad
    % \Sigma^{\mathrm{hom}}_{\mathrm{PI}}=0\le \Sigma^{\mathrm{hom}}_{\mathrm{IS}} \leq \Sigma^{\mathrm{hom}}_{\mathrm{AB}} \,.\]
\end{theorem}
Theorem~\ref{thm: local_VAR_compare} replicates Theorem~\ref{thm: VR_two_arm}. The difference is that 
now we generalize $\Sigma_{\Gamma}^{\mathrm{het}}$ to variance associated with the crucial subset $\gS^\ft$ rather than single crucial state $s^1$ in SST. This demonstrates that the effectiveness of IS diminishes as the number of states that can share information (i.e. the homogeneous part) decreases. In the extreme case of a  general treatment where $\gS^\ft=\gS$, meaning all states are treated and no states to share information, IS will reduce to A/B testing. Finally, we can derive a similar generalization to local treatment with multiple test arms as Theorem~\ref{thm: VR_n_arm}. We delay the details to the Appendix~\ref{app: local_n_arm}.

\section{Information Sharing Always Improves Efficiency for Differentiable Estimators}\label{sec: general-estimator}

In the previous sections, we have established the optimal asymptotically unbiased estimators for local treatments in MDPs. In practice, however, decision-makers can have
many other design choices. For example, a slight
bias may be acceptable if it substantially reduces variance. Moreover, beyond long-term cumulative outcomes, practitioners may also be interested in other long-term treatment effects, such as average or point outcomes. This raises a fundamental question: does information sharing always improve performance regardless of the estimating approach or estimand? To address this, we introduce a broad class of estimators, \textit{differentiable estimators}, which includes many widely used methods and, importantly, always benefits from information sharing.

\subsection{Differentiable Estimators}

We now define the differential estimators in the vanilla A/B testing without information sharing. For any piece of data we collect $X_i=(s_i,a_i, s_{i+1}, r_i)$ and $a\in \{\ft,\fc\}$, we denote
    \begin{align*}
    \mK^a_\AB(X_i) \coloneqq 2\mE_{s_i,s_{i+1}}\mathbf{1}[a_i=a]\,,\qquad
    \vR^a_\AB(X_i) \coloneqq 2\ve_{s_i} r_i\mathbf{1}[a_i=a]\,,
    \end{align*}
    where $\mE_{s_i,s_{i+1}}\in\R^{n\times n}$ is a matrix with all entries equal to zero except for a one at the $(s_i,s_{i+1})$ entry and $\ve_{s_i}\in\R^n$ is the standard basis vector with a one in the $s_i$-th position and zeros elsewhere. Using these definitions, we can have the IPW estimators for $\Diag(\vmu_{1/2}) \mP^a$ and $\Diag(\vmu_{1/2})\vr^a$,
\begin{align*}
    \overline{ \mK^a_\AB} \coloneqq \frac{1}{T}\sum_{i=1}^T \mK^a_\AB (X_i)\,,\qquad
    \overline{\vR^a_\AB} \coloneqq \frac{1}{T}\sum_{i=1}^T \vR^a_\AB(X_i)\,.
\end{align*}

\begin{definition}[Differentiable Estimator]\label{def: diff-estimator}
    An estimator $\widehat{\vDelta}^{\textrm{AB}}$ is called a differentiable estimator if 
    \begin{equation}\label{eq: general est}
    \widehat{\vDelta}^{\textrm{AB}}=f(\overline{ \mK^\ft_\AB},\overline{ \mK^\fc_\AB},\overline{\vR^\ft_\AB},\overline{\vR^\fc_\AB})\,,
    \end{equation}
where $f$ is any function that is differentiable w.r.t to $\overline{ \mK^\ft_\AB},\overline{ \mK^\fc_\AB},\overline{\mR^\ft_\AB},\overline{\mR^\fc_\AB}$. All the differentiable estimators constitute the class $\gF$.
\end{definition}

Note that although our previous ATE $\Delta$ is defined as the difference between two value functions, the estimand $\vDelta^{\AB}$ of differentiable estimator $\widehat{\vDelta}^{\AB}$ can be more general. It can also represent, for example, the difference in average rewards, the difference in finite-horizon rewards, or even the difference in advantage functions. Moreover, the function $f$ need not yield unbiased estimators, and we impose no restrictions beyond differentiability.

% We first highlight that now $\Delta$ can be any value beyond the precious defined difference of value function. It can incorporate the difference between average reward, that between finite reward or the advantage function etc.  

Furthermore, we highlight that Definition \ref{def: diff-estimator} and \Eqref{eq: general est} subsumes many existing and widely used estimators for MDPs, including:

\begin{itemize}
    \item Naive estimator of \cite{farias2022markovian};
    \item Difference-in-Q (DQ) estimator of  \cite{farias2022markovian};
    \item Off-policy evaluation estimator in \cite{farias2022markovian};
    \item Longitudinal surrogate model of \cite{huang2023estimating};
    \item Our unbiased A/B testing estimators in Algorithm \ref{alg: ab_testing}.
\end{itemize}
This demonstrates that our framework provides a unifying perspective on several strands of the recent literature.
\begin{remark}
    The fact that our \emph{model-based} A/B testing estimator in Algorithm \ref{alg: ab_testing} belongs to $\gF$ has deeper theoretical implications. In fact, every differentiable \emph{model-based} estimator falls within $\gF$. More concretely, suppose the MDP is parameterized by $(\mP^\ft,\mP^\fc,\vr^\ft,\vr^\fc)$. A differentiable \emph{model-based} estimator first obtains the estimators $(\widehat{\mP}^\ft,\widehat{\mP}^\fc,\widehat{\vr}^\ft,\widehat{\vr}^\fc)$ via Algorithm \ref{alg: model-based}, and then computes the estimate of ATE as $\widehat{\vDelta}^{\textrm{AB}}=g(\widehat{\mP}^\ft,\widehat{\mP}^\fc,\widehat{\vr}^\ft,\widehat{\vr}^\fc)$,
% \begin{equation*}
% \widehat{\Delta}^{\textrm{AB}}=g(\widehat{\mP}^\ft,\widehat{\mP}^\fc,\widehat{\vr}^\ft,\widehat{\vr}^\fc)\,,
% \end{equation*}
where $g$ is some functional differentiable w.r.t $\widehat{\mP}^\ft,\widehat{\mP}^\fc,\widehat{\vr}^\ft,\widehat{\vr}^\fc$. We denote the class of all differentiable model-based estimators as $\gM$. The following proposition establishes the inclusion $\gM\subseteq \gF$ by explicitly bridging  $\widehat{\mP}^\ft,\widehat{\mP}^\fc,\widehat{\vr}^\ft,\widehat{\vr}^\fc$ to $\overline{ \mK^\ft_\AB},\overline{ \mK^\fc_\AB},\overline{\vR^\ft_\AB},\overline{\vR^\fc_\AB}$.
\begin{proposition}\label{prop: gm in gf}
    Given the trajectory $\tau = \{(s_i,a_i,s_{i+1}, r_i)\}_{i\in[T]}$ sampled under policy $\pi$ and assume all states are visited at least once in $\tau$, then it holds $\gM\subseteq \gF$.
\end{proposition}
\proof{Proof of Proposition \ref{prop: gm in gf}.}
    For $\widehat {\mP^\ft}$, notice that $\widehat {\mP^\ft}_{ij} = \overline{\mK^\ft_\AB}(i,j)/(\sum_k \overline{\mK^\ft_\AB}(i,k))$. Then $\widehat {\mP^\ft}$ is actually a function of $\overline{\mK^\ft_\AB}$. Similarly, to estimate $\widehat {\vr^\ft}$, we have $\widehat {\vr^\ft}(i) = \overline{\mR^\ft_\AB}(i)/(\sum_k \overline{\mK^\ft_\AB}(i,k))$.\Halmos
\endproof
\end{remark}

\subsection{Variance Reduction for Differentiable Estimators}
Applying the information-sharing idea to differentiable estimators, we again observe a linear reduction in variance for the non-treated states, without introducing additional bias. For clarity, we focus on the SST case, and the extension to general local treatments follows naturally. In particular, information sharing modifies the estimation of model parameters at every state except $s^1$. In particular, for $X_i=(s_i,a_i, s_{i+1},r_i)$ and $a\in \{\ft,\fc\}$, we have
    \begin{equation*}
    \mK^a_\IS(X_i) \coloneqq \bigg(2\mathbf{1}[a_i=a, s_i=s^1] + \mathbf{1}[s_i\ne s^1]\bigg)\mE_{s_i,s_{i+1}}\,,\quad
    \vR^a_\IS(X_i) \coloneqq \bigg(2\mathbf{1}[a_i=a, s_i=s^1] + \mathbf{1}[s_i\ne s^1]\bigg)\ve_{s_i} r_i\,.
    \end{equation*}
    And similarly, we denote $\overline{ \mK^a_\IS} \coloneqq \frac{1}{T}\sum_{i=1}^T \mK^a(X_i)$ and $
    \overline{\vR^a_\IS} \coloneqq \frac{1}{T}\sum_{i=1}^T \vR^a(X_i)$.
Finally, we plug in the estimators into the same $f$ and obtain our information sharing estimator,
\[\widehat{\vDelta}^{\textrm{IS}}=f(\overline{ \mK^\ft_\IS},\overline{ \mK^\fc_\IS},\overline{\vR^\ft_\IS},\overline{\vR^\fc_\IS})\,.\]
An immediate observation is that both $\overline{ \mK^a_\IS}$ and $\overline{ \mK^a_\AB}$ converge almost surely to $\Diag(\vmu_{1/2}) \mP^a$ for $a\in \{\ft,\fc\}$. Similarly, $\overline{\vR^a_\IS}$ and $\overline{\vR^a_\AB}$ converge almost surely to $\Diag(\vmu_{1/2})\vr^a$. Since $f$ is differentiable (and thus continuous), $\mK^a_\AB(X_i)$ and $\mK^a_\IS(X_i)$ have exactly the same asymptotic bias. More formally, and of greater importance, we have the following variance analysis.
\begin{theorem}\label{thm: general_est_variance compare}
    For any differentiable estimators $\widehat{\vDelta}^{\Gamma}=f(\overline{ \mK^\ft_\Gamma},\overline{ \mK^\fc_\Gamma},\overline{\vR^\ft_\Gamma},\overline{\vR^\fc_\Gamma})$ where $\Gamma \in \{\AB,\IS\}$, $\widehat{\vDelta}^{\AB}$ and $\widehat{\vDelta}^{\IS}$ are both asymptotic normal with the same asymptotic bias. Furthermore, their asymptotic variance can be decomposed as $\Sigma_\Gamma=\Sigma^{\mathrm{het}}_\Gamma + \Sigma^{\mathrm{hom}}_\Gamma+\Sigma^{\mathrm{cov}}_\Gamma$ where \begin{equation}\notag%\label{eq: general estimator}
        \Sigma^{\mathrm{het}}_{\mathrm{IS}} = \Sigma^{\mathrm{het}}_{\mathrm{AB}}\,,\quad \Sigma^{\mathrm{cov}}_{\mathrm{IS}} = \Sigma^{\mathrm{cov}}_{\mathrm{AB}}\,,\textrm{ and } \quad\Sigma^{\mathrm{hom}}_{\mathrm{IS}} \preceq \Sigma^{\mathrm{hom}}_{\mathrm{AB}}.
        \end{equation}
\end{theorem}
Following Theorem \ref{thm: VR_two_arm}, the term $\Sigma^{\mathrm{het}}_{\Gamma}$ captures the variance associated with the non-treated states, while $\Sigma^{\mathrm{hom}}_{\Gamma}$ reflects the variance of the state targeted by the SST. The new term introduced here is $\Sigma^{\mathrm{cov}}_{\Gamma}$, which represents the covariance across time. Because the system operates under an MDP, future states depend on the past states and actions. Compared with Theorem \ref{thm: VR_two_arm}, we can learn that selecting an appropriate functional form of $f$ can eliminate temporal correlation.

For general estimators to A/B/n testing, we can also rigorously show that information sharing can linearly reduce part of the variance. The asymptotic variance can be similarly decomposed as $\Sigma_\Gamma(i,j)=\Sigma^{\mathrm{het}}_\Gamma(i,j) + \Sigma^{\mathrm{hom}}_\Gamma(i,j)+\Sigma^{\mathrm{cov}}_\Gamma(i,j)$ where \begin{equation}\notag
        \Sigma^{\mathrm{het}}_{\mathrm{IS}}(i,j) = \Sigma^{\mathrm{het}}_{\mathrm{AB}}(i,j)\,,\quad \Sigma^{\mathrm{cov}}_{\mathrm{IS}}(i,j) = \Sigma^{\mathrm{cov}}_{\mathrm{AB}}(i,j)\,,\textrm{ and } \quad\Sigma^{\mathrm{hom}}_{\mathrm{IS}}(i,j) \preceq \frac{2}{n}\Sigma^{\mathrm{hom}}_{\mathrm{AB}}(i,j).
        \end{equation}
We defer the formal statement and proofs to Appendix~\ref{app: General Estimator}.

\section{Simulations}\label{sec: exp}
\subsection{Synthetic Data: Customer Lifetime Value}
We revisit the example of customer lifetime value (Example~\ref{exp: CLV}), with a synthetic transition matrix detailed in Appendix~\ref{app: CLV_Exp}. We assume that under the control policy, when the customer makes a purchase, the company will receive a net income of $30$ while the income in all other states is $0$. Specifically, the transition matrix and the reward under the control policy are set as follows.
\[\mP^\fc=\begin{bmatrix}
 0.6 & 0.4 &  &  & \\
 0.5 &  & 0.5 &  & \\
 0.4 &  &  & 0.6 & \\
 0.3 &  &  &  & 0.7\\
 0.1 &  &  &  & 0.9
\end{bmatrix}\,,\qquad \vr^\fc=\begin{bmatrix}
 30\\
 0\\
 0\\
 0\\
0
\end{bmatrix}\,.\]

We consider the following two types of treatment:
%The treatment is a SST policy that the company sends a coupon when the customer is only at \emph{highest} recency ($r=5$).
\begin{enumerate}
    \item \textbf{SST.} The company considers to offer coupons when the customer reaches the highest recency (recency = 5). We evaluate five different candidate coupon values, each associated with a different treatment effect. These treatments lead to a decrease in reward at state $s^5$ by values of $2, 2.5, 3, 3.5, 4$, respectively. Correspondingly, they increase the probability of transitioning to state $s^1$ by $0.10, 0.15, 0.20, 0.25, 0.30$, respectively.
    
    %The company offers coupons when the customer is at highest recency (recency=5). We consider 5 different candidate coupon values. The five different treatments lead to the decrease in reward at state $s^5$ by $2.5, 3, 3.5, 4. 4.5$ and the increase the probability transiting to state $s^1$ by $0.15, 0.20, 0.25, 0.30, 0.35$, respectively.

    %The treatment affects state $s^5$ with a decrease in reward (we set the coupon costs $2+0.5n$ for the $n$-th test arm) and increase in the probability transiting to state $s^1$ (increased by $0.1+0.05n$ for the $n$-th test arm)
    
    \item \textbf{Local Treatment.} The company plans to offer coupons whenever the customer's recency is high (recency$>$2). We still consider five different candidate values. They uniformly decrease the rewards under states $s^3,s^4,s^5$  by $2, 2.5, 3, 3.5, 4$, respectively and increase the probability transiting to state $s^1$ by $0.10, 0.11, 0.12, 0.13, 0.14$, respectively.

   % The treatment affects state $(s^3,s^4,s^5)$ with decreases in reward (the coupon costs $2+0.5n$ at all three states for the $n$-th test arm) and increases in the probability transiting to state $s^1$ at three states (increased by $0.1+0.01n$ at all three states for the $n$-th test arm). 
    
\end{enumerate}
%For each type of treatment, we consider $5$ candidate coupon values (advertising cost), i.e. $n=5$ test arms (the coupon value (advertising cost) is $0$ for the control arm). 

We additionally introduce another benchmark for comparison, called \textit{perfect information} (PI), where the decision-makers have the access to the perfect information of the control arm. The PI estimators can always serve as the lower bound of the variance (i.e., the necessary variance). The detailed description of the PI estimators and their properties are provided in Appendix \ref{sec: PI}.

\begin{figure}[ht]
\centering
\begin{tabular}{l l }
        \includegraphics[height=.35\textwidth]{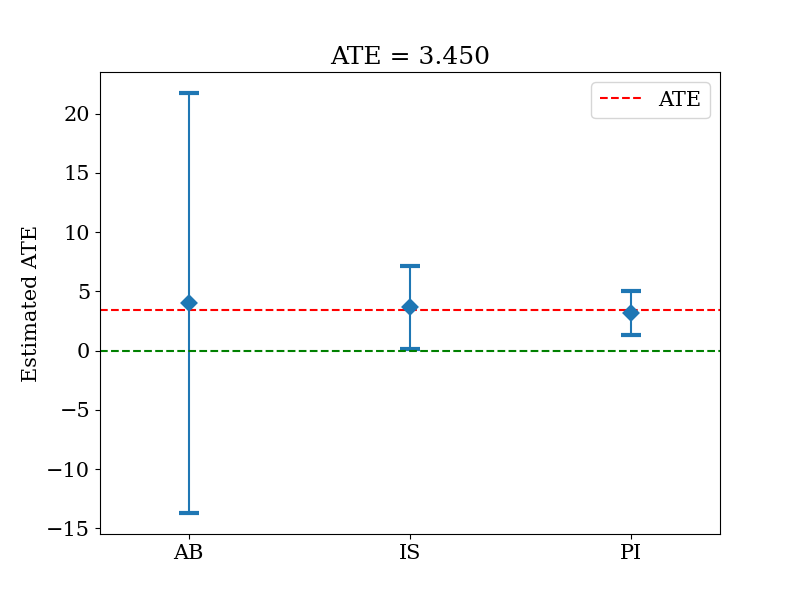}
        &
        \includegraphics[height=.35\textwidth]{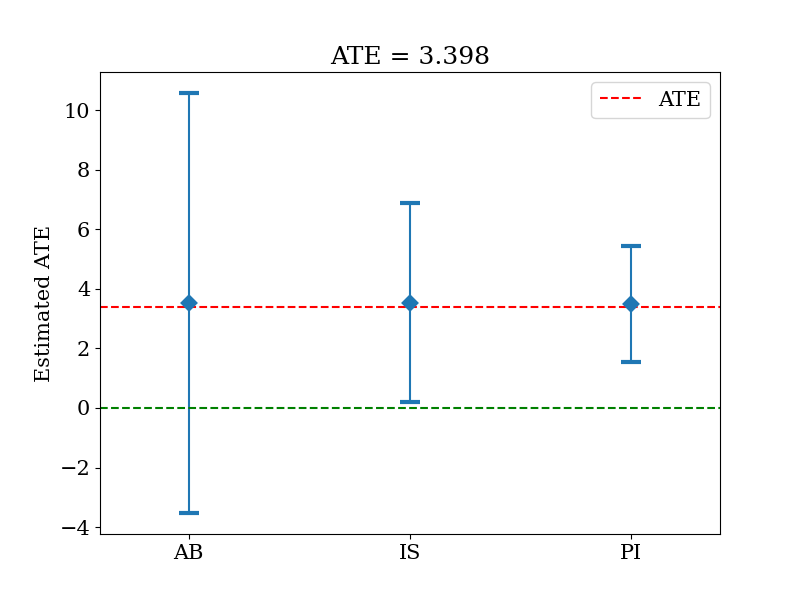}
    \end{tabular}
    \caption{Synthetic simulations for Example~\ref{exp: CLV}. Diamonds and caps represent the empirical mean and standard error respectively for each estimator. {\it Left}: the result for SST policy with $n=5$ test arms and experiment horizon $T=350$. {\it Right}: the result for Local Treatment policy with $n=5$ test arms and experiment horizon $T=2000$.}
    \label{fig: CLV_Exp}
\end{figure}

We run experiments for each candidate value under both cases treatment possible values, aiming to identify the optimal coupon value which are 4 and 2. In each experiment, we run all three estimators over 1,000 independent trajectories and record their empirical mean and standard error. The experiment horizon is adjusted so that the IS estimator's standard error is smaller than its empirical mean. In Figure \ref{fig: CLV_Exp}, we present the simulation results.

%Figure~\ref{fig: CLV_Exp} summarizes the simulation results, we immediately see:

\noindent\textbf{Unbiasedness across all estimators:} The empirical means for all three estimators in both experiments demonstrate negligible bias. This aligns with our theoretical results of the consistency of all three estimators.
%This aligns with our theoretical prediction of asymptotic normality for all three estimators.

\noindent\textbf{Variance reduction by IS:} In both examples, we observe a significant reduction in variance by at least half through the application of IS. The variance of IS is closer to the variance provided by PI.  This illustrates the effectiveness of IS in overcoming the lower bound of original A/B testing by utilizing the local structure. Also, the more states the information can be shared, the closer the variance of IS estimator is to the PI estimator.

%\noindent\textbf{Local Treatment is Harder than SST:} 
While ATEs in both experiments are close, the local treatment case requires a substantially longer horizon for the standard error of IS estimator to decay comparing with the SST case. This discrepancy aligns with the intuition that the SST allows information sharing for more states than the local treatment case. This observation showcases that the efficacy of IS increases with the number of shared states.

\subsection{A Real-World Sepsis Simulator}\label{sec: exp_sepsis}
To evaluate our method in a more realistic setting, we employ a simulator of sepsis treatment for intensive care unit (ICU) patients (\citealt{komorowski2018artificial,oberst2019counterfactual, tran2023inferring}). Our focus is on experimenting with various combinations of treatments applicable to patients in critical conditions. The primary objective is to understand the long-term effects of these treatment strategies.

\textbf{The Environment:} The physiological status of patients can be captured by a underlying MDP. The state space encompasses discrete measurements of vital signs (heart rate, blood pressure, oxygen concentration, and glucose levels), categorized into levels (e.g., low, normal, high), along with records of previous treatments. The size of the state space size 720. The reward function assigns values of -1, 0, or +1 based on the number of abnormal vital signs exhibited by a patient. In the absence of treatment intervention, patient signals fluctuate according to specified transition probabilities, reflecting the dynamic nature of the health condition.

\textbf{The Local Treatment:} When patients enter a critical state—characterized by all signals being abnormal—three treatment options become available: antibiotics, vasopressors, and mechanical ventilation. This results in $n=8$ test arms, corresponding to the possible combinations of these treatments. The application of these interventions alters the transition probabilities of the various signals. In our experimental setting, the total number of treated states amounts to 128, representing approximately 18\% of the overall state space.

\textbf{The Simulation:} Unlike the previous synthetic example, real-world constraints limit our ability to conduct extended experiments on one individual patient. Instead, we adopt a common practice in sepsis studies: assuming patients with similar contexts follow the same MDP. We pool observation data from a group of non-diabetic patients to estimate the average long-term effect of treatments. Our simulator generates observed trajectories of length 100 for each patient. These patients are initialized according to a state distribution learned a priori.

We run the simulations for $N=250$ patients over $100$ independent runs and then demonstrate the inference results of each estimators. We additionally introduce two more benchmarks, the Naive and \textit{Difference-in-Q} (DQ) estimators, from \cite{farias2022markovian}, which is originally designed for the general treatment policies on MDPs with two test arms. For the Naive estimators, we naively take the average on the rewards collected under the treatment policy and the control policy, ignoring the interference of state transitions. The DQ estimator strikes a balance between bias and variance under general and global treatment policies in MDPs. In Appendix~\ref{app: Sepsis_Exp}, we generalize the DQ theory and provide the expression of DQ estimator for multiple test arms.

\begin{figure}[h]
\centering
\begin{tabular}{l l }
        \includegraphics[width=.45\textwidth, height=.35\textwidth]{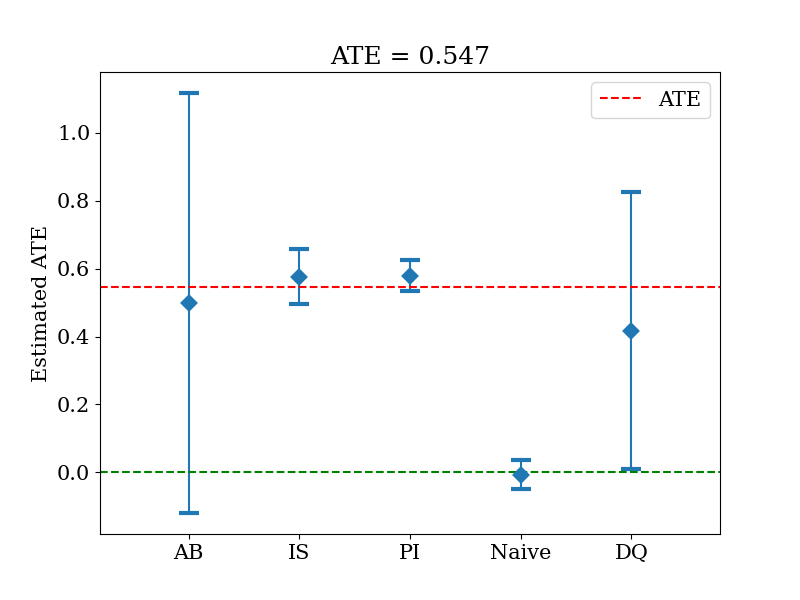}
        &
        \includegraphics[height=.35\textwidth]{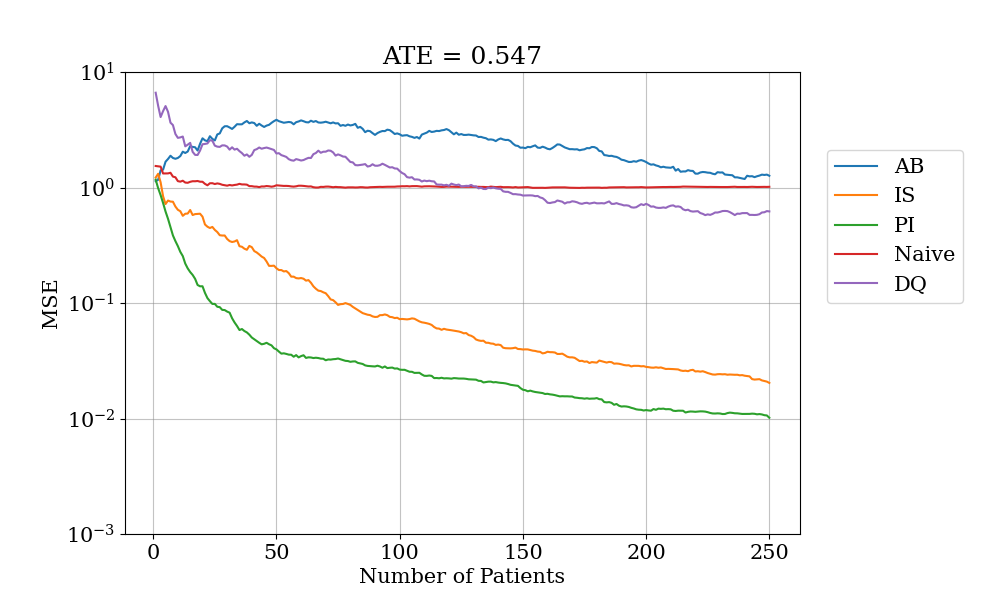}
    \end{tabular}
    \caption{Sepsis Simulator. Diamonds and caps represent the empirical mean and standard error respectively for each estimator. {\it Left}: Estimated ATE the best-performing arm among $n=8$ tested arms at patients $N=250$. {\it Right}: MSE of each estimator vs. $N$ the number of patients. IS, PI dominate over all experiment horizon.}
    \label{fig: Sepsis_Exp}
\end{figure}

Figure~\ref{fig: Sepsis_Exp} summarizes the simulation results as well as the Mean Square Error (MSE) curve with the number of patients. We have the following observations.

\noindent\textbf{Variance Reduction of IS:} We observe a significant reduction in variance through the application of IS while still obtaining negligible bias comparing with the A/B testing estimator. The variance of IS is even close to the PI estimator. The IS and PI estimators also enjoy fast decay of MSE compared to all other estimators. This confirms the effectiveness of IS and demonstrates the power of leveraging treatment structure to obtain striking efficiency improvement. %Moreover, aligning with our intuition, PI consistently outperforms IS, confirming its capability in further variance reduction with more information.

\noindent\textbf{Unbiasedness of IS:} Although the Naive estimator has smaller variance than IS, its bias is non-negligible. Furthermore, the Naive estimator struggles to exclude the null effect, potentially resulting in insufficient statistical power. In contrast, the IS and PI estimators are centered around the true ATE.

\noindent\textbf{DQ exhibits bias-variance trade-offs:} DQ demonstrates a notable bias-variance trade-off when compared to A/B testing and Naive estimators. DQ achieves variance reduction at the cost of introducing small bias, resulting in a faster decay of MSE. However, DQ fails to leverage the inherent local structures of the treatments, leading to significantly higher MSE compared to IS and PI. As we have discussed that DQ belongs to the general differentiable estimators, we can expect that Is can improve performance over the original DQ formulation.

\section{Discussion}\label{sec: discussion}
\subsection{Fairness and Welfare: The Monotonicity of SST}
MDPs are complex systems with interconnected states. Therefore, out of fairness and welfare concern, it would be the best for the decision-makers to ensure that a new treatment improves long-term outcomes for customers across all states, rather than benefiting some at the expense of others. Fortunately, we have a theorem that guarantees the ``monotonicity" of SST policies. 
\begin{theorem}[Monotonicity of Single-State Treatment Effect]\label{thm: monotonicity}
Consider SST policy $\pi^\ft$ with crucial state $s^1$. If there exists $s\in\gS$ such that $\Delta(s)>0$, then for all $s\ne s^1$, \[0\leq \Delta(s)< \gamma \Delta(s^1)\,.\]Vice versa, if there exists $s\in\gS$ such that $\Delta(s)<0$, then $\gamma \Delta(s^1)< \Delta(s)\leq 0$ for all $s\ne s^1\,.$
\end{theorem}
This ensures that the treatment effect remains consistently positive (or negative) for all states before and after the intervention. Furthermore, when the treatment effect is negative, all states except the crucial state will experience a welfare loss strictly less than $\gamma \Delta(s^1)$, where $\gamma\in[0,1)$ is the discount factor. Notably, this monotonicity theorem holds true regardless of the underlying Markov chain's structure.

Technically, we provide two different proofs for Theorem~\ref{thm: monotonicity} in Appendix~\ref{app: Monotonicity}. The first proof is based on an important observation that the treatment effect $\vDelta$ is an eigenvector of a special matrix associated with the transition probability under the control policy. The second proof offers another perspective to characterize $\vDelta$ using the advantage function, derived from the well-knwon \emph{Performance Difference Lemma} (\citealt{Kakade2002ApproximatelyOA, Agarwal2019ReinforcementLT}) in RL theory. We believe these techniques may be of independent interest to the RL literature.

% \subsection{Distribution Shift of Perfect Information}
\subsection{Other Applications of Information Sharing}
While this paper has mainly focused on A/B testing on a non-episodic MDP under the tabular setting, we believe it's interesting directions to extend our framework to various other RL settings.
%This approach naturally lends itself to generalization across other common RL settings.
% \newline\noindent\textbf{Average/Total Reward.} Under different definitions of the value function, the information sharing idea is still applicable whenever the new treatment is localized.
\newline\noindent\textbf{Function Approximations.} Modern practical reinforcement learning often faces the \emph{curse of dimensionality}, i.e. a prohibitively large or infinite state space. This may render our previous tabular methods of fitting value functions unrealistic. However, one can resort to any estimation methods designed for function approximation while still leveraging the information sharing idea. For example, consider a parameterized class of value functions $V_{\Theta}=\{V_{\vtheta}\colon \vtheta\in\Theta\}$, where common function approximations include linear (\citealt{tsitsiklis96analysis,jalaj21afinite}) or neural networks (\citealt{cai19neural}). Given a trajectory $\tau = \{(s_i,a_i,s_{i+1}, r_i)\}_{i\in[T]}$, typical TD algorithms solve
\[\vtheta = \argmin_{\vtheta\in\Theta} \sum_{s\in\gS} \pa{\sum_{s_i=s} r(s_i,a_i)+\gamma  V_{\vtheta}(s_{i+1})- V_{\vtheta}(s_i) }^2,\]
which can be viewed as a parameterized version of \Eqref{eq: LSTD}. When it comes to A/B testing, it's then evident that one can split trajectories to fit the value function by solving this optimization problem. Additionally, information can still be shared for local treatments, allowing the construction of richer datasets and improving efficiency.
\newline\noindent\textbf{Episodic Setting. }Another commonly used framework is the episodic MDP. In this scenario, multiple independent samples of the MDP can be obtained. In section~\ref{sec: exp_sepsis}, each patients can be viewed as one such independent MDP. We thus empirically demonstrate that IS is also effective for episodic MDPs and brings about substantial variance reduction. We leave it to future work to derive rigorous theoretical guarantees for information sharing under episodic MDPs.
\newline\noindent\textbf{Policy Optimization. }Most RL literature on finding optimal policy (e.g. \citealt{azar2017minimax,jin2018qlearning,li2024q}) does not impose structural assumption on the action set. Given our modeling of treatment locality, a natural question is whether this structure can also be leveraged to improve the sample efficiency for policy optimization.

% For instance, the company in Example~\ref{exp: CLV} might assume that certain customer groups share the same transition matrix. This allows for the observation of multiple trajectories, enabling the inference of the value function by aggregating data from these various episodes. Given this context, it's natural to consider extending the concept of Information Sharing as a more sophisticated approach to data pooling in these scenarios. This extension could potentially lead to more efficient and accurate value function estimation by leveraging similarities across different episodes while accounting for their unique characteristics.

\section{Concluding Remarks}\label{sec: conclusion}
This paper systematically explores how treatment structure can be utilized to improve the estimation of life-time effects. By exploiting treatment locality, our proposed estimators achieve substantial variance reduction, linear in the number of test arms, without sacrificing unbiasedness. We further show that a broad class of differentiable estimators can universally benefit from information sharing. These results not only establish new theoretical benchmarks but also highlight the practical value of local interventions for efficient experimentation in dynamic systems.

Finally, we want to highlight the trade-off between \emph{Robustness} and \emph{Efficiency}. The efficiency gain from information sharing relies heavily on the local structure of the new treatment. If the new local treatment on some states also influences transitions and rewards of the other states, our information sharing approach could introduce bias. In contrast, the inference methods that do not utilize the local structure, as introduced in Section~\ref{sec: general-ab-testing}, are always robust and valid, though potentially at the cost of reduced efficiency.

%Finally, we want to highlight the idea of a \emph{Robustness} v.s. \emph{Efficiency} trade-off. The efficiency gain from information sharing relies heavily on the local structure of the new treatment. If the new local treatment on some states also influences the transitions and the rewards, our information sharing idea would introduce bias. In contrast, the inference methods without utilizing the local structure introduced in Section~\ref{sec: general-ab-testing} are always robust and valid, albeit potentially at the cost of reduced efficiency.

%Our work incorporates the idea of a \emph{Robustness} v.s. \emph{Efficiency} trade-off. In practice, the effectiveness of our methods depends on the accuracy of the treatment's locality. If the understanding of locality is correct, our methods can substantially improve efficiency. However, if we share information for states that have different transitions or rewards under treatment and control policies, our methods may introduce bias. In contrast, conducting A/B testing without considering the treatment structure always ensures an unbiased estimation, albeit potentially at the cost of reduced efficiency.
% \section{Extension: Multiple Treatments}
% \section{Numerical Results}
% \section{Discussion and Concluding Remarks}

	% References here (outcomment the appropriate case)
	
	% CASE 1: BiBTeX used to constantly update the references
	%   (while the paper is being written).
	\bibliographystyle{informs2014} % outcomment this and next line in Case 1
	%\bibliography{<your bib file(s)>} % if more than one, comma separated
	
	% CASE 2: BiBTeX used to generate mypaper.bbl (to be further fine tuned)
	%\input{mypaper.bbl} % outcomment this line in Case 2
	
	%If you don't use BiBTex, you can manually itemize references as shown below.

	\bibliography{ref.bib}
	\clearpage

	% Appendix here
	% Options are (1) APPENDIX (with or without general title) or
	%             (2) APPENDICES (if it has more than one unrelated sections)
	% Outcomment the appropriate case if necessary
	%
	\ECSwitch
	
	%\ECHead{}

	\begin{center}
		\large{Online Appendix for ``Experimenting on Markov Decision Processes with Local Treatments''}
	\end{center}

    \section{Notation}
% Let $[N]\coloneqq\{1,2,\ldots, N\}$ for positive integer $N$. Given a trajectory $\{(s_i,a_i,s_{i+1})\}_{i\in[T]}$, we denote $I(T-i)\coloneqq\{i\mid s_i=x,i\in[T]\}.$ For simplicity, we shorthand $I(T)\coloneqq I(T-1).$

Let $\ve_i\in\R^n$ be a vector with all entries 0 but $i$-th entry 1 and $\mE_{ij}\in\R^{n\times n}$ be a matrix with all elements 0 but $i,j$-th element 1. For any vector $x$ and matrix $\mA$, the matrix norm $\|x\|_\mA\coloneqq \sqrt{x^\top \mA x}\,.$ For any matrix $\mA=(\va_1,\va_2,\ldots,\va_n)\in \R^{m\times n}$, we define the vectorization operator \[\vect (\mA)\coloneqq \begin{bmatrix}
 \va_1\\
 \va_2\\
 \vdots\\
\va_n
\end{bmatrix}\in\R^{mn}\] as the concatenation of its columns.

    \section{Fixed Policy Evaluation Algorithms} \label{app: algorithms}
We formally give the two policy evaluation algorithms for reference.
\begin{center}
\begin{algorithm}
\caption{Model-based Estimation}
\begin{algorithmic}[1]\label{alg: model-based}
\REQUIRE{The obtained trajectory $\tau = \{(s_i,a_i,s_{i+1}, r_i)\}_{i\in[T]}$; Discount factor $\gamma$.}
\STATE $N(s,a) \gets \sum_{i=1}^T \mathbb{I}\{s_i=s,a_i=a\}$, $N(s,a,s') \gets \sum_{i=1}^T \mathbb{I}\{s_i=s,a_i=a,s_{i+1}=s'\}$, $\forall s, s^\prime\in \gS, a\in\gA$
\STATE $\widehat r(s)\gets (\sum_{i=1}^T r_i \mathbb{I}\{s_i=s\})/(N(s)\vee 1)$, $\forall s\in \gS$ 
\STATE  $\widehat P(s^\prime\mid s,a) = N(s,a,s^\prime)/(N(s,a)\vee 1)$, $\forall s, s^\prime\in \gS, a\in\gA$
% \FOR{$i$ in $T$}
%     \STATE Execute $\pi$ and obtain $(s_i,a_i,s_{i+1}, r_i)$.
%     \STATE $N(s_i,a_i) \gets N(s_i)+1$.
%     \STATE $N(s_i,a_i, s_{i+1})\gets N(s_i,a_i, s_{i+1})$
%     \STATE $\widehat r(s_i,a_i)\gets \widehat r(s_i,a_i) + (r_i-\hat r(s_i,a_i))/N(s_i,a_i)$
% \ENDFOR
% \STATE $\widehat P(s^\prime\mid s,a) = N(s,a,s^\prime)/N(s,a)$ for all $(s,a,s^\prime)$.
\STATE $\widehat \mV^{\textnormal{MB}}\gets (\mI-\gamma \widehat \mP)^{-1}\widehat \vr$.
\RETURN $\widehat \mV^{\textnormal{MB}}$.
\end{algorithmic}
\end{algorithm}
\end{center}
\begin{center}
\begin{algorithm}
\caption{Model-free Estimation (Temporal Difference Learning)}
\begin{algorithmic}[1]\label{alg: model-free}
\REQUIRE{The obtained trajectory $\tau = \{(s_i,a_i,s_{i+1}, r_i)\}_{i\in[T]}$; Discount factor $\gamma$.}
%\STATE Execute policy $\pi$ for $T$ epochs and obtain trajectory $\tau = \{(s_i,a_i,s_{i+1}, r_i)\}_{i\in[T]}$.
\STATE Solve the following optimization problem \Eqref{eq: LSTD}, \[
     \min_{\widehat \mV}\sum_{s\in\gS} \pa{\sum_{s_i=s} r(s_i,a_i)+\gamma \widehat V(s_{i+1})-\widehat V(s_i) }^2.\]
%\[\min_{\widehat \mV}\sum_{s\in\gS} \pa{\sum_{i,s_i=s} r(s_i,a_i)+\gamma \widehat V(s_{i+1})-\widehat V(s_i) }^2\,.\]
\RETURN $\widehat \mV^{\textnormal{TD}}$.
\end{algorithmic}
\end{algorithm}
\end{center}

    \section{Proofs of Asymptotic Normality}
In this section, we will discuss the proof framework for asymptotic normality on Markov chains. We will start with the warm-up from of fixed policy inference in Proposition~\ref{prop: MB_Normal}. Then for A/B testing, we will directly prove Theorem~\ref{thm: general_est_variance compare} for general differentiable estimators, which implies the results for unbiased estimators in Theorem~\ref{thm: AB_Testing} and~\ref{thm: info_sharing}.

\subsection{Proof of Proposition~\ref{prop: MB_Normal}}\label{app: one_policy}
We introduce the following technical assumption on Markov chain, which is commonly used in literature (e.g. \cite{farias2022markovian,hu2022switchback,jones04CLT}).
\begin{definition}[Uniform Ergodicity]\label{def: uniform ergodic}
    A Markov chain with the state transition matrix $\mP$ is \emph{uniformly ergodic} if it has stationary distribution $\vmu$ and there exists some constant $C\geq 0$ and $0\leq\lambda<1$ such that for all 
    $k\in\mathbb N$ and all $s\in\gS$, 
    \[D_{\mathrm{TV}}(\mP^k(s,\cdot), \vmu)\leq C\lambda^k\,,\]
    where $D_{\mathrm{TV}}(\cdot,\cdot)$ is the total variation distance.
\end{definition}
Now we present the detailed version of Proposition~\ref{prop: MB_Normal}. 

\begin{theorem}[Asymptotic Normality]\label{thm: CLT_TD}
Assuming the Markov chain induced by policy $\pi$ is uniformly ergodic, with the input of a trajectory generated by a fixed policy $\pi$, it holds that as $T\to\infty$,
    %, i.e. \[\widehat \mV = (\mI-\widehat \mP)^{-1}\vr\,.\]
%    Then, 
    \[\sqrt{T}(\widehat \mV^{\textnormal{MB}} - \mV^\pi)\xrightarrow{d} \gN(\vzero, \Sigma_{\mathrm{MB}}),\]
     where\[\Sigma_{\mathrm{MB}} = (\mI-\gamma \mP^\pi)^{-1} \Diag_{s\in\gS}\pa{\sigma^2_\pi(s)/\mu^\pi(s) + \norm{\gamma\mV^\pi}_{\Sigma_{p^\pi(s)}}^2/\mu^\pi(s)} (\mI-\gamma \mP^\pi)^{-\top}\,.\]
\end{theorem}

\paragraph{Step 1: Markov Chain CLT}
To start with, for $X_i=(s_i,a_i, s_{i+1}, r_i)$, define
    \begin{align*}
    K(X_i) \coloneqq \mE_{s_i,s_{i+1}}\,, \qquad
    R(X_i) \coloneqq \ve_{s_i} r_i\,.
    \end{align*}
    And denote
    \begin{align*}
    \overline{ K_T} \coloneqq \frac{1}{T}\sum_{i=1}^T K(X_i)\,, \qquad
    \overline{R_T} \coloneqq \frac{1}{T}\sum_{i=1}^T R(X_i)\,.
    \end{align*}
    We also have$
    \overline{ K_T} \xrightarrow{a.s.} \mK_{\vmu^{\pi}} \coloneqq \Diag_{s\in\gS}({\vmu^{\pi}}(s)) \mP^\pi$ and $
    \overline{R_T} \xrightarrow{a.s.} \vr_{{\vmu^{\pi}}} \coloneqq \Diag_{s\in\gS}({\vmu^{\pi}}(s))\vr^\pi$.
    Then by Markov CLT in Lemma~\ref{thm: Multi_Markov_CLT} it holds
    \[\sqrt{T}\pa{\begin{bmatrix}
 \vect(\overline{ K_T})\\
\overline{R_T}
\end{bmatrix} - \begin{bmatrix}
\vect(\mK_{\vmu^{\pi}})\\
\vr_{\vmu^{\pi}}
\end{bmatrix}}\xrightarrow{d} \gN(\vzero, \Sigma)\,,\]for some covariance matrix $\Sigma$. 

\paragraph{Step 2: Delta Method}
To estimate $\widehat {\mP^\pi}$, notice that $\widehat {\mP^\pi}_{ij} = \overline{K_T}(i,j)/(\sum_k \overline{K_T}(i,k))$. Then $\widehat {\mP^\pi}$ is actually a function of $\overline{K_T}$. Similarly, to estimate $\widehat {\vr^\pi}$, we have $\widehat {\vr^\pi}(i) = \overline{R_T}(i)/(\sum_k \overline{K_T}(i,k))$. Recall the model-based estimation, we have $\widehat\mV^{\mathrm{MB}} = (\mI-\widehat {\mP^\pi})^{-1}\widehat {\vr^\pi} = g(\overline{ K_T},
\overline{R_T})$. We then apply delta method and obtain
\[\sqrt{T}(g(\overline{ K_T},
\overline{R_T}) - g(\mK_{\vmu^{\pi}}, \vr_{\vmu^{\pi}}))\xrightarrow{d}\gN(\vzero, \nabla g^\top \Sigma \nabla g)\,.\]

\paragraph{Step 3: Linearization} In general, $\Sigma$ is hard to compute. We thus apply the linearization technique introduced in \cite{farias2022markovian}. Specifically, define the linearized part
\begin{align*}
G(X_i)&=\nabla g(\mK_{\vmu^{\pi}}, \vr_{\vmu^{\pi}})^\top \begin{bmatrix}
 \vect(K(X_i))\\
R(X_i)
\end{bmatrix}\\
&= \underbrace{\sum_{a=1}^n \left.\frac{\partial g}{\partial r_a} \right\vert_{\mK_{\vmu^{\pi}}, \vr_{\vmu^{\pi}}} R(X_i)(a)}_{\texttt{Linear-Reward}}
+ \underbrace{\sum_{a,b}\left.\frac{\partial g}{\partial K(a,b)} \right\vert_{\mK_{\vmu^{\pi}}, \vr_{\vmu^{\pi}}} K(X_i)(a,b)}_{\texttt{Linear-Transition}}\,.
\end{align*}
Then by Lemma~\ref{lem: Linearization}, we have 
\[\Var\pa{\frac{1}{T}\sum_{i=1}^TG(X_i)} = \nabla g^\top \Sigma \nabla g\,.\]
It's easy to see that the \texttt{Linear-Reward} as 
\[\sum_{a=1}^n \left.\frac{\partial g}{\partial R(a)} \right\vert_{\mK_{\vmu^{\pi}}, \vr_{\vmu^{\pi}}} R(X_i)(a) = (\mI-\gamma \mP^\pi)^{-1}\Diag_{s\in\gS}(1/{\vmu^{\pi}}(s))R(X_i)\,.\]
For the \texttt{Linear-Transition} term, it holds
\begin{align*}
    &\sum_{a,b}\left.\frac{\partial g}{\partial K(a,b)} \right\vert_{\mK_{\vmu^{\pi}}, \vr_{\vmu^{\pi}}} K(X_i)(a,b)\\
    =&\sum_{a,b}\left.\frac{\partial (\mI-\gamma\widehat {\mP^\pi})^{-1}\vr}{\partial K(a,b)} \right\vert_{\mK_{\vmu^{\pi}}, \vr_{\vmu^{\pi}}} K(X_i)(a,b) + \sum_{a,b}\left.\frac{\partial (\mI-\gamma\mP^\pi)^{-1}\widehat \vr}{\partial K(a,b)} \right\vert_{\mK_{\vmu^{\pi}}, \vr_{\vmu^{\pi}}} K(X_i)(a,b)\\
    =&  \sum_{a,b} - (\mI-\gamma\mP^\pi)^{-1} \left.\frac{\partial (\mI-\gamma \mP^\pi))}{\partial K(a,b)} \right\vert_{\mK_\mu} \mV^\pi K(X_i)(a,b) -\frac{1}{\mu^\pi(s_i)}(\mI-\gamma\mP^\pi)^{-1}\begin{bmatrix}
        \vzero\\
        r(s_i)\\
        \vzero
    \end{bmatrix}\\
%     = & \gamma (\mI-\gamma\mP^\pi)^{-1} \pa{\sum_{a,b} -\frac{1}{\mu^\pi(s_a)} \begin{bmatrix}
%   &   \vzero   & \\
%  \mathbb{P}(s_1\mid s_a) \dots & \mathbb{P}(s_b\mid s_a)-1 \dots & \mathbb{P}(s_n\mid s_a) \\
%   &   \vzero   &
% \end{bmatrix} (K(X_i)(a,b) - \mK_\mu^\pi(a,b))} (\mI-\gamma\mP^\pi)^{-1}  \vr\\
 = & \gamma (\mI-\gamma\mP^\pi)^{-1} \pa{\sum_{a,b} \left.\frac{\partial \mP^\pi}{\partial K(a,b)} \right\vert_{\mK_\mu}  K(X_i)(a,b) } \mV^\pi-\frac{1}{\mu^\pi(s_i)}(\mI-\gamma\mP^\pi)^{-1}\begin{bmatrix}
        \vzero\\
        r(s_i)\\
        \vzero
    \end{bmatrix}\\
 = & -\frac{\gamma}{\mu^\pi(s_i)}  (\mI-\gamma\mP^\pi)^{-1} \begin{bmatrix}
  &   \vzero   & \\
 \mathbb{P}(s^1\mid s_i) \dots & \mathbb{P}(s^{i+1}\mid s_i)-1 \dots & \mathbb{P}(s^K\mid s_i) \\
  &   \vzero   &
\end{bmatrix} \mV^\pi-\frac{1}{\mu^\pi(s_i)}(\mI-\gamma\mP^\pi)^{-1}\begin{bmatrix}
        \vzero\\
        r(s_i)\\
        \vzero
    \end{bmatrix}\,.
\end{align*}
Then we have
\begin{align*}
G(X_i) &= \underbrace{-\frac{\gamma}{\mu^\pi(s_i)}  (\mI-\gamma\mP^\pi)^{-1} \begin{bmatrix}
  &   \vzero   & \\
 \mathbb{P}(s^1\mid s_i) \dots & \mathbb{P}(s^{i+1}\mid s_i)-1 \dots & \mathbb{P}(s^K\mid s_i) \\
  &   \vzero   &
\end{bmatrix} \mV^\pi}_{G_1}\\
&\qquad +\underbrace{\frac{1}{\mu^\pi(s_i)}(\mI-\gamma\mP^\pi)^{-1}\begin{bmatrix}
        \vzero\\
        r_i-r(s_i)\\
        \vzero
    \end{bmatrix}}_{G_2}\,.
\end{align*}
It remains to compute $\Sigma_G$. Note that $\E\br{G(X_k)\mid X_1} = \vzero$ for $k\geq 2$, thus we have $\Cov[G(X_1), G(X_k)]= \Cov[G(X_k), G(X_1)] = \vzero$. Then it holds 
\begin{align*}
    \Sigma_G & = \Var (G(X_1)) + \sum_{i=1}^\infty \Cov[G(X_1), G(X_k)] + \sum_{i=1}^\infty \Cov[G(X_k), G(X_1)]\\
    & = \Var (G(X_1))\\
    &= \Var (G_1(X_1)) + \Var (G_2(X_1))\,,
\end{align*}
with the last equation due to the fact that $\E[G_2(X_1)\mid X_1]=\vzero\,.$

To see $\Var(G_2(X_1))$, we have
\begin{align*}
    \Var(G_2(X_1))  = (\mI-\gamma \mP^\pi)^{-1} \Diag_{s\in\gS} \pa{\sigma_\pi^2(s)/{\vmu^{\pi}}(s)} (\mI-\gamma \mP^\pi)^{-\top}\,.
\end{align*}
For $\Var(G_2(X_1))$, it holds
\begin{align*}
    \Var(G_2(X_1)) &= (\mI-\gamma \mP^\pi)^{-1} \Var(-\frac{\gamma}{\mu^\pi(s_i)}\begin{bmatrix}
  &   \vzero   & \\
 \mathbb{P}(s^1\mid s_i) \dots & \mathbb{P}(s^{i+1}\mid s_i)-1 \dots & \mathbb{P}(s^K\mid s_i) \\
  &   \vzero   &
\end{bmatrix} \mV^\pi)(\mI-\gamma \mP^\pi)^{-\top}\\
&= (\mI-\gamma \mP^\pi)^{-1} \E_{s_1\sim \vmu}\br{\frac{\gamma^2}{\mu^\pi(s_i)^2} \Var_{s_2\sim \vp(\cdot\mid s_1)}\pa{\begin{bmatrix}
    \vzero\\
    V^\pi(s_2)\\
    \vzero
\end{bmatrix}} } (\mI-\gamma \mP^\pi)^{-\top}\\
& = (\mI-\gamma \mP^\pi)^{-1} \Diag_{s\in\gS}\pa{\norm{\gamma\mV^\pi}_{\Sigma_p(s)}^2/{\vmu^{\pi}}(s)} (\mI-\gamma \mP^\pi)^{-\top}\,.
\end{align*}

Hence the final covariance matrix is given by 
\[\Sigma_{\mathrm{MB}} = (\mI-\gamma \mP^\pi)^{-1} \Diag_{s\in\gS}\pa{\sigma_\pi^2(s)/{\vmu^{\pi}}(s) + \norm{\gamma\mV^\pi}_{\Sigma_p(s)}^2/{\vmu^{\pi}}(s)} (\mI-\gamma \mP^\pi)^{-\top}\,.\]

\subsection{Proof of Proposition~\ref{thm: general_est_variance compare}}\label{app: General Estimator}
We notice that for both estimation $\overline{ K^a_\Gamma},\overline{ R^a_\Gamma}$ defined in section~\ref{sec: general-estimator}, it holds $a\in\{t,c\}$
    \begin{align*}
    \overline{ K^a_\Gamma} \xrightarrow{a.s.} \mK_{\vmu_{1/2}}^a = \Diag(\vmu_{1/2}) \mP^a \,,\qquad
    \overline{R^a_\Gamma} \xrightarrow{a.s.} \vr^a_{\vmu_{1/2}} = \Diag(\vmu_{1/2})\vr^a \,.
    \end{align*}
Thus, it is clear that $\widehat{\Delta}^{\textrm{AB}}$ and $\widehat{\Delta}^{\textrm{IS}}$ shares the same asymptotic bias. We then focus on the asymptotic variance. Let \[\vu_\Gamma(X_i)\coloneqq\begin{bmatrix}
 \vect( K^\ft_\Gamma(X_i))\\
R^\ft_\Gamma(X_i)\\
\vect( K^\fc_\Gamma(X_i))\\
R^\fc_\Gamma(X_i)
\end{bmatrix}\,.\] By Markov chain CLT, we can have \[\sqrt{T}\left(\frac{1}{T}\sum_{i=1}^T\vu(X_i)-\E[\vu(X_1)]\right)\xrightarrow{d}\gN(0,\Sigma_\vu)\] where $\Sigma_\vu=\Var_{X_1\sim\mu} \br{\vu(X_1)} +  \sum_{i=1}^\infty \Cov_{X_1\sim\mu}\br{\vu(X_1), \vu(X_{1+i})} + \sum_{i=1}^\infty \Cov_{X_1\sim\mu}\br{\vu(X_{1+i}), \vu(X_{1})}$. Given that $f$ is differentiable, we can apply delta method, which can lead to that $\widehat{\Delta}^{\Gamma}$ is asymptotically normal for both $\Gamma\in\{\textrm{AB},\textrm{IS}\}$. The variance is given by 
\[\Sigma_f^\Gamma= \nabla f^\top\Sigma_\vu^\Gamma\nabla f\,.\]
It suffices to compare $\Sigma_f^\textrm{AB}$ and $\Sigma_f^\textrm{IS}$. For $a_i=a\in\{\ft,\fc\}$, we denote \[\mL^a_i\coloneqq\mL^a(X_i)\coloneqq \begin{bmatrix}
    \vect{(\mE_{s_i,s_{i+1}})}\\
    \ve_{s_i}r_i
\end{bmatrix}\] 
and we can rewrite
\[\vu_\AB(X_i)=\begin{bmatrix}
    2\mathbf{1}[a_i=\ft] \mL^\ft_i(s^1)\\
    2\mathbf{1}[a_i=\ft] \mL^\ft_i(s^{-1})\\
    2\mathbf{1}[a_i=\fc] \mL^\fc_i(s^1)\\
    2\mathbf{1}[a_i=\fc] \mL^\fc_i(s^{-1})
\end{bmatrix}\,,\qquad \vu_\IS(X_i)=\begin{bmatrix}
    2\mathbf{1}[a_i=\ft] \mL^\ft_i(s^1)\\
     \mL^\fc_i(s^{-1})\\
    2\mathbf{1}[a_i=\fc] \mL^\fc_i(s^1)\\
    \mL^\fc_i(s^{-1})
\end{bmatrix}\,.\]
For the original A/B testing where we do not share data, we have
\begin{align*}
    \Var(\vu_\AB(X))=\begin{bmatrix}
        2\Var(\mL^\ft)+(\E \mL^\ft)^2 & -\E\mL^\ft (\E\mL^\fc)^\top\\
        -\E\mL^\fc (\E\mL^\ft)^\top & 2\Var(\mL^\fc)+(\E \mL^\fc)^2
    \end{bmatrix}
\end{align*}
and for $k\geq 2$,
\begin{align*}
    % \Cov(\vu_\AB(X_1),\vu_\AB(X_k))=\begin{bmatrix}
    %     \E[\mL^\ft_1 \mL^{\ft^\top}_k]-\E[\mL^\ft_1]\E[\mL^{\ft}_k]^\top & \E[\mL^\ft_1 \mL^{\fc^\top}_k]-\E[\mL^\ft_1]\E[\mL^{\fc}_k]^\top\\
    %     \E[\mL^\fc_1 \mL^{\ft^\top}_k]-\E[\mL^\fc_1]\E[\mL^{\ft}_k]^\top & \E[\mL^\fc_1 \mL^{\fc^\top}_k]-\E[\mL^\fc_1]\E[\mL^{\fc}_k]^\top
    % \end{bmatrix}
    \Cov(\vu_\AB(X_1),\vu_\AB(X_k))_{\ft,  \ft}=\E[\mL^\ft_1 \mL^{\ft^\top}_k\mid a_1=\ft]-\E[\mL^\ft_1]\E[\mL^{\ft}_k]^\top\,.
\end{align*}
Then for information-sharing, we further denote $\vl^a_j = \mL^a(s^j)$ and we have
\begin{align*}
    \Var(\vu_\IS(X))=\begin{bmatrix}
        2\Var(\vl^\ft_1)+(\E \vl^\ft_{1})^2 & \Cov(\vl^\ft,\vl_{-1}) & -\E\vl^\ft_1 \E\vl^\fc_1 & \Cov(\vl^\ft_1,\vl_{-1})\\
        \Cov(\vl_{-1}, \vl^\ft) & \Var(\vl_{-1}) & \Cov(\vl_{-1}, \vl^\fc) & \Var(\vl_{-1})\\
        -\E\vl^\fc_1 \E\vl^\ft_1 & \Cov(\vl^\fc,\vl^{-1}) & 2\Var(\vl^\ft_1)+(\E \vl^\fc_{1})^2 & \Cov(\vl^\fc,\vl^{-1})\\
        \Cov(\vl^{-1},\vl^\ft_1) & \Var(\vl_{-1}) & \Cov(\vl_{-1},\vl^\fc) & \Var(\vl^{-1})
    \end{bmatrix}\,,
\end{align*}
and for $k\geq 2$,
\begin{align*}
    % \Cov(\vu_\IS(X_1),\vu_\IS(X_k))=\begin{bmatrix}
    %     \Cov(\mL^\ft_1, \mL^\ft_k) & \Cov(\mL^\ft_1, \mL^\fc_k)\\
    %     \Cov(\mL^\fc_1, \mL^\ft_k) & \Cov(\mL^\fc_1, \mL^\fc_k)
    % \end{bmatrix}
\Cov(\vu_\IS(X_1),\vu_\IS(X_k))_{\ft,\ft}=\begin{bmatrix}
        \Cov(2\mathbf{1}[a_1=\ft] \mL^\ft_1(s^1), 2\mathbf{1}[a_k=\ft] \mL^\ft_k(s^1)) & \Cov(2\mathbf{1}[a_1=\ft] \mL^\ft_1(s^1), \mL^\fc_k(s^{-1}))\\
        \Cov(\mL^\fc_1(s^{-1}), 2\mathbf{1}[a_k=\ft] \mL^\ft_k(s^1)) & \Cov(\mL^\fc_1(s^{-1}), \mL^\fc_k(s^{-1}))
    \end{bmatrix}
\end{align*}
Note that 
\begin{align*}
    \Cov(\mL^\fc_1(s^{-1}), 2\mathbf{1}[a_k=\ft] \mL^\ft_k(s^1))&=\E[\mL^\fc_1(s^{-1})\mL^\ft_k(s^1)^\top]-\E[\mL^\fc_1(s^{-1})]\E[\mL^\ft_k(s^1)]^\top\\
    &=\mu(s_1\in s^{-1})\E[\mL^\fc_1(s^{-1})\mL^\ft_k(s^1)^\top\mid s_1\in s^{-1}]-\E[\mL^\fc_1(s^{-1})]\E[\mL^\ft_k(s^1)]^\top\\
    &\overset{(i)}{=} \mu(s_1\in s^{-1})\E[\mL^\fc_1(s^{-1})\mL^\ft_k(s^1)^\top\mid s_1\in s^{-1}, a_1=t]-\E[\mL^\fc_1(s^{-1})]\E[\mL^\ft_k(s^1)]^\top\\
    &= \E[\mL^\fc_1(s^{-1})\mL^\ft_k(s^1)^\top\mid a_1=t]-\E[\mL^\fc_1(s^{-1})]\E[\mL^\ft_k(s^1)]^\top\,,
\end{align*}
where $(i)$ follows from the definition of SST such that $P(s^\prime\mid s,\ft)=P(s^\prime\mid s,\fc)$ for all state $s\in s^{-1}$. Similarly, one can verify that the two covariance for $\AB$ and $\IS$ are the same, i.e. $\Cov(\vu_\AB(X_1),\vu_\AB(X_k))=\Cov(\vu_\IS(X_1),\vu_\IS(X_k))$. In other words, information-sharing will not shave off these covariance and we only need to focus on $\Var(\vu_\AB(X))$ and $\Var(\vu_\IS(X))$.
Notice that $\E[\vl^a_1\vl_{-1}^\top]=\vzero$, we have
\begin{align*}
    \Var(\vu_\AB(X))-\Var(\vu_\IS(X))=\begin{bmatrix}
        0&0&0&0\\
        0&\E[\vl_{-1}\vl_{-1}^\top] &0& -\E[\vl_{-1}\vl_{-1}^\top]\\
        0&0&0&0\\
        0&-\E[\vl_{-1}\vl_{-1}^\top]&0&\E[\vl_{-1}\vl_{-1}^\top]
    \end{bmatrix}\,.
\end{align*}
Hence we conclude that all the row and columns and rows associated with the crucial state $s^1$ turns out to be $0$. In other words, variance associated with $s^1$ cannot be reduced. This matches the previous result in Theorem~\ref{thm: VR_two_arm} where we focus on a specific unbiased estimator.

For the variance associated with all other state $s^{-1}$, one can easily verify \[\Var(\vu_\AB(X))-\Var(\vu_\IS(X))\succeq 0\,.\]
This demonstrates that information sharing can always reduce variance for any differentiable estimators. Finally, we conclude our results in the following theorem.
% \begin{theorem}
%     Let $\widehat{\Delta}^{\Gamma}=f(\overline{ K^\ft_\Gamma},\overline{ K^\fc_\Gamma},\overline{R^\ft_\Gamma},\overline{R^\fc_\Gamma})$ for $\Gamma \in \{\AB,\IS\}$ be any differentiable estimator of any ATE $\Delta$. Then $\widehat{\Delta}^{\AB}$ and $\widehat{\Delta}^{\IS}$ are both asymptotic normal with the same asymptotic bias. Furthermore, their asymptotic variance can be decomposed as $\Sigma_\Gamma=\Sigma^{\mathrm{het}}_\Gamma + \Sigma^{\mathrm{hom}}_\Gamma+\Sigma^{\mathrm{cov}}_\Gamma$ where \begin{equation}\label{eq: general estimator}
%         \Sigma^{\mathrm{het}}_{\mathrm{IS}} = \Sigma^{\mathrm{het}}_{\mathrm{AB}}\,,\quad \Sigma^{\mathrm{cov}}_{\mathrm{IS}} = \Sigma^{\mathrm{cov}}_{\mathrm{AB}}\,,\textrm{ and } \quad\Sigma^{\mathrm{hom}}_{\mathrm{IS}} \preceq \Sigma^{\mathrm{hom}}_{\mathrm{AB}}.
%         \end{equation}
% \end{theorem}
We can similarly extend our result for general estimators to A/B/n testing,
\begin{theorem}
    Let $\widehat{\Delta}^{\Gamma}_{i,j}=f(\overline{ K^i_\Gamma},\overline{ K^j_\Gamma},\overline{R^i_\Gamma},\overline{R^j_\Gamma})$ for $\Gamma \in \{\mathrm{ABn},\mathrm{ISn}\}$ be any differentiable estimator of any ATE $\Delta_{i,j}$. Then $\widehat{\Delta}^{\mathrm{ABn}}_{i,j}$ and $\widehat{\Delta}^{\mathrm{ISn}}_{i,j}$ are both asymptotic normal with the same asymptotic bias. Furthermore, their asymptotic variance can be decomposed as $\Sigma_\Gamma(i,j)=\Sigma^{\mathrm{het}}_\Gamma(i,j) + \Sigma^{\mathrm{hom}}_\Gamma(i,j)+\Sigma^{\mathrm{cov}}_\Gamma(i,j)$ where \begin{equation}\label{eq: general estimator ABn}
        \Sigma^{\mathrm{het}}_{\mathrm{IS}}(i,j) = \Sigma^{\mathrm{het}}_{\mathrm{AB}}(i,j)\,,\quad \Sigma^{\mathrm{cov}}_{\mathrm{IS}}(i,j) = \Sigma^{\mathrm{cov}}_{\mathrm{AB}}(i,j)\,,\textrm{ and } \quad\Sigma^{\mathrm{hom}}_{\mathrm{IS}}(i,j) \preceq \frac{2}{n}\Sigma^{\mathrm{hom}}_{\mathrm{AB}}(i,j).
        \end{equation}
\end{theorem}

\subsection{Proof of Theorem~\ref{thm: AB_Testing}}\label{app: proof_ab_testing}
We first carefully state the detailed version of Theorem~\ref{thm: AB_Testing} and then show the proof.

\begin{theorem}[Asymptotic Normality of A/B Testing]\label{thm: app_AB_Testing}
    Assuming the Markov chain induced by the experiment policy $\pi^{1/2}$ is uniformly ergodic with stationary distribution $\vmu_{1/2}\coloneqq (\mu_{1/2}(s^1),\cdots,\mu_{1/2}(s^K))^\top$, it holds that as $T\to\infty$,
    \[\sqrt{T}\pa{\widehat \vDelta ^{\mathrm{AB}} - \vDelta}\xrightarrow{d}\gN(\vzero, \Sigma_{\mathrm{AB}})\,,\]
    where 
    \begin{align*}
        \Sigma_{\mathrm{AB}}& = 2(\mI-\gamma \mP^\ft)^{-1} \Diag_{s\in\gS}\pa{\frac{\sigma_\ft^2(s)}{\mu_{1/2}(s)} + \frac{\norm{\gamma\mV^\ft}_{\Sigma_{p^\ft}(s)}^2}{\mu_{1/2}(s)}} (\mI-\gamma \mP^\ft)^{-\top} \\
&\qquad + 2(\mI-\gamma \mP^\fc)^{-1} \Diag_{s\in\gS}\pa{\frac{\sigma_\fc^2(s)}{\mu_{1/2}(s)} +\frac{ \norm{\gamma\mV^\fc}_{\Sigma_{p^\fc}(s)}^2}{\mu_{1/2}(s)}} (\mI-\gamma \mP^\fc)^{-\top}\,.
    \end{align*} 
\end{theorem}

Following the Step 2 in Appendix~\ref{app: one_policy}, for $a\in\{\ft,\fc\}$, define 
    \begin{align*}
        G^a(X_i)&\coloneqq \nabla g^a(\mK^a_{\vmu^{1/2}}, \vr^a_{\vmu^{1/2}})^\top \begin{bmatrix}
 \vect(K^a(X_i))\\
R^a(X_i)
\end{bmatrix}\\
        &=\underbrace{-2\mathbf{1}[a_i=a]\frac{\gamma}{\mu_{1/2}(s_i)}  (\mI-\gamma\mP^a)^{-1} \begin{bmatrix}
  &   \vzero   & \\
 P^a(s^1\mid s_i) \dots & P^a(s_{i+1}\mid s_i)-1 \dots & P^a(s^K\mid s_i) \\
  &   \vzero   &
\end{bmatrix} \mV^a}_{G_1^a}\\
&\qquad +\underbrace{\frac{2\mathbf{1}[a_i=a]}{\mu_{1/2}(s_i)}(\mI-\gamma\mP^a)^{-1}\begin{bmatrix}
        \vzero\\
        r_i-r(s_i,a)\\
        \vzero
    \end{bmatrix}}_{G_2^a}\,.
    \end{align*}
Following the similar framework as argued in Appendix~\ref{app: one_policy}, one can verify that linearized part is $G(X_i)= G^t(X_i)-G^c(X_i)$.
Again, it can be verified again that $\E\br{G(X_k)\mid X_1} = \vzero$ for $k\geq 2$, thus we have $\Cov[G(X_1), G(X_k)]= \Cov[G(X_k), G(X_1)] = \vzero$. Then it holds 
$\Sigma_G  = \Var (G(X_1)).$

To see $\Var\pa{G(X_1)}$, it holds
\begin{align*}
    \Var\pa{G(X_1)} &=  \Var\pa{G^t(X_1) - G^c(X_1)}\\
& = 2(\mI-\gamma \mP^\ft)^{-1} \Diag_{s\in\gS}\pa{\sigma_\ft^2(s)/\mu_{1/2}(s) + \norm{\gamma\mV^\ft}_{\Sigma_{p^\ft}(s)}^2/\mu_{1/2}(s)} (\mI-\gamma \mP^\ft)^{-\top} \\
&\qquad + 2(\mI-\gamma \mP^\fc)^{-1} \Diag_{s\in\gS}\pa{\sigma_\fc^2(s)/\mu_{1/2}(s) + \norm{\gamma\mV^\fc}_{\Sigma_{p^\fc}(s)}^2/\mu_{1/2}(s)} (\mI-\gamma \mP^\fc)^{-\top}\,.
\end{align*}

Hence the final covariance matrix is given by 
\[\sigma_{\mathrm{AB}}^2(s^1) = \ve_1^\top\Sigma_G \ve_1 = 2 \sum_{s\in\gS}\frac{\rho^\ft_{s^1}(s)^2}{\mu_{1/2}(s)}\Big(\sigma_\ft^2(s) + \norm{\gamma\mV^\ft}_{\Sigma_{p^\ft}(s)}^2\Big) + 
    2 \sum_{s\in\gS}\frac{\rho^\fc_{s^1}(s)^2}{\mu_{1/2}(s)}\pa{\sigma_\fc^2(s) + \norm{\gamma\mV^\fc}_{\Sigma_{p^\fc}(s)}^2} \,.\]

\subsection{Proof of Theorem~\ref{thm: info_sharing}}\label{app: CLT_IS}
The asymptotic variance $\Sigma_\IS$ is given by
\begin{align*}
        \Sigma&_{\mathrm{IS}}  = 2(\mI-\gamma \mP^\ft)^{-1} 
        \begin{bmatrix}
            \frac{\sigma_\ft^2(s^1)}{\mu_{1/2}(s^1)} + \frac{\norm{\gamma\mV^\ft}_{\Sigma_{p^\ft}(s^1)}^2}{\mu_{1/2}(s^1)} &\\
            &\vzero 
        \end{bmatrix}
         (\mI-\gamma \mP^\ft)^{-\top} \\
&\qquad + 2(\mI-\gamma \mP^\fc)^{-1} 
\begin{bmatrix}
    \frac{\sigma_\fc^2(s^1)}{\mu_{1/2}(s^1)} +\frac{ \norm{\gamma\mV^\fc}_{\Sigma_{p^\fc}(s^1)}^2}{\mu_{1/2}(s^1)}&\\
    &\vzero
\end{bmatrix}
 (\mI-\gamma \mP^\fc)^{-\top}\\
        &\qquad + \sum_{s\ne s^1} \frac{1}{\mu_{1/2}(s)}\pa{\gamma\vrho^\ft(s) \mV^{\ft^\top} -  \gamma\vrho^\fc(s) \mV^{\fc^\top}}\Sigma_{p^\fc(s)} \pa{\gamma \mV^\ft\vrho^\ft(s)^\top -  \gamma \mV^\fc\vrho^\fc(s)^\top}\\
&\qquad + \sum_{s\ne s^1} \frac{\sigma_\fc^2(s)}{\mu_{1/2}(s)} \pa{\vrho^\ft(s)-\vrho^\fc(s)} \pa{\vrho^\ft(s)-\vrho^\fc(s)}^\top\,,
\end{align*}
where ${\vrho^a}(s) = [\rho^a_{s^1}(s), \rho^a_{s^2}(s),\ldots,\rho^a_{s^K}(s)]^\top$ for any $s\in\gS$ and $a\in\{\ft,\fc\}\,.$

Following the Step 2 in Appendix~\ref{app: one_policy}, for $a\in\{\ft,\fc\}$, define 
    \begin{align*}
        G^a(X_i)&\coloneqq \nabla g^a(\mK^a_{\vmu^{1/2}}, \vr^a_{\vmu^{1/2}})^\top \begin{bmatrix}
 \vect(K^a(X_i))\\
R^a(X_i)
\end{bmatrix}\\
        &= \underbrace{-\bigg(2\mathbf{1}[a_i=a, s_i=x] + \mathbf{1}[s_i\ne s^1]\bigg)\frac{\gamma}{\mu_{1/2}(s_i)}  (\mI-\gamma\mP^a)^{-1} \begin{bmatrix}
    \vzero    \\
 \sum_sP^a(s\mid s_i)V^a(s)-V^a(s_{i+1})\\
   \vzero   
\end{bmatrix} }_{G_1^a}\\
&\qquad +\underbrace{\frac{2\mathbf{1}[a_i=a, s_i=s^1] + \mathbf{1}[s_i\ne s^1]}{\mu_{1/2}(s_i)}(\mI-\gamma\mP^a)^{-1}\begin{bmatrix}
        \vzero\\
        r_i-r(s_i)\\
        \vzero
    \end{bmatrix}}_{G_2^a}\,.
    \end{align*}
    Following similar argument of unbiased A/B testing, one can verify that linearized part is $G(X_i)= G^\ft(X_i)-G^\fc(X_i)$. It can be verified again that $\E\br{G(X_k)\mid X_1} = \vzero$ for $k\geq 2$, thus we have $\Cov[G(X_1), G(X_k)]= \Cov[G(X_k), G(X_1)] = \vzero$. Then it holds $\Sigma_G  = \Var (G(X_1))$

To see $\Var\pa{G(X_1)}$, it holds
\begin{align*}
    \Var\pa{G(X_1)} &=  \Var\pa{G^\ft(X_1) - G^\fc(X_1)}\\
& = 2\frac{\sigma_\ft^2(s^1)+\norm{\gamma\mV^\ft}_{\Sigma_{p^\ft(s^1)}}^2}{\mu_{1/2}(s^1)} {\vrho^\ft}(s^1) {\vrho^\ft}(s^1)^\top\\
&\qquad + 2\frac{\sigma_\fc^2(s^1)+\norm{\gamma\mV^\fc}_{\Sigma_{p^\fc(s^1)}}^2}{\mu_{1/2}(s^1)}{\vrho^\fc}(s^1) {\vrho^\fc}(s^1)^\top\\
&\qquad + \sum_{s\ne s^1} \frac{1}{\mu_{1/2}(s)}\pa{\gamma\vrho^\ft(s) \mV^{t^\top} -  \gamma\vrho^\fc(s) \mV^{\fc^\top}}\Sigma_{p^\fc(s)} \pa{\gamma \mV^\ft\vrho^\ft(s)^\top -  \gamma \mV^\fc\vrho^\fc(s)^\top}\\
&\qquad + \sum_{s\ne s^1} \frac{\sigma_\fc^2(s)}{\mu_{1/2}(s)} \pa{\vrho^\ft(s)-\vrho^\fc(s)} \pa{\vrho^\ft(s)-\vrho^\fc(s)}^\top\,,
\end{align*}
where ${\vrho^a}(s) = [\rho^a_{s^1}(s), \rho^a_{s_1}(s),\ldots,\rho^a_{s_n}(s)]^\top$ for any $s\in\gS$ and $a\in\{t,c\}\,.$

Hence the final covariance matrix is given by 
\begin{align*}
        \sigma_{\mathrm{IS}}^2 =& \ve_1^\top\Sigma_G \ve_1 \\
        =&2 \pa{\frac{\rho^\ft_{s^1}(s^1)^2}{\mu_{1/2}(s^1)}\Big(\sigma_\ft^2(s^1) + \norm{\gamma\mV^\ft}_{\Sigma_{p^\ft}(s^1)}^2\Big) +  \frac{\rho^\fc_{s^1}(s^1)^2}{\mu_{1/2}(s^1)}\pa{\sigma_\fc^2(s^1) + \norm{\gamma\mV^\fc}_{\Sigma_{p^\fc}(s^1)}^2} } \\
        &\qquad +  \sum_{s\ne s^1} \frac{1}{\mu_{1/2}(s)}\pa{\pa{\rho^\ft_{s^1}(s)-\rho^\fc_{s^1}(s)}^2 \sigma_\fc^2(s) + \norm{\gamma\rho^\ft_{s^1}(s)\mV^\ft - \gamma\rho^\fc_{s^1}(s)\mV^\fc}_{\Sigma_{p^\fc}(s)}^2}\,.
    \end{align*}

% \end{proof}

\subsubsection{Variance Comparison: Proof of Theorem~\ref{thm: VR_two_arm}}\label{app: proof_VR_two_arm}
    The first equation $\Sigma^{\mathrm{het}}_{\mathrm{IS}} = \Sigma^{\mathrm{het}}_{\mathrm{AB}}$ is straightforwards. For the second inequality, we have for any $s\in\gS\setminus s^1$,
    \begin{align*}
        \norm{\gamma\rho^\ft_{s^1}(s)\mV^\ft - \gamma\rho^\fc_{s^1}(s)\mV^\fc}_{\Sigma_{p^\fc}(s)}^2 &\leq 2\norm{\gamma\rho^\ft_{s^1}(s)\mV^\ft}_{\Sigma_{p^\fc}(s)}^2 + 2\norm{ \gamma\rho^\fc_{s^1}(s)\mV^\fc}_{\Sigma_{p^\fc}(s)}^2\,,
        % & = 2\norm{\gamma\rho^\ft_{s^1}(s)\mV^\ft}_{\Sigma_{p^\ft}(s)}^2 + 2\norm{ \gamma\rho^\fc_{s^1}(s)\mV^\fc}_{\Sigma_{p^\fc}(s)}^2\,,
    \end{align*}
    since $\Sigma_{p^\fc}(s)$ is positive semi-definite. And
    \[(\rho^\ft_{s^1}(s)-\rho^\fc_{s^1}(s))^2< 2\rho^\ft_{s^1}(s)^2+2\rho^\fc_{s^1}(s)^2\,,\]
    since $\rho^\ft_{s^1}(s),\rho^\fc_{s^1}(s)>0$ for ergodic chains. Finally, under SST we have $\Sigma_{p^\ft}(s)=\Sigma_{p^\fc}(s)$ and $\sigma_\ft^2(s)=\sigma_\fc^2(s)$, giving the result.

    For the equality to hold, it's easy to see one necessary condition is that all reward of state $s\in\gS\setminus s^1$ are fixed, i.e. $\sigma_\fc^2(s)=0$ for all $s\ne s^1$.

    On the other hand, for $\norm{\gamma\rho^\ft_{s^1}(s)\mV^\ft + \gamma\rho^\fc_{s^1}(s)\mV^\fc}_{\Sigma_{p^\fc}(s)}^2 = 0$ to hold, we have for all $s\ne s^1$, there exists some constant $\alpha_s\in\mathbb{R}$ such that $\rho^\ft_{s^1}(s)\mV^\ft + \rho^\fc_{s^1}(s)\mV^\fc=\alpha_s \mathbf{1}$, where $\mathbf{1}$ is the all-ones vector. We can write this constraint as
    \begin{align}\label{eq: condition of equality}
        (\mV^\ft,\mV^\fc) \begin{pmatrix}
        \rho^\ft_{s^1}(s^{2:K})\\
        \rho^\fc_{s^1}(s^{2:K})
    \end{pmatrix} = \mathbf{1} \cdot (\alpha_{2}, \alpha_{3}, \ldots, \alpha_{K})\,.
    \end{align}
    where $\rho^a_{s^1}(s^{2:K})\coloneqq[\rho^\ft_{s^1}(s^2)  ,\rho^\ft_{s^1}(s^3) , \ldots , \rho^\ft_{s^1}(s^K)]$ for $a\in\{\ft,\fc\}.$
    
    A trivial case is $\mV^\ft=\mV^\fc = \vzero$, which corresponds to the trivial MDP with all reward zeros. When MDP is non-trivial, by Sylvester’s rank inequality, we derive a necessary condition condition for \Eqref{eq: condition of equality} to hold: either  $\mathrm{rank}(\mV^\ft,\mV^\fc) = 1 $ or $\mathrm{rank}(
        \rho^\ft_{s^1}(s^{2:K})^\top,
        \rho^\fc_{s^1}(s^{2:K})^\top) = 1 $. 
    The second one implies the linear dependence of $\rho$. 
    
    We are then interested in the case where the second condition is not satisfied, which implies  $\mathrm{rank}(\mV^\ft,\mV^\fc) = 1 $.  That is, there exists some constant $\beta\ne 0$ such that $\mV^\ft = \beta\mV^\fc$. Recall $\rho^\ft_{s^1}(s)\mV^\ft + \rho^\fc_{s^1}(s)\mV^\fc=(\rho^\ft_{s^1}(s)\beta + \rho^\fc_{s^1}(s))\mV^\fc = \alpha_s \mathbf{1}$. Then it implies $\mV^\fc = \phi \mathbf{1}$ for some $\phi\ne 0$. Since in this case $\mV^\ft$ and $\mV^\fc$ are linear dependent, we have $\vDelta = \phi^\prime \mathbf{1}$. This will contradict the Monotonicity Theorem~\ref{thm: monotonicity} unless $\phi^\prime=0$, that effectively corresponds to the A/A testing case. And in A/A testing, we have $\rho^\ft_{s^1}(s^{2:K}) = \rho^\fc_{s^1}(s^{2:K})$, which contradicts the condition that the second condition does not hold.

    In conclusion, the necessary condition is given by the linear dependence of $\rho$.
    \section{Proof of Asymptotic Efficiency}
In this section, we will discuss the proof framework for asymptotic efficiency via constrained Cramer-Rao Bound. We will again start with the warm-up from of fixed policy inference in Proposition~\ref{prop: MB_Normal}. We then discuss the asymptotic results in A/B testing.

\subsection{Proof of Proposition~\ref{prop: CRB_TD}}\label{app: CRB_TD}
    The MDP system is controlled by parameters $\vtheta = (\mP^\pi,\vr^\pi)$. Since we assume the initial distribution is $\vmu^\pi$, the observations $(s_i,a_i,s_{i+1},r_i)$ in $\tau$ can be seen as i.i.d. By Constrained Cramer-Rao Bound, Lemma~\ref{lem: cramer_rao_td} shows for one single observation $X =(s,a,s^\prime,r)$, we have
    \[\mathrm{CCRB}(\mP^\pi, \vr^\pi) = \begin{bmatrix}
        \Diag_{s\in\gS}(\Sigma_{p}(s)/\vmu(s))&\\
        & \Diag_{s\in\gS}(\sigma_\pi^2(s)/\vmu(s))
    \end{bmatrix}\,.\]
    % where $\Sigma_{p}/\mu = [\Sigma_{p}(s_1)/\mu^\pi(s_1), \ldots,\Sigma_{p}(s_n)/\mu^\pi(s_n)]^\top$ and $\sigma_r^2/\mu = [\sigma_r^2(s_1)/\mu^\pi(s_1), \ldots, \sigma_r^2(s_n)/\mu^\pi(s_n)]^\top$.

    Also note that 
    \begin{align*}
        \frac{\partial \mV^\pi}{\partial P(i,j)} &=  \frac{\partial (\mI-\gamma\widehat \mP^\pi)^{-1}\vr^\pi}{\partial P(i,j)}\\
        & = -\gamma (\mI-\gamma\mP^\pi)^{-1} \mE_{i,j} \mV^\pi\,.
    \end{align*}
    and \begin{align*}
        \frac{\partial \mV^\pi}{\partial r(i)} &=  (\mI-\gamma \mP^\pi)^{-1}\frac{\partial \vr^\pi}{\partial r(i)}\\
        & =  (\mI-\gamma\mP^\pi)^{-1} \ve_{i} \,.
    \end{align*}

    By Lemma~\ref{lem: Delta_CCRB}, the Constrained Cramer-Rao Bound for $\mV^\pi$ is given by
    \begin{align*}
        \mathrm{CCRB}(\mV^\pi) &= \nabla \mV^{\pi ^\top} \frac{1}{T}\mathrm{CCRB}(\mP^\pi,\vr^\pi) \nabla \mV^\pi\\
         &= \frac{1}{T}(\mI-\gamma \mP^\pi)^{-1} \Diag_{s\in\gS}\pa{ \sigma_\pi^2(s)/\vmu(s)+ \norm{\gamma\mV^\pi}_{\Sigma_p(s)}^2/\vmu(s)} (\mI-\gamma \mP^\pi)^{-\top}\\
         & = \frac{1}{T}\Sigma_{\mathrm{MB}}\,.
        \end{align*}

% \end{proof}

\begin{lemma}\label{lem: cramer_rao_td}
    For any unbiased estimator of Transition Matrix $\mP$ and mean reward vector $\vr$, it holds
    \[\mathrm{CCRB}(\mP, \vr) = \begin{bmatrix}
        \Diag_{s\in\gS}(\Sigma_{p}(s)/\vmu(s))&\\
        & \Diag_{s\in\gS}(\sigma^2(s)/\vmu(s))
    \end{bmatrix}\,.\]
    % where $\Sigma_{p}/\mu = [\Sigma_{p}(s_1)/\mu^\pi(s_1), \ldots,\Sigma_{p}(s_n)/\mu^\pi(s_n)]^\top$ and $\sigma_r^2/\mu = [\sigma_r^2(s_1)/\mu^\pi(s_1), \ldots, \sigma_r^2(s_n)/\mu^\pi(s_n)]^\top$.
\end{lemma}
\proof{Proof.}
    For the first row $\mP(1) = (P(1,i))_{i\in[n]}$ and reward $\vr(1)$, it holds for observation $X = (s_1 = 1,a_1,s_2=s,r_1)$,
    \begin{align*}
        p(X) & = \mu(1)^{\mathbf1\br{s_1=1}}\prod_{i\in [n]} P(1,i)^{\mathbf1\br{s=i}} P(r(s_1,a_1)=r_1)^{\mathbf1\br{s_1=1}}\\
        \log p(X) & = \mathbf1\br{s_1=1}\log \mu(1) + \sum_{i\in[n]} \mathbf1\br{s=i} \log P(1,i) \\
        &\qquad +\mathbf1\br{s_1=1} \log P(r(s_1,a_1)=r_1)\\
        \frac{\partial\log p(X)}{\partial P(1,i)}  &=  \mathbf1\br{s=i} \frac{1}{P(1,i)}\\
        \mI(\mP(1),\vr(1)) &= -\E_{X}[\nabla^2 \log p(X)]\\
        &= \mu(1)\begin{bmatrix}
        \Diag(1/\mP(1))&\\
        & 1/\sigma^2(s)
    \end{bmatrix}\,.
    \end{align*}
    The constraint for $\mP(1),\vr(1)$ is $f_1(\mP(1),\vr(1)) = \sum_{i\in[n]} P(1,i)-1=0$, then we have 
    \[\nabla f_1 = [1,1,\ldots, 1,0]\,.\]
    Following Lemma~\ref{lem: CCRB}, we construct 
    \[\mU_1= \begin{bmatrix}
  1&1  &\ldots  &1 &\\
  -1&  &  &&\\
  &  -1&  & &\\
  &  &  \ddots&& \\
  &  &  & &-1&\\
  &  &  &  &  & 1\end{bmatrix}\,.\]
It can be verified that
\[\mathrm{CCRB}(\mP(1),\vr(1))= \mU_1 (\mU_1^\top \mI(\mP(1)) \mU_1)^{-1} \mU_1^\top = \begin{bmatrix}
        \Sigma_p(1)/\vmu(1)&\\
        & \sigma^2(s)/\vmu(1)
    \end{bmatrix}\,.\]
One fuhrer notice that 
\[\mI(\mP,\vr) = \Diag(\mI(\mP(i),\vr(i))\,,\quad \nabla f = \begin{bmatrix}
  \nabla f_1&  &  & \\
  & \nabla f_1 &  &\\
  &  & \ddots & \\
  &  &  & \nabla f_1
\end{bmatrix}\,,\quad \mU = \begin{bmatrix}
  \mU_1&  &  & \\
  & \mU_1 &  &\\
  &  & \ddots & \\
  &  &  & \mU_1
\end{bmatrix}\,,\]
and the final Constrained Cramer-Rao Bound is 
\[\mathrm{CCRB}(\mP, \vr) = \begin{bmatrix}
        \Diag_{s\in\gS}(\Sigma_{p}(s)/\vmu(s))&\\
        & \Diag_{s\in\gS}(\sigma^2(s)/\vmu(s))
    \end{bmatrix}\,.\]
\Halmos
\endproof

\subsection{Proof of Theorem~\ref{thm: CRB_AB}}\label{app: CRB_AB}
% \begin{theorem}[Asymptotic Efficiency of A/B Testing]
%     Assume the dataset $\gD=\{(s_i,a_i,s_{i+1},r_i)\mid i\in [0,T]\}$ is sampled from MDP under mixed policy $\mu_{1/2}$ with initial distribution $\mu_{1/2}$. Then for any unbiased estimator of ATE $\vDelta$ with covariance matrix $\Sigma$, it holds\[\Sigma_{\mathrm{AB}}\preceq\Sigma\,. \]
% \end{theorem}
% \begin{proof}
    The MDP system is controlled by parameters $\vtheta = (\mP_0,\mP_1, \vr_0,\vr_1)$. Since we assume the initial distribution is $\vmu_{1/2}$, the observations $(s_i,a_i,s_{i+1},r_i)$ in $\tau$ can be seen as i.i.d. A similar argument to Lemma~\ref{lem: cramer_rao_td} can show
    \[\mathrm{CCRB}(\vtheta) = 2\Diag(\Sigma_{p^0}/\mu_{1/2}, \Sigma_{p^1}/\mu_{1/2}, \sigma_0^2/\mu_{1/2}, \sigma_1^2/\mu_{1/2})\,,\]
    where $\Sigma_{p^i}/\mu_{1/2} = [\Sigma_{p^i}(s^1)/\mu_{1/2}(s^1), \Sigma_{p^i}(s^2)/\mu_{1/2}(s^2),\ldots,\Sigma_{p^i}(s_K)/\mu_{1/2}(s^K)]^\top$ and $\sigma_0^2/\mu_{1/2} = [\sigma_0^2(s^1)/\mu_{1/2}(s^1), \sigma_0^2(s^2)/\mu_{1/2}(s^2),\ldots, \sigma_0^2(s^K)/\mu_{1/2}(s^K)]^\top$ for $i=0,1$. 

    Also note that 
    \begin{align*}
        \frac{\partial \vDelta}{\partial P_0(i,j)} &=
        -\frac{\partial \mV_0}{\partial P_0(i,j)}\\
        &=-\frac{\partial (\mI-\gamma\widehat \mP_0)^{-1}\vr_0}{\partial P_0(i,j)}\\
        & = \gamma (\mI-\gamma\mP_0)^{-1} \mE_{i,j} \mV_0\,.
    \end{align*}
    and \begin{align*}
        \frac{\partial \vDelta}{\partial r(i)} &=
        -\frac{\partial \mV_0}{\partial r_0(i)}\\
        &=  -(\mI-\gamma \mP)^{-1}\frac{\partial \vr_0}{\partial r_0(i)}\\
        & =  -(\mI-\gamma\mP)^{-1} \ve_{i} \,.
    \end{align*}
    And similar results hold for $\frac{\partial \vDelta}{\partial P_1(i,j)}$ and $\frac{\partial \vDelta}{\partial r_1(i)}$. By Lemma~\ref{lem: Delta_CCRB}, the Constrained Cramer-Rao Bound for $\mV$ is given by
    \begin{align*}
        \mathrm{CCRB}(\vDelta) &= \nabla \vDelta ^\top \frac{1}{T}\mathrm{CCRB}(\vtheta) \nabla \vDelta\\
         &= \frac{2}{T}(\mI-\gamma \mP^\ft)^{-1} \Diag_{s\in\gS}\pa{\sigma_\ft^2(s)/\mu_{1/2}(s) +\norm{\gamma\mV^\ft}_{\Sigma_{p^\ft}(s)}^2/\mu_{1/2}(s)} (\mI-\gamma \mP^\ft)^{-\top} \\
&\qquad + \frac{2}{T}(\mI-\gamma \mP^\fc)^{-1} \Diag\pa{\sigma_\fc^2/\mu_{1/2}(s) +\norm{\gamma\mV^\fc}_{\Sigma_{p^\fc}(s)}^2/\mu_{1/2}(s)} (\mI-\gamma \mP^\fc)^{-\top}\\
         & = \frac{1}{T}\Sigma_{\mathrm{AB}}\,.
        \end{align*}

\subsection{Proof of Theorem~\ref{thm: CRB_IS}}
% \begin{theorem}[Asymptotic Efficiency of Information-Sharing]
%     Assume the treatment is SPT and the dataset $\gD=\{(s_i,a_i,s_{i+1},r_i)\mid i\in [0,T]\}$ is sampled from MDP under mixed policy $\mu_{1/2}$ with initial distribution $\mu_{1/2}$. Then for any unbiased estimator of ATE $\vDelta$ with covariance matrix $\Sigma$, it holds\[\Sigma_{\mathrm{IS}}\preceq\Sigma\,. \]
% \end{theorem}
% \begin{proof}
    We denote $s_{i:j}\coloneqq (s_i,s_{i+1},\ldots,s_j)$. The MDP system is controlled by parameters $\vtheta = (\mP_1(s^1),\mP_0, r(s^1,1), \vr_0)$. Since we assume the initial distribution is $\vmu_{1/2}$, the observations $(s_i,a_i,s_{i+1},r_i)$ in $\tau$ can be seen as i.i.d. A similar argument to Lemma~\ref{lem: cramer_rao_td} can show
    \begin{align*}
        \mathrm{CCRB}(\vtheta) &= \Diag(2\Sigma_{p_0}(s^1)/\mu_{1/2}(s^1), 2\Sigma_{p_1}(s^1)/\mu_{1/2}(s^1), \Sigma_{p_1}(s_{2:n})/\mu_{1/2}(s_{2:n}),\\
        &\qquad 2\sigma_\ft^2(s^1)/\mu_{1/2}(s^1), 2\sigma_\fc^2(s^1)/\mu_{1/2}(s^1), \sigma_\fc^2(s_{2:n})/\mu_{1/2}(s_{2:n}))\,.
    \end{align*}
    % \[\mathrm{CCRB}(\theta) = \Diag(2\Sigma_{p_0}(s^1)/\mu_{1/2}(s^1), 2\Sigma_{p_1}(s^1)/\mu_{1/2}(s^1), \Sigma_{p_1}(s_{2:n})/\mu_{1/2}(s_{2:n}))\,.\]

    Also note that for state $s^1$, 
    \begin{align*}
        \frac{\partial \vDelta}{\partial P_0(1,j)} &=
        -\frac{\partial \mV_0}{\partial P_0(1,j)}\\
        &=-\frac{\partial (\mI-\gamma\widehat \mP_0)^{-1}\vr_0}{\partial P_0(1,j)}\\
        & = \gamma (\mI-\gamma\mP_0)^{-1} \mE_{1,j} \mV_0\,.
    \end{align*}
    And similar results hold for $\frac{\partial \vDelta}{\partial P_1(1,j)}$.

    For state $s\ne s^1$, 
    \begin{align*}
        \frac{\partial \vDelta}{\partial P(s,j)} &=
        \frac{\partial \mV_1}{\partial P(s,j)}-\frac{\partial \mV_0}{\partial P(s,j)}\\
        &=\frac{\partial (\mI-\gamma\widehat \mP_1)^{-1}\vr_1}{\partial P_1(s,j)}-\frac{\partial (\mI-\gamma\widehat \mP_0)^{-1}\vr_0}{\partial P_0(s,j)}\\
        & = -\gamma (\mI-\gamma\mP_1)^{-1} \mE_{s,j} \mV_1+\gamma (\mI-\gamma\mP_0)^{-1} \mE_{s,j} \mV_0\,.
    \end{align*}One can also calculate the gradient for the reward part.
    
    By Lemma~\ref{lem: Delta_CCRB}, the Constrained Cramer-Rao Bound for $\mV$ is given by
    \begin{align*}
        \mathrm{CCRB}(\vDelta) &= \nabla \vDelta ^\top \mathrm{CCRB}(\theta) \nabla \vDelta\\
         &= 2\frac{\sigma_\ft^2(s^1)+\norm{\gamma\mV^\ft}_{\Sigma_{p^\ft(s^1)}}^2}{\mu_{1/2}(s^1)} {\vrho^\ft}(s^1) {\vrho^\ft}(s^1)^\top + 2\frac{\sigma_\fc^2(s^1)+\norm{\gamma\mV^\fc}_{\Sigma_{p^\fc(s^1)}}^2}{\mu_{1/2}(s^1)}{\vrho^\fc}(s^1) {\vrho^\fc}(s^1)^\top\\
&\qquad + \sum_{s\ne s^1} \frac{1}{\mu_{1/2}(s)}\pa{\gamma\vrho^\ft(s) \mV^{t^\top} -  \gamma\vrho^\fc(s) \mV^{c^\top}}\Sigma_{p^\fc(s)} \pa{\gamma \mV^\ft\vrho^\ft(s)^\top -  \gamma \mV^\fc\vrho^\fc(s)^\top}\\
&\qquad + \sum_{s\ne s^1} \frac{\sigma_\fc^2(s)}{\mu_{1/2}(s)} \pa{\vrho^\ft(s)-\vrho^\fc(s)} \pa{\vrho^\ft(s)-\vrho^\fc(s)}^\top\\
         & = \Sigma_{\mathrm{IS}}\,.
        \end{align*}
% \end{proof}

\subsubsection{Extension to multiple test arms}\label{app: ISN_Lower_Bound}
\begin{theorem}[Asymptotic Efficiency of Information-Sharing with $n$ Test Arms]
    Assume the treatment is SST with $n$ test arms. If running mixed policy $\pi^{1/n}$ with initial distribution $\mu_{1/n}$, for any unbiased estimator of ATE $\vDelta_{i,j}$ for any two test arms $(\pi^i,\pi^j)$ with covariance matrix $\Sigma$, it holds, $\Sigma_{\mathrm{ISn}}(i,j)\preceq T\Sigma$.
\end{theorem}
The proof follows the same manner of the previous section.
    \section{Generalization to Local Treatment}\label{app: local_treament}
For clarity, we denote $\gS^\fc\coloneqq \gS\setminus\gS^\ft$. 
\begin{theorem}[Asymptotic Normality of A/B Testing with Local Treatment]
     Assuming the Markov Chain induced by policy $\pi^\ft,\pi^{1/2}$ are both uniformly ergodic, with Local Treatment on state $\gS^\ft \subseteq \gS$, it holds as $T\to\infty$, for both $\Gamma \in \{\mathrm{AB},\mathrm{IS}, \mathrm{PI}\}$ and any state $s\in \gS$,
    \[\sqrt{T}\pa{\widehat \Delta^{\Gamma}(s) - \Delta(s)}\xrightarrow{d}\gN(0, \sigma_{\Gamma}^2(s))\,,\]
    where 
    \begin{align*}
        \sigma_{\mathrm{AB}}^2(s) & = \underbrace{2\sum_{s^\prime\in \gS^\ft}\frac{\rho^{\ft}_{s}(s^\prime)^2}{\mu_{1/2}(s^\prime)}\pa{\sigma_\ft^2(s^\prime) +\norm{\gamma\mV^{\ft}}_{\Sigma_{p^{\ft}}(s^\prime)}^2} + 2 \sum_{s^\prime\in \gS^\ft}\frac{\rho^{\fc}_{s}(s^\prime)^2}{\mu_{1/2}(s^\prime)}\pa{\sigma_\fc^2(s^\prime) +\norm{\gamma\mV^{\fc}}_{\Sigma_{p^{\fc}}(s^\prime)}^2} }_{\Sigma_{\mathrm{AB}}^{\mathrm{het}}}\\
        &\qquad + \underbrace{ 2\sum_{s^\prime\in \gS^\fc}\frac{\rho^{\ft}_{s}(s^\prime)^2}{\mu_{1/2}(s^\prime)}\pa{\sigma_\ft^2(s^\prime) +\norm{\gamma\mV^{\ft}}_{\Sigma_{p^{\ft}}(s^\prime)}^2} + 2 \sum_{s^\prime\in \gS^\fc}\frac{\rho^{\fc}_{s}(s^\prime)^2}{\mu_{1/2}(s^\prime)}\pa{\sigma_\fc^2(s^\prime) +\norm{\gamma\mV^{\fc}}_{\Sigma_{p^{\fc}}(s^\prime)}^2}
        }_{\Sigma_{\mathrm{AB}}^{\mathrm{hom}}},
    \end{align*}
    and
    \begin{align*}
        \sigma_{\mathrm{IS}}^2(s) = &\underbrace{2\sum_{s^\prime\in \gS^\ft}\frac{\rho^{\ft}_{s}(s^\prime)^2}{\mu_{1/2}(s^\prime)}\pa{\sigma_\ft^2(s^\prime) +\norm{\gamma\mV^{\ft}}_{\Sigma_{p^{\ft}}(s^\prime)}^2} + 2 \sum_{s^\prime\in \gS^\ft}\frac{\rho^{\fc}_{s}(s^\prime)^2}{\mu_{1/2}(s^\prime)}\pa{\sigma_\fc^2(s^\prime) +\norm{\gamma\mV^{\fc}}_{\Sigma_{p^{\fc}}(s^\prime)}^2} }_{\Sigma_{\mathrm{IS}}^{\mathrm{het}}}\\
        &\qquad +  \underbrace{\sum_{s^\prime\in\gS^\fc} \frac{1}{\mu_{1/2}(s^\prime)}\pa{\big(\rho^\ft_{s}(s^\prime)-\rho^\fc_{s}(s^\prime)\big)^2 \sigma_\fc^2(s^\prime) + \norm{\gamma\rho^\ft_{s}(s^\prime)\mV^\ft - \gamma\rho^\fc_{s}(s^\prime)\mV^\fc}_{\Sigma_{p^\fc}(s^\prime)}^2}}_{\Sigma_{\mathrm{IS}}^{\mathrm{hom}}}\,.
    \end{align*}
    and 
    \[\sigma_{\mathrm{PI}}^2(s)  = \underbrace{\sum_{s^\prime\in \gS^\ft}\frac{\rho^{\ft}_{s}(s^\prime)^2}{\mu^\ft(s^\prime)}\pa{\sigma_\ft^2(s^\prime) +\norm{\gamma\mV^{\ft}}_{\Sigma_{p^{\ft}}(s^\prime)}^2} }_{\Sigma_{\mathrm{PI}}^{\mathrm{het}}}\,.\]
\end{theorem}
The proof can be easily adapted from the proof of Theorem~\ref{thm: AB_Testing}, \ref{thm: info_sharing} and \ref{thm: PI}.

\subsection{Local Treatment with Multiple Test Arms}\label{app: local_n_arm}
We first define the model for local treatment with $n$ test arms, which can be extended from the model of SST with $n$ test arms in Section~\ref{sec: multiple-arms-IS}. In particular, there exists a pre-determined crucial subset $\gS^\ft\subseteq \gS$. For each treatment policy $\pi^i\in \Pi\coloneqq\{\pi^j\}_{j\in[n]}$, $\pi^i(s)\equiv \ft^i$ for all $s\in\gS^\ft$ and $\pi^i(s)\equiv \ft^1$ for all $s\in\gS\setminus \gS^\ft$. Additionally, each local treatment policy is associated with its own transition matrix $\mP^{\ft^i}$ and mean reward $\vr^{\ft^i}$.

The objective is still to infer the ATE between any pair of policies $(\pi^i,\pi^j)$, i.e. $\vDelta_{i,j}=\mV^i-\mV^j$. We apply similar ideas of A/B testing, IS and PI and obtain the following generalization of Corollary~\ref{coro: linear_Variance},
\begin{corollary}
    Assuming the Markov Chain induced by the mixed policy $\pi^{1/(n-1)},\pi^{1/n}$ are both uniformly ergodic, with Local Treatment on state $\gS^\ft\subseteq\gS$, it holds as $T\to\infty$, for $\Gamma \in \{\mathrm{ABn},\mathrm{ISn},\mathrm{PIn}\}$ and any $s\in\gS$,
    \[\sqrt{T}\pa{\widehat \Delta_{i,j} ^{\Gamma}(s) - \Delta_{i,j}(s)}\xrightarrow{d}\gN(0, \sigma_{\Gamma,(i,j)}^2(s))\,,\]
    where 
    \begin{align*}
        \sigma_{\mathrm{ABn},(i,j)}^2(s) & = \underbrace{n\sum_{s^\prime\in\gS^\ft}\frac{\rho^i_s(s^\prime)^2}{\mu_{1/n}(s^\prime)}\pa{\sigma_i^2(s^\prime) + \norm{\gamma\mV^i}_{\Sigma_{p^i}(s^\prime)}^2} +  \sum_{s^\prime\in\gS^\ft}\frac{\rho^j_s(s^\prime)^2}{\mu_{1/n}(s^\prime)}\pa{\sigma_j^2(s^\prime) + \norm{\gamma\mV^j}_{\Sigma_{p^j}(s^\prime)}^2}  }_{\Sigma_{\mathrm{ABn}}^{\mathrm{het}}(i,j)}\\
        &\qquad + \underbrace{n \sum_{s^\prime\in\gS^\fc}\frac{\rho^{i}_s(s^\prime)^2}{\mu_{1/n}(s^\prime)}\pa{\sigma_i^2(s^\prime) +\norm{\gamma\mV^{i}}_{\Sigma_{p^{i}}(s^\prime)}^2} + n \sum_{s^\prime\in\gS^\fc}\frac{\rho^{j}_{s}(s^\prime)^2}{\mu_{1/n}(s^\prime)}\pa{\sigma_j^2(s^\prime) +\norm{\gamma\mV^{j}}_{\Sigma_{p^{j}}(s^\prime)}^2}
        }_{\Sigma_{\mathrm{ABn}}^{\mathrm{hom}}(i,j)},
    \end{align*}
    and
    \begin{align*}
        \sigma_{\mathrm{ISn},(i,j)}^2(s) = &\underbrace{n \sum_{s^\prime\in\gS^\ft}\frac{\rho^i_s(s^\prime)^2}{\mu_{1/n}(s^\prime)}\pa{\sigma_i^2(s^\prime) + \norm{\gamma\mV^i}_{\Sigma_{p^i}(s^\prime)}^2} +  \sum_{s^\prime\in\gS^\ft}\frac{\rho^j_s(s^\prime)^2}{\mu_{1/n}(s^\prime)}\pa{\sigma_j^2(s^\prime) + \norm{\gamma\mV^j}_{\Sigma_{p^j}(s^\prime)}^2} }_{\Sigma_{\mathrm{ISn}}^{\mathrm{het}}(i,j)} \\
        &\qquad +  \underbrace{\sum_{s^\prime\in\gS^\fc} \frac{1}{\mu_{1/n}(s^\prime)}\pa{\big(\rho^i_s(s^\prime)-\rho^j_s(s^\prime)\big)^2 \sigma_1^2(s^\prime) + \norm{\gamma\rho^i_s(s^\prime)\mV^i - \gamma\rho^j_s(s^\prime)\mV^j}_{\Sigma_{p^1}(s^\prime)}^2}}_{\Sigma_{\mathrm{ISn}}^{\mathrm{hom}}(i,j)}\,.
    \end{align*}
    and 
   \begin{align*}
\sigma_{\mathrm{PIn},(i,j)}^2(s) = &\mathbf{1}[i\ne1](n-1)\sum_{s^\prime\in\gS^\ft}\frac{\rho^i_{s}(s^\prime)^2}{\mu_{1/(n-1)}(s^\prime)}\pa{\sigma_i^2(s^\prime)+ \norm{\gamma\mV^i}^2_{\Sigma_{p^i}(s^\prime)}}  \\
&+\mathbf{1}[j\ne1](n-1)\sum_{s^\prime\in\gS^\ft}\frac{\rho^j_{s}(s^\prime)^2}{\mu_{1/(n-1)}(s^\prime)}\pa{\sigma_j^2(s^\prime)+ \norm{\gamma\mV^j}^2_{\Sigma_{p^j}(s^\prime)}}\\
&=:\Sigma_{\mathrm{PIn}}^{\mathrm{het}}(i,j)
\end{align*}
\end{corollary}
Similarly, we can derive the following generalization of Theorem~\ref{thm: VR_n_arm},
\begin{corollary}
    Consider local treatment of $\gS^\ft\subseteq \gS$ with $n$ test arms $\{\pi^i\}_{i\in[n]}$, it holds for any pair of test arms $(\pi^i,\pi^j)$, 
    \[\Sigma^{\mathrm{het}}_{\mathrm{IS}} = \Sigma^{\mathrm{het}}_{\mathrm{AB}} \quad\textrm{and}\quad \Sigma^{\mathrm{hom}}_{\mathrm{PI}}=0\leq \Sigma^{\mathrm{hom}}_{\mathrm{IS}} \leq \frac{2}{n}\Sigma^{\mathrm{hom}}_{\mathrm{AB}}\,.\]
\end{corollary}
This corollary justifies that the linear variance reduction of IS with the number of test arms keeps to take effect for local treatments. 
    % \subsection{Oracle for Control}
% Estimation of state-visitation measure given trajectory $\tau = \{(S_i,A_i,R_i)\}_{i\in[T]}\cup \{S_{T+1}\}$, 
% \begin{equation}
%     \widehat {\rho^t_x}(x) \gets \sum_{i=0}^T \gamma^i \frac{\sum_{j=1}^{T-i+1}\mathbf{1}[S_{j+i}=x\mid S_j=x]}{\sum_{j=1}^{T-i+1}\mathbf{1}[S_j=x]}\,.
% \end{equation}
% Then by Lemma~(\ref{lem: performance}), we have the following Algorithm~\ref{alg: OSC}. 
% \begin{center}
% \begin{algorithm}
% \caption{One-step Estimator with Oracle}
% \begin{algorithmic}[1]\label{alg: OSC}
% \REQUIRE{Control and Treatment $\pi^c,\pi^t$; Experiment horizon $T$; Oracle $V^c$; Discount factor $\gamma$.}
% \STATE  Execute policy $\pi^t$ for $T$ epochs and obtain trajectory $\tau = \{(S_i,A_i,R_i)\}_{i\in[T]}\cup \{S_{T+1}\}$.
% \STATE $$\widehat r(x,t)\gets \frac{\sum_{i=1}^T \mathbf{1}[S_i=x]r_i}{\sum_{i=1}^T \mathbf{1}[S_i=x]}\,.$$
% \STATE For all $s\in \gS\,$, $$\widehat P(s\mid x,t)\gets \frac{\sum_{i=1}^T \mathbf{1}[S_{i+1}=s\mid S_i=x]}{\sum_{i=1}^T \mathbf{1}[S_i=x]}\,.$$ 
% \STATE $$\widehat{A^c}(x,t) \gets \widehat{r}(x,t) + \gamma \sum_{s^\prime}\widehat{P}(s^\prime\mid x,t) V^c(s^\prime)-V^c(x)\,.$$
% \STATE $$\widehat {\rho^t_x}(x) \gets \sum_{i=0}^T \gamma^i \frac{\sum_{j=1}^{T-i+1}\mathbf{1}[S_{j+i}=x\mid S_j=x]}{\sum_{j=1}^{T-i+1}\mathbf{1}[S_j=x]}\,.$$
% \RETURN $$\widehat{\Delta}(x) \gets \widehat {\rho^t_x}(x) \widehat{A^c}(x,t)$$
% \end{algorithmic}
% \end{algorithm}
% \end{center}

\section{Estimation with Perfect Information about the Control Arm}\label{sec: PI}

 In this section, we explore the information-sharing concept further by considering scenarios where we have \emph{Perfect Information (PI)} about the control arm. Typically, the control arm represents a default policy that may have been operational for an extended period within the system, allowing the organization to accumulate a substantial dataset and develop a thorough understanding of this baseline. Extending Example~\ref{exp: CLV}, the company might already possess ample historical data on customer behaviors, enabling it to accurately determine the customers' transition matrices when not offering coupon. We assume the historical data accurately characterizes the customer's behavior (the same transition) at present. In other words, there is no distribution shift between the offline dataset and the current trial. Access to extensive historical user data is a common strategy in marketing, utilized to optimize selling policies (\citealt{simester06dynamic,jeremy24targeting}), predict future purchases (\citealt{jacobs16model}), or gauge customer responses (\citealt{das23using}). Similarly, the assumption of having an offline dataset is prevalent in other domains involving MDPs, such as offline reinforcement learning (\citealt{levine2020offline}), reinforcement learning with human feedback (RLHF, \citealt{chang2024dataset}), and healthcare applications (\citealt{yu21reinforce, hu24fast}).

Formally, we assume that the experimenter has access to both the precise transition matrix $\mP^\fc$ and the exact mean reward $\vr^\fc$ of control policy $\pi^\fc$. With this perfect information, she can directly computing $\mV^\fc$ using the Bellman equation. Consequently, her primary focus during experimental periods becomes gathering data on the new treatment policy $\pi^\ft$, applying it consistently. Moreover, since $\pi^\ft$ is SST, a substantial amount of information from $\vr^\fc$ and $\mP^\fc$, especially those associated with the non-crucial states, can be shared when estimating $\mV^\ft$. More specifically, after consistently applying $\pi^\ft$, she only needs to estimate the transitions and mean reward associated with the crucial state $s^1$, which is 
\begin{equation*}
    \widehat{\mP}^{\ft}_{\textrm{PI}}(s^1,s)=\frac{\sum_{i=1}^T \mathbf{1}[s_i=s^1, s_{i+1}=s] }{1 \vee \sum_{i=1}^T \mathbf{1}[s_i=s^1]}\, \forall s\in\mathcal{S} \quad \textrm{ and } \quad \widehat{r}_{\textrm{PI}}^\ft(s^1)= \frac{\sum_{i=1}^T \mathbf{1}[s_i=s^1] r_i }{1 \vee \sum_{i=1}^T \mathbf{1}[s_i=s^1]}.
\end{equation*}
%\[\gD^\ft_\PI = \{(s_i,a_i,s_{i+1}, r_i): (s_i,a_i)=(s^1,\ft),i\in [T]\}\,,\] and 
For all the other components in $\widehat{\mP}_{\textrm{PI}}^{\ft}$ and $\widehat{\vr}_{\textrm{PI}}^{\ft}$, we directly plug in those in $\mP^\fc$ and $\vr^\fc$, i.e., $\widehat{\mP}^{\ft}_{\textrm{PI}}(s,s')=\mP^\fc(s,s')$ and $\widehat{r}_{\textrm{PI}}^\ft(s)= r^\fc(s)$ for all $s\neq s^1$ and any $s'\in\mathcal{S}$.  Applying $\widehat{\mP}_{\textrm{PI}}^{\ft}$ and $\widehat{\vr}_{\textrm{PI}}^{\ft}$ to the Bellman equation generates an estimator $\widehat{\mV}^\ft_{\textrm{PI}}$. Then, the difference between $\widehat{\mV}^\ft_{\textrm{PI}}$ and ${\mV}^\fc$ yields an estimator $\widehat \vDelta^\PI$ of the treatment effects.   The following theorem presents the central limit theorem of our estimator. The detailed proof is delayed to Appendix~\ref{app: perfect_info}. 
\begin{theorem}[Asymptotic Normality of Perfect Information]\label{thm: PI}
    Assuming the Markov chain induced by the SST policy $\pi^\ft$ is uniformly ergodic, it holds that as $T\to\infty$,
    \[\sqrt{T}\pa{\widehat \vDelta^\PI  - \vDelta}\xrightarrow{d}\gN(\vzero, \Sigma_{\PI})\,,\]
    where 
\begin{align*}
    \Sigma_{\PI} = (\mI-\gamma \mP^\ft)^{-1} 
    \begin{bmatrix}
    \frac{\sigma_\ft^2(s^1)}{\mu^\ft(s^1)} + \frac{\norm{\gamma\mV^\ft}_{\Sigma_{p^\ft}(s^1)}^2}{\mu^\ft(s^1)}&\\
    &\vzero
    \end{bmatrix}
    (\mI-\gamma \mP^\ft)^{-\top}\,.
\end{align*}

Specifically, for the crucial state $s^1$, it holds that as
$T\to\infty$,
\begin{equation*}
    \sqrt{T}\pa{\widehat \Delta^\PI(s^1)-\Delta(s^1)}\xrightarrow{d} \gN(0,\sigma^2_{\mathrm{PI}}(s^1))\,,
\end{equation*}
where 
\[\sigma^2_{\mathrm{PI}}(s^1) = \frac{\rho^\ft_{s^1}(s^1)^2}{\mu^\ft(s^1)}\pa{\sigma_\ft^2(s^1)+ \norm{\gamma\mV^\ft}^2_{\Sigma_{p^\ft}(s^1)}}\,.\]
\end{theorem}
 Since we are implementing a fixed policy $\pi^\ft$ during the experimental periods, it is logical that the asymptotic variance $\Sigma_\PI$ and $\sigma^2_\PI(s^1)$ exhibit similar structures with $\Sigma_{\mathrm{MB}}$ and $\sigma^2_{\mathrm{MB}}(s^1)$ from the model-based estimator under a fixed policy discussed in Section~\ref{sec: inference-fixed-policy}. Unlike the approach in that section, which requires estimating transitions for every state, here we only need to estimate the transitions and rewards associated with the crucial state $s^1$. With PI, $\sigma^2_{\textrm{PI}}(s^1)$ only keeps the variance linked to the crucial state $s^1$.
 %In particular, the variance associated with the crucial state $s^1$ is the same for PI and model-based estimation, while the variance of all other states, the homogeneous part $\Sigma^{\mathrm{hom}}_{\mathrm{PI}}$, are effectively reduced to zero with PI. 
 This reduction aligns with our expectations, given the complete information on the transitions and rewards for non-crucial states now at our disposal.
 
 We now would like to compare the variance with A/B testing, both with and without information sharing, as discussed in Sections \ref{sec: general-ab-testing} and \ref{sec: IS}. In addition to the perfect information about the control arm, as we are running $\pi^\ft$ all the time instead of the mixed policy like A/B testing, we naturally accumulate more data specific to $\pi^\ft$. Hence it's natural to anticipate a further reduction of variance associated with crucial state $s^1$. To facilitate a detailed comparison and ensure consistency in our analysis, we introduce two notations, albeit redundant, $\Sigma^{\mathrm{het}}_{\mathrm{PI}}$ and $\Sigma^{\mathrm{hom}}_{\mathrm{PI}}$. These notations are designed to differentiate the parts of $\sigma^2_{\mathrm{PI}}(s^1)$ related to the crucial states $s^1$ and all other states, respectively. Thus, $\Sigma^{\mathrm{het}}_{\mathrm{PI}}=\sigma^2_{\mathrm{PI}}(s^1)$ and $\Sigma^{\mathrm{hom}}_{\mathrm{PI}}=0$ since $\sigma^2_{\mathrm{PI}}(s^1)$ is only related to $s^1$. Formally, we have the following theorem characterizing the variance reduction of PI, whose detailed proof is delayed to Appendix~\ref{app: PI_analysis}.
\begin{theorem}\label{thm: PI_Analysis}
     Consider SST with treatment policy $\pi^\ft$ and control policy $\pi^\fc$, it holds \[\Sigma^{\mathrm{het}}_{\mathrm{PI}}< \brk{1 \land \frac{1+\delta L}{2}}\Sigma^{\mathrm{het}}_{\mathrm{AB}}\quad \textrm{and}\quad
    \Sigma^{\mathrm{hom}}_{\mathrm{PI}}=0\le \Sigma^{\mathrm{hom}}_{\mathrm{IS}} \leq \Sigma^{\mathrm{hom}}_{\mathrm{AB}} \,,\]
    where $\delta = D_{\mathrm{TV}}(\mP^\ft(s^1,\cdot),\mP^\fc(s^1,\cdot))$ and $L\coloneqq (\ln (C) +1)/(1-\lambda)$. The constants $C,\lambda$ are defined in the uniform ergodicity of the Markov chain (Definition \ref{def: uniform ergodic}) induced by $\pi^{1/2}$.
\end{theorem}
The parameter $\delta$ measures the difference between the transition probabilities under the treatment and the control. One of the reasons that such a parameter takes an effect is that the denominators in $\Sigma^{\mathrm{het}}_{\mathrm{AB}}$ and $\Sigma^{\mathrm{het}}_{\mathrm{PI}}$ are $\mu_{1/2}(s^1)$ and $\mu^{\ft}(s^1)$ respectively, which are the stationary probabilities of the crucial state $s^1$. A little difference in the transition probability may have an significant impact on the reciprocal of the stationary probability. Also, recall that $\Sigma^{\mathrm{het}}_{\mathrm{AB}}=\Sigma^{\mathrm{het}}_{\mathrm{IS}}$, and Theorem~\ref{thm: PI_Analysis} confirms that PI can always strictly outperform A/B testing even with information sharing. Moreover, if the shift of transition probability $\delta$ is small or the Markov chain mixes fast enough, Theorem~\ref{thm: PI_Analysis} indicates that PI can approximately halve the variance compared to A/B testing for the heterogeneous component. To reiterate,  PI eliminates uncertainty stemming from all non-crucial states.

% \begin{proof}
%     Delayed to Appendix~\ref{app: perfect_info}. 
% \end{proof}

% \begin{center}
% \begin{algorithm}
% \caption{Perfect Information Estimation for Single-Point Experiments}
% \begin{algorithmic}[1]\label{alg: PI}
% \REQUIRE{Policy $\pi^t$; Experiment horizon $T$; Discount factor $\gamma$; Control transition $\mP^c$ and Control reward $\vr^c$.}
% \STATE Execute policy $\pi^t$ for $T$ epochs and obtain datasets \[\gD^t = \{(s_i,a_i,s_{i+1}, r_i)\mid (s_i,a_i)= (x,t), i\in [T]\}\,.\] 
% \STATE Combine $\gD^t, \mP^c, \vr^c$ and call Algorithm~\ref{alg: model-based} or Algorithm~\ref{alg: model-free} for $\widehat {\mV^t}$ and $\widehat {\mV^c}\,.$
% \RETURN $\widehat {\mV^t}(x) - \widehat {\mV^c}(x)\,.$
% \end{algorithmic}
% \end{algorithm}
% \end{center}

% \subsection{Comparison with Value Inference}
% The variance is refined as \[\sigma^2_{\mathrm{PI}} = \frac{\rho^t_x(x)^2}{\mu(x)}\norm{\gamma\mV^t}^2_{\Sigma_{p}(x)}\,.\]
% Recall that if we ignore the offline information and estimate online directly, the variance is given by
% \begin{align*}
% \sigma^2_{\mathrm{TD}} &= \sum_{s\in\gS}\frac{\rho_x(s)^2}{\mu(s)}\norm{\gamma\mV^t}_{\Sigma_{p}(s)}^2 \\
% &= \sigma^2_{\mathrm{PI}} + \sum_{s\ne x} \frac{\rho_x(s)^2}{\mu(s)}\norm{\gamma\mV^t}_{\Sigma_{p}(s)}^2\,.
% \end{align*}

%\subsection{Analysis of Variance Reduction for Perfect Information}

Similar ideas can also be generalized to SST with multiple test arms, where the perfect information about control arm is available. Following the notations in Section \ref{sec: multiple-arms-IS}, assume there are $n$ test arm with the first arm serving as the control. Consequently, sampling is required only for the $n-1$ test arms excluding the control arm, such that $\pi^{1/(n-1)}(a\mid s)\coloneqq \frac{1}{n-1}\sum_{i\ne 1}\pi^{i}(a\mid s)$ for any $(s,a)$ pair. For all the test arms, we estimate only $\mP^{\ft^i}(s^1,\cdot)$ and $r^{\ft^i}(s^1)$ using sample means, while inheriting the other values from the known control arm. Using the Bellman equation, we derive the estimator $\widehat{\mV}^{\ft^i}$ for each SST, leading to the estimator of $\Delta_{i,j}(s^1)$ by taking the difference. In Corollary \ref{coro: ABn-PI}, we establish the asymptotic normality for the estimation of $\Delta_{i,j}(s^1)$, confirming the statistical robustness of our approach under these settings.
\begin{corollary} \label{coro: ABn-PI}
    Assuming the Markov chain induced by the mixed policy $\pi^{1/(n-1)},\pi^{1/n}$ are both uniformly ergodic, it holds that as $T\to\infty$,
    \[\sqrt{T}\pa{\widehat \Delta^{\mathrm{PIn}}_{i,j}(s^1)  - \Delta_{i,j}(s^1)}\xrightarrow{d}\gN(0, \sigma_{\mathrm{PIn},(i,j)}^2(s^1))\,,\]
    where $\sigma_{\mathrm{PIn},(i,j)}^2(s^1)$ is given in Appendix~\ref{app: PIn_variance}. Furthermore, we introduce $\Sigma^{\mathrm{het}}_{\mathrm{PIn}}(i,j)$ and $\Sigma^{\mathrm{hom}}_{\mathrm{PIn}}(i,j)$ to denote the terms in $\sigma_{\mathrm{PIn},(i,j)}^2(s^1)$ associated with $s^1$ and all the other states respectively. We have
    \[\Sigma^{\mathrm{het}}_{\mathrm{PIn}}(i,j)< \brk{1 \land \frac{n-1}{n}\pa{1+\frac{2\delta L}{n}} }\Sigma^{\mathrm{het}}_{\mathrm{ABn}}(i,j)\quad \textrm{and}\quad
    \Sigma^{\mathrm{hom}}_{\mathrm{PIn}}(i,j)=0\,,\]
    where $\delta = \max_iD_{\mathrm{TV}}(\mP^i(s^1,\cdot),\mP^1(s^1,\cdot))$ and $L\coloneqq(\ln (C) +1)/(1-\lambda)$. The constants $C,\lambda$ are related to the uniform ergodicity of the Markov chain induced by $\pi^{1/n}$ from Definition \ref{def: uniform ergodic}.
\end{corollary}
With $n$ test arms, PI continues to yield asymptotically normal estimators and significant variance reduction. Moreover, the term $ \brk{1\land \frac{n-1}{n}\pa{1+\frac{2\delta L}{n}} }$ is almost non-decreasing with increasing $n$. This pattern indicates that as $n$ grows, the relative benefit of variance reduction for  $\Sigma^{\mathrm{het}}_{\mathrm{PIn}}(i,j)$ diminishes. Such an observation aligns with the intuition that as the number of test arms grows, the data available for each test arm decreases, potentially becoming a limiting factor in further reducing variance.

\subsection{Proofs of Theorem~\ref{thm: PI}}\label{app: perfect_info}
For $X_i=(s_i,a_i,s_{i+1},r_i)$, define
\begin{align*}
    Z(X_i)&\coloneqq \mathbf{1}[s_i=s^1]\ve_{s_{i+1}}\,,\\
    R(X_i)&\coloneqq \mathbf{1}[s_i=s^1]r_i\,,
\end{align*}
and denote $Y(X_i)\coloneqq [Z(X_i),R(X_i)]^\top$.

    We have the following CLT
\[\sqrt{T}\pa{\overline{Y_T} - \begin{bmatrix}
 P(s^1\mid s^1)\\
 P(s^2\mid s^1)\\
 \vdots\\
 P(s^K\mid s^1)\\
r(s^1,\ft)
\end{bmatrix}}\xrightarrow{d}\gN(\vzero, \frac{1}{\mu^\ft(s^1)}\Sigma_Y)\,,\]
where the covariance matrix is given by 
\[\Sigma_Y = \begin{bmatrix}
  \Diag (\vp_{s^1})-\vp_{s^1}\vp_{s^1}^\top & 0\\
  0&\sigma_\ft^2(s^1)
\end{bmatrix} = \begin{bmatrix}
  \Sigma_{p}(s^1) & 0\\
  0&\sigma_\ft^2(s^1)
\end{bmatrix}\,,\]
with $\vp_{s^1} = [
 P(s^1\mid s^1),
 P(s^2\mid s^1),
 \ldots,
 P(s^K\mid s^1)]^\top$ and $\sigma_\ft^2(s^1)$ the variance of reward. 

To apply the Delta Method, let \[g(\overline{Y_T})=\ve_1^\top \pa{\mI-\gamma\begin{bmatrix}
   \overline{Z_T}^\top\\
  P^\fc(\gS\setminus s^1)
\end{bmatrix}}^{-1}\begin{bmatrix}
   \overline{R_T}\\
  r^\fc(\gS\setminus s^1)
\end{bmatrix}\,.\]

Taking the gradient gives
\begin{equation*}\label{eq: gradient_Oracle}
\nabla g(\overline{Y})^\top = \br{\ve_1^\top\frac{\partial (\mI-\gamma \widehat\mP^\ft)^{-1}}{\partial P(1)}\vr^\ft\,,\quad \rho^\ft_{s^1}(s^1)}^\top\,.
\end{equation*}

One can verify that
\begin{align*}
\ve_1^\top\frac{\partial (\mI-\gamma \widehat\mP^\ft)^{-1}}{\partial P(1,i)}\vr_t 
& = \gamma \ve_1^\top (\mI-\gamma\mP^\ft)^{-1} \mE_{1i} (\mI-\gamma\mP^\ft)^{-1} \vr^\ft\\
& = \gamma \rho^\ft_{s^1}(s^1) V^\ft(i)\,.
\end{align*}

Then the final variance is given by
\[\sigma^2 = \frac{\rho^\ft_{s^1}(s^1)^2}{\mu^\ft(s^1)}\pa{\norm{\gamma\mV^\ft}^2_{\Sigma_{p}(s^1)}+\sigma_\ft^2(s^1)}\,.\]

% \end{proof}

\subsubsection{PI with Multiple Test Arms}\label{app: PIn_variance}
The variance of \textrm{PIn} is given by
\begin{align*}
\sigma_{\mathrm{PIn},(i,j)}^2(s^1) = &\mathbf{1}[i\ne1](n-1)\frac{\rho^i_{s^1}(s^1)^2}{\mu_{1/(n-1)}(s^1)}\pa{\sigma_i^2(s^1)+ \norm{\gamma\mV^i}^2_{\Sigma_{p^i}(s^1)}}  \\
&+\mathbf{1}[j\ne1](n-1)\frac{\rho^j_{s^1}(s^1)^2}{\mu_{1/(n-1)}(s^1)}\pa{\sigma_j^2(s^1)+ \norm{\gamma\mV^j}^2_{\Sigma_{p^j}(s^1)}}\,.    
\end{align*}

\subsection{Asymptotic Relative Efficiency: Proof of Theorem~\ref{thm: PI_Analysis}}\label{app: PI_analysis}
% \begin{theorem}
%     Under Assumption~\ref{assump: no_variance} and Assumption~\ref{assump: uniform ergodic}, for A/B testing with two test arms, it holds \[\Sigma^{\mathrm{hom}}_{\mathrm{PI}}=0\,, \]
%     and\[\Sigma^{\mathrm{het}}_{\mathrm{PI}}\leq \frac{1+\delta L}{2} \Sigma^{\mathrm{het}}_{\mathrm{AB}}\]
%     where $\delta = D_{\mathrm{TV}}(\mP^\ft(x,\cdot),\mP^c(x,\cdot))$ and $L$ is a constant given by $L=(\ln (C) +1)/1-\lambda$.
% \end{theorem}
% \begin{proof}
    Recall \[\Sigma^{\mathrm{het}}_{\mathrm{AB}} = 2\pa{\frac{\rho^\ft_{s^1}(s^1)^2}{\mu_{1/2}(s^1)}\norm{\gamma\mV^\ft}_{\Sigma_{p^\ft}(s^1)}^2 +  \frac{\rho^\fc_{s^1}(s^1)^2}{\mu_{1/2}(s^1)}\norm{\gamma\mV^\fc}_{\Sigma_{p^\fc}(s^1)}^2 }\,,\]
    and \[\Sigma^{\mathrm{het}}_{\mathrm{PI}}=\frac{\rho^\ft_{s^1}(s^1)^2}{\mu_1(s^1)}\norm{\gamma\mV^\ft}^2_{\Sigma_{p^\ft}(s^1)}\,.\]
    By Lemma~\ref{lem: pertub}, it holds
    \begin{align*}
        \mu_1^\top &= \mu_{1/2}^\top + \mu_1^\top (\mP_1-\mP_{1/2})(\mI-\mP_{1/2})^\sharp\\
        &= \mu_{1/2}^\top + \mu_1^\top \begin{bmatrix}
            \vdelta(s^1)^\top\\
            \vzero
        \end{bmatrix}(\mI-\mP_{1/2})^\sharp\\
        & = \mu_{1/2}^\top + 
            \mu_1(s^1)\vdelta(s^1)^\top(\mI-\mP_{1/2})^\sharp\,,
    \end{align*}
    where $\vdelta(s^1) = [P^1(s|s^1)-P^{1/2}(s|s^1)]_{s\in\gS}^\top$ and note that the treatment only happens on state $s^1$.
    
    Multiply by $\ve_1$ to both sides, 
    \begin{equation}\label{eq: PI1}
      \mu_{1/2}(s^1) = \mu_1(s^1)\pa{1-\vdelta(s^1)^\top(\mI-\mP_{1/2})^\sharp \ve_1}\,.  
    \end{equation}
    By Lemma~\ref{lem: pertub_bound}, it holds
    $\mu_{1/2}(s^1)< \mu_1(s^1) (1+\delta L)\,.$
    And this implies $\Sigma^{\mathrm{het}}_{\PI}< \frac{1+\delta L}{2}\Sigma^{\mathrm{het}}_{\AB}$.

    Similarly, one can derive 
    \[\mu_{1/2}(s^1) = \mu_0(s^1)\pa{1+\vdelta(s^1)^\top(\mI-\mP_{1/2})^\sharp \ve_1}\,.\]
    Subtract with \eqref{eq: PI1} and obtain
    \[\vdelta(s^1)^\top(\mI-\mP_{1/2})^\sharp \ve_1 = \frac{\mu_1(s^1)-\mu_0(s^1)}{\mu_1(s^1)+\mu_0(s^1)}\in(-1,1)\,.\]
    This indicates $\mu_{1/2}(s^1)< 2\mu_1(s^1)$ and $\Sigma^{\mathrm{het}}_{\PI}< \Sigma^{\mathrm{het}}_{\AB}$. 
 
    When there are $n$ arms, one can verify 
    \begin{align*}
        \mu_{1/n}(s^1) &= \mu_{1/(n-1)}(s^1)\pa{1+\frac{1}{n-1}\vdelta^\top(\mI-\mP_{1/n})^\sharp \ve_1}\\
        \mu_{1/n}(s^1) &= \mu_{1}(s^1)\pa{1-\vdelta^\top(\mI-\mP_{1/n})^\sharp \ve_1}\,,
    \end{align*}
    where $\vdelta = \frac{1}{n}\sum_{i=2}^n \vp^1-\vp^i$ and 
    \[\vdelta^\top(\mI-\mP_{1/n})^\sharp \ve_1 = \frac{\mu_1(s^1)-\mu_{1/(n-1)}(s^1)}{\mu_1(s^1)+1/(n-1)\mu_{1/(n-1)}(s^1)} < 1\,,\]
    finishing the proof.
    \section{Additional Proofs}
\subsection{Proof of Proposition \ref{prop: TD_mb}}
    Let $N(s^i)>0$ be the number of visitations to state $s^i\in\gS$ in $\tau$. Then the optimization problem in Algorithm~\ref{alg: model-free} can be rewritten as
    \begin{align}\label{eq: opt_rewrite}
        \min_{\widehat \mV} \norm{\Diag(N(s^1), N(s^2),\ldots, N(s^K)) \cdot \pa{\widehat \mV - \pa{\widehat \vr + \gamma \widehat \mP \widehat \mV}}}_2^2\,,
    \end{align}
    % where $\widehat \vr \in \mathbb{R}^n := (\frac{\sum_{s_j=s_1}r(s_j,a_j)}{N(s_1)},\cdots,\frac{\sum_{s_j=s_n}r(s_j,a_j)}{N(s_n)})^\top$ and $\widehat \mP \in \mathbb{R}^{n\times n}$ is with its entry $\widehat \mP \in \mathbb{R}^{n\times n}$ 
    where $\widehat \vr \in\R^K$ is the empirical average reward with $\widehat \vr(s) = \sum_{i=1}^T\mathbf{1}[s_i=s]r(s_i,a_i)/N(s)$ for $s\in\gS$ and $\widehat\mP\in\R^{K\times K}$ is the empirical transition matrix with $\widehat \mP(s,s^\prime) = \sum_{i=1}^T\mathbf{1}[s_i=s, s_{i+1}=s^\prime]/N(s)$ for all $s,s^\prime\in\gS$.

    It is worth noting that any estimation $\widehat \mP$ is always a transition matrix. Thus the matrix $(\mI-\gamma \widehat\mP)$ is always non-singular. Therefore one can verify that the unique solution to \Eqref{eq: opt_rewrite} is $\widehat \mV^{\textnormal{TD}} = (\mI-\gamma \widehat\mP)^{-1}\widehat \vr\,,$which matches $\widehat \mV^{\textnormal{MB}}$.

\subsection{Proofs of Theorem~\ref{thm: monotonicity}}\label{app: Monotonicity}
\subsubsection{Proof 1: a Pure Algebraic Proof}
    For $s\ne s^1$, we have
    \begin{align}
        \Delta(s) &= V^t(s)-V^c(s)\nonumber\\
        &= r(s,c) + \gamma \sum_{s^\prime \in \gS} \mP^c(s^\prime|s,c) V^t(s^\prime) - r(s,c) - \gamma \sum_{s^\prime \in \gS} \mP^c(s^\prime|s,c) V^c(s^\prime)\nonumber\\
        &= \gamma \sum_{s^\prime \in \gS} \mP^c(s^\prime|s,c) (V^t(s^\prime)-V^c(s^\prime))\label{eq: advs}\,.
    \end{align}
    Denote $S^\prime = S\setminus s^1$, $\vDelta=(\Delta(s^1),\Delta(S^\prime))^\top$ and $\mP^c=\begin{pmatrix}
\mP^c(s^1) \\\mP^c(S^\prime)
\end{pmatrix}$. Then following \Eqref{eq: advs}, we have 
\begin{align}\label{eq: Geigen}
\vDelta=     \begin{pmatrix}
\ve_1^\top \\\gamma \mP^\fc(S^\prime)
\end{pmatrix} \vDelta\,.
\end{align}
Further denote $\gG=\begin{pmatrix}
\ve_1^\top \\\gamma \mP^\fc(S^\prime)
\end{pmatrix}\,,$ and \Eqref{eq: Geigen} indicates that $\gG$ has eigenvalue $1$.

% Notice that \[\textnormal{det}(\gG-I)=\textnormal{det}\begin{pmatrix}
% 0 \\\gamma P(S^\prime)- \mI
% \end{pmatrix}=0\,,\] which indicates that $\gG$ has eigenvalue $1$. 

On the other hand, it is known that the eigenvalues $\{\lambda_i\}_{i\in[|S|]}$ satisfy \[\max_{i\in[|S|]}|\lambda_{i}| \leq \|\gG\|_{\infty}=\max_i\sum_j \gG_{ij}=1\,.\]
This implies that $1$ is the spectral radius of $\gG$. Notice that $\gG$ is non-negative matrix, then by Perron-Frobenius Theorem \cite{meyer01}, it admits a non-negative eigenvector of $1$. 

It remains to show that $\textnormal{dim}(\textnormal{ker}(\gG-I))=1\,.$ Since $\gG$ and $\gG^\top$ share exactly the same characteristic polynomial, 
this is equivalent to $\textnormal{dim}(\textnormal{ker}(\gG^\top-I))=1\,.$ Assume for some $\vh=(h(s^1),\ldots,h(s^K))^\top\in \R^{K},\vh\ne \vzero$ such that 
\begin{align}
    \vh&=\gG^\top \vh\nonumber\\
    &= \begin{pmatrix}
 \ve_1 & \mP^\fc(S^\prime)^\top
\end{pmatrix} \vh\,.\label{eq: Gtopeigen}
\end{align}
 Notice that \Eqref{eq: Gtopeigen} holds for any $h(s^1)\in\R$. Then by denoting $\vh^\prime=(h(s^2), \ldots, h(s^K))^\top$, we can refine \Eqref{eq: Gtopeigen} as 
\begin{align}
    \vh^\prime &= \begin{bmatrix} 
    \gamma P_{22} & \dots & \gamma P_{K2} \\
    \vdots & \ddots & \\
    \gamma P_{2K} &        & \gamma P_{KK} 
    \end{bmatrix} \vh^\prime \,.
\end{align}
We further denote $\gT\coloneqq\begin{bmatrix} 
     P_{22} & \dots &  P_{K2} \\
    \vdots & \ddots & \\
     P_{2K} &        &  P_{KK} 
    \end{bmatrix}$, then we have

\begin{align}
    \frac{1}{\gamma} \vh^\prime = \gT \vh^\prime\,. 
\end{align}
However, the eigenvalues $\{\mu_i\}_{i\in[|S^\prime|]}$ of $\gT$ must satisfy \[\max_{i\in[|S^\prime|]}|\mu_{i}| \leq \|\gT\|_{1}=\max_i\sum_j \gT_{ji}\leq1<\frac{1}{\gamma}\,.\] This implies $\vh^\prime = \vzero\,.$ and hence $\textnormal{dim}(\textnormal{ker}(\gG^\top-I))=\textnormal{dim}(\textnormal{ker}(\gG-I))=1\,.$ Recall $\Delta(s^1)> 0$, then $\Delta(s)\geq 0$ for all $s\in S$.

Moreover, let $\Delta(s_i)=\max_{s\ne s^1}\Delta(s)$. It holds that 
\begin{align}
    \Delta(s_i)&=\gamma \sum_{j\in [K]}\mP^\fc(i,j) \Delta(s^j)\nonumber\\
    &\leq \gamma \mP^\fc(i,1)\Delta(s^1)+\gamma (1-\mP^\fc(i,1)) \Delta(s_i)\label{eq: maxAsi}\,.
\end{align}
If $\mP^\fc(i,1)=0$, then \Eqref{eq: maxAsi} shows $\Delta(s_i)\leq \gamma \Delta(s_i)$, which implies $\Delta(s_i)=0$. It's hence evident that $\Delta(s)< \gamma \Delta(s^1)$ for all $s\ne s^1$. Otherwise if $\mP^\fc(i,1)>0$, then 
\begin{align}
    \Delta(s_i)&\leq \gamma \mP^\fc(i,1)\Delta(s^1)+\gamma (1-\mP^\fc(i,1)) \Delta(s_i)\nonumber\\
    &< \gamma \mP^\fc(i,1)\Delta(s^1)+ (1-\mP^\fc(i,1)) \Delta(s_i)\,,\nonumber\\
    \Longrightarrow \quad &\Delta(s_i)<\gamma \Delta(s^1)\,,
\end{align}
which completes the proof for $\Delta(s^1)>0$. Similar proof can be derived for $\Delta(s^1)<0$ by setting $\Delta(s_i)=\min_{s\ne s^1}\Delta(s)\,.$ 

\subsubsection{Proof 2: Performance Difference Lemma}
We draw insights from reinforcement learning theory and apply the well-known \emph{performance difference lemma} \cite{Kakade2002ApproximatelyOA} to our single-point treatment scenario

\begin{lemma}[Performance Difference Lemma of SST]\label{lem: performance}
    Given a SST policy $\pi^\ft$ on crucial state $s^1$ and control policy $\pi^\fc$, for all $s\in\gS$,
    \begin{equation*}
        \Delta(s) = \rho^\ft_s(s^1) A^\fc(s^1,\ft) = -\rho^\fc_s(s^1) A^\ft(s^1,\fc)\,.
    \end{equation*}
    where $A^\pi(s^1,a) = Q^\pi(s^1,a)-V^\pi(s^1)$ is the advantage function at state $s^1$.
\end{lemma}
This lemma clearly characterizes the treatment effect using the advantage function. In particular, the treatment effect of all states all determined by the advantage function at state $s^1$, multiplied by some non-negative measure. Consequently, the monotonicity property follows directly from this relationship.
    \section{Technical Tools}
% We first provide several results for matrix derivatives, see e.g. section 2.8.1 in \cite{Petersen2008}.
% \begin{lemma}[Chain Rule of Matrix-to-Scalar]\label{lem: chain_rule_1}
%     Let $\vx\in\R^n,f:\R^n\to\R^{n\times n}, g:\R^{n\times n}\to \R$ and $y= g(f(\vx))$. Then it holds
%     \begin{align*}
%         \frac{\partial y}{\partial x_i} = \angl{\frac{\partial y}{\partial f(\vx)}, \frac{\partial f(\vx)}{\partial x_i}}_F\,.  
%     \end{align*}
% \end{lemma}

% \begin{lemma}[Chain Rule of Matrix-to-Matrix]\label{lem: chain_rule_2}
%     Let $\vx\in\R^n,f:\R^n\to\R^{n\times n}, g:\R^{n\times n}\to \R^{n\times n}$ and $\mY= g(f(\vx))$. Then it holds
%     \begin{align*}
%         \frac{\partial \mY}{\partial x_i} = \frac{\partial \mY}{\partial f(\vx)}\odot_F \frac{\partial f(\vx)}{\partial x_i}\,.  
%     \end{align*}
% \end{lemma}

% \begin{lemma}[Derivative of Matrix Inverse]\label{lem: derivative_matrix_inverse}
% Let $\vx\in\R^n$ and $\mA\in\R^{n\times n}$ be any non-singular matrix, then it holds
% \begin{align*}
%     \frac{\partial \mA^{-1}}{\partial \vx} = \brk{-\mA^{-1}\frac{\partial \mA}{\partial x_i} \mA^{-1}}_{i}\,.
% \end{align*}    
% \end{lemma}
% \begin{proof}
%     By definition, \[\mA\cdot \mA^{-1}=\mI\,.\]
%     Taking the derivative of $A_{ij}$ to both sides, 
%     \[\frac{\partial \mA}{\partial x_i}\mA^{-1} + \mA \frac{\partial \mA^{-1}}{\partial A_{ij}} = \mathbf 0\,.\]
%     Rearranging terms gives \[\frac{\partial \mA^{-1}}{\partial A_{ij}} = -\mA^{-1}\frac{\partial \mA}{\partial x_i} \mA^{-1}\,.\]
% \end{proof}

\subsection{Markov Chain Central Limit Theorem}

\begin{lemma}[Markov Chain CLT, Corollary 5 in \cite{jones04CLT}]
    Let $X=\{X_i\}_{[n]}$ be a uniformly ergodic Markov Chain w.r.t $\mu$ and $f$ be a Borel function with $\E_\mu f^2 < \infty$. Then under any initial distribution, it holds 
    \begin{equation}
        \sqrt{n}(\frac{1}{n}\sum_{i=1}^n f(X_i) - \E_\mu f) \xrightarrow{d} \gN(0,\sigma_f^2)\,,
    \end{equation}
    where $\sigma_f^2\coloneqq \Var_{X_1\sim\mu} \br{f(X_1)} + 2 \sum_{i=1}^\infty \Cov_{X_1\sim\mu}\br{f(X_1), f(X_{1+i})}\,.$
\end{lemma}

\begin{lemma}[Multivariate Markov Chain CLT, \cite{vats17output}]\label{thm: Multi_Markov_CLT}
    Let $X=\{X_i\}_{i\in[n]}$ be a uniformly ergodic Markov Chain w.r.t $\mu$ and $f\colon \R \to \R^p$ be a Borel function. Under mild technical conditions, it holds 
    \begin{equation}
        \sqrt{n}(\frac{1}{n}\sum_{i=1}^n f(X_i) - \E_\mu f) \xrightarrow{d} \gN(0,\Sigma_f)\,,
    \end{equation}
    where $\Sigma_f\coloneqq \Var_{X_1\sim\mu} \br{f(X_1)} +  \sum_{i=1}^\infty \Cov_{X_1\sim\mu}\br{f(X_1), f(X_{1+i})} + \sum_{i=1}^\infty \Cov_{X_1\sim\mu}\br{f(X_{1+i}), f(X_{1})}\,.$
\end{lemma}

\begin{lemma}[Linearization, Adapted from Lemma 6 in \cite{farias2022markovian}]\label{lem: Linearization}
    Consider $\{X_i\coloneqq(s_i,a_i,s_{i+1},r_i)\}_{i\in[T]}$ be sampled from a uniformly ergodic Markov Chain with stationary distribution $\mu$. Let $Y\colon \gX \to \R^m$ and $g\colon \R^m\to\R^n$ such that as $T\to\infty$, 
    \[\sqrt{T}(g(\frac{1}{T}\sum_{i=1}^TY(X_i)) - g(Y_{\mu}))\xrightarrow{d}\gN(\vzero, \Gamma)\,,\]
    for some covariance matrix $\Gamma$. Define $G(X_i)=\nabla g(\mY_{\mu})^\top Y(X_i)$, then it holds
    \[\Var\pa{\frac{1}{T}\sum_{i=1}^TG(X_i)} = \Gamma\,.\]
\end{lemma}

% One such technical condition for the above two CLT to hold is \emph{uniformly ergodic} and $\E_\mu [f^2]<\infty$, see Theorem 9 in \cite{jones04CLT}. 

\subsection{Perturbation Bounds}
\begin{lemma}[Stationary Distribution Perturbation, \cite{Meyer1980TheCO,farias2022markovian}]\label{lem: pertub}
    Let $\mu_1,\mu_0$ be the respective stationary distribution of $\mP_1,\mP_0$. It holds\[\mu_1^\top = \mu_0^\top + \mu_1^\top (\mP_1-\mP_0)(\mI-\mP_0)^\sharp\,.\] 
\end{lemma}

\begin{lemma}[Adapted from Lemma 2, \cite{farias2022markovian}]\label{lem: pertub_bound}
    For uniform ergodic chain $\mP$ with constant $C,\lambda$ defined in Definition~\ref{def: uniform ergodic}, it holds \[\norm{(\mI-\mP)^\sharp}_{1,\infty}< \frac{2\ln(C)+2}{1-\lambda}\,.\]
\end{lemma}

\subsection{Constrained Cramer-Rao Bounds}
\begin{lemma}[Theorem 1, \cite{stonica98on}]\label{lem: CCRB}
    Let $\vtheta\in\mathbb{R}^n$ satisfying $f(\vtheta)=\vzero$ be the parameters of the system, where $f:\R^n\to\R^k$. Then for any unbiased estimator $\widehat\vtheta$ satisfying $f(\widehat\vtheta)=\vzero$, under some regularity conditions\footnote{The regularity conditions are required for the interchange of certain
integration and differentiation operators}, it holds
\[\Cov (\widehat\vtheta) \succeq \mathrm{CCRB}(\vtheta)\coloneqq \mU (\mU^\top \mI(\vtheta) \mU)^{-1} \mU^\top\,,\]where $\mI(\vtheta)$ is the Fisher Information and $\mU$ is a matrix whose column space is an orthogonal complement of the row space of $\nabla f(\theta)$, i.e. 
\[\nabla f(\vtheta)\mU =0,\qquad \mathrm{rank}(\mU) = n-k\,.\]
\end{lemma}

\begin{lemma}[Corollary 3.10, \cite{Moore2010ATO}]\label{lem: Delta_CCRB}
    If $f(\vtheta)=\vzero$, then for any unbiased estimator of $\bm\alpha = g(\vtheta)$, it holds
    \[\Cov (\widehat {\bm\alpha}) \succeq \mathrm{CCRB}(\bm \alpha)\coloneqq\nabla g(\vtheta)^\top \mathrm{CCRB}(\vtheta)\nabla g(\vtheta)\,. \]
\end{lemma}

    \section{Experiment Details}
All implementations can be found on Github.
\subsection{Customer Lifetime Value}\label{app: CLV_Exp}
We simulate a synthetic example of Example~\ref{exp: CLV}. The code is available on Github. For both SST and local treatment, the control transition matrix and the reward is given by
\[\mP^\fc=\begin{bmatrix}
 0.6 & 0.4 &  &  & \\
 0.5 &  & 0.5 &  & \\
 0.4 &  &  & 0.6 & \\
 0.3 &  &  &  & 0.7\\
 0.1 &  &  &  & 0.9
\end{bmatrix}\,,\qquad \vr^\fc=\begin{bmatrix}
 30\\
 0\\
 0\\
 0\\
0
\end{bmatrix}\,.\]
\begin{enumerate}
    \item \textbf{SST.} The treatment affects state $s^5$ with a decrease in reward (we set the coupon costs $2+0.5n$ for the $n$-th test arm) and increase in the probability transiting to state $s^1$ (increased by $0.1+0.05n$ for the $n$-th test arm).  
    \item \textbf{Local Treatment.} The treatment affects state $(s^3,s^4,s^5)$ with decreases in reward (the coupon costs $2+0.5n$ at all three states for the $n$-th test arm) and increases in the probability transiting to state $s^1$ at three states (increased by $0.1+0.01n$ at all three states for the $n$-th test arm). 
\end{enumerate}
The discounted factor $\gamma$ is set to be $1/1.2\approx0.83$ aligned with \cite{pfeifer2000modeling}.

\subsection{Sepsis Simulator}\label{app: Sepsis_Exp}
\noindent\textbf{Environment.} The state transition follows from complex fluctuation and treatment impacts. Hence the transition matrices are learned via Monte-Carlo samplings. The learning process will generate the transition matrices for all possible actions at all states as well as the initial state distribution of patients. We assume the initial state distribution is known a prior to the estimators, which is a common for discounted reward RL. The discount factor is set to 0.9 aligned with \cite{tran2023inferring}.

\noindent\textbf{Algorithms.} For A/B testing, IS and PI, the transition matrices are learned by model-based methods. However, the large state space may hinder the visitation of all states. To mitigate this issue, we initialize all transition matrix as uniform distribution.

The Naive estimator follows from the generalization of DQ theory to discounted reward case, given by\[\widehat \vDelta^{\textrm{Naive}}_{ij}\coloneqq \pa{\mI-\gamma \widehat\mP_{1/n}}^{-1}\pa{\widehat\vr_i-\widehat\vr_j}\,.\]

The DQ estimator is originally designed for A/B testing. We generalize the DQ estimator to multiple test arms via Taylor expansion,
\[\widehat{\mV}^{\rm{DQ}}_i=(\mI-\gamma \mP_{1/n})^{-1}\br{\mI+\gamma (\mP_i-\mP_{1/n}) (\mI-\gamma \mP_{1/n})^{-1}}\vr_i\,.\]
and $\widehat{\Delta}_{ij}^{\rm{DQ}}=\widehat{\mV}^{\rm{DQ}}_i-\widehat{\mV}^{\rm{DQ}}_j$. Notice that when $n=2$ we recover the original DQ estimator in \cite{farias2022markovian,farias2023correcting} and one can show similar $\gO(\delta^2)$ bias bounds for our generalized estimator. 

\noindent\textbf{Results.} The MSE curve is the average of 100 independent runs and each MSE is given by $\sum_{i=1}^{100}\pa{\hat{\rm{ATE}}-\rm{ATE}}^2/\rm{ATE}^2$ where $\rm{ATE}$ is the true ATE.

\subsection{Bias-Variance Trade-off}\label{app: DQ}
\cite{farias2022markovian} introduces the DQ estimator for the average reward setting, achieving notable bias-variance trade-offs. \cite{farias2023correcting} extends this work, generalizing the DQ theory to other reward settings and proposing a unified framework for Taylor series expansions. We review the DQ theory for the discounted reward setting.

We begin by examining the naive estimator, which can be viewed as the zeroth-order expansion, given by,
\[\widehat \vDelta^{\textrm{Naive}}\coloneqq \pa{\mI-\gamma \widehat\mP_{1/2}}^{-1}\pa{\widehat\vr(\ft)-\widehat\vr(\fc)}\,.\]
By Corollary 1 in \cite{farias2023correcting}, the bias for the naive estimator is bounded by $\gO(\delta/(1-\gamma)^2)$. Then, the DQ estimator is defined as the first order expansion,
\[\widehat \vDelta^{\textrm{DQ}}\coloneqq \pa{\mI-\gamma \widehat\mP_{1/2}}^{-1}\pa{\widehat \mQ^{\pi_{1/2}}(\ft)- \widehat \mQ^{\pi_{1/2}}(\fc) }\,,\]
where the $\mQ^{\pi_{1/2}}(a) = \widehat \vr(a) + \gamma \widehat \mP(a) \widehat V^{\pi_{1/2}}$ for $a\in\{\ft,\fc\}$. The bias of DQ is bounded by $\gO(\delta^2/(1-\gamma)^3)$. When $\delta$ is small enough, DQ can outperform naive estimator for the bias. DQ intuitively trades a small bias for substantially lower variance compared to unbiased estimators. This intuition is well-justified in the average reward setting in \cite{farias2022markovian}. However, the performance of naive and DQ estimators in terms of variance remains unclear in the discounted reward setting. 

Moreover, we can apply the idea of information sharing to the learning of Q-functions to reduce variance. Specifically, information sharing enables us to obtain better estimation of $\widehat \vr(a)$ and $\widehat \mP(a)$ for $a\in\{\ft,\fc\}$. We call this estimator DQ-IS, taking advantage of orthogonal ideas of DQ and IS.

\begin{figure}[h]
    \centering
    \includegraphics[width=0.5\linewidth]{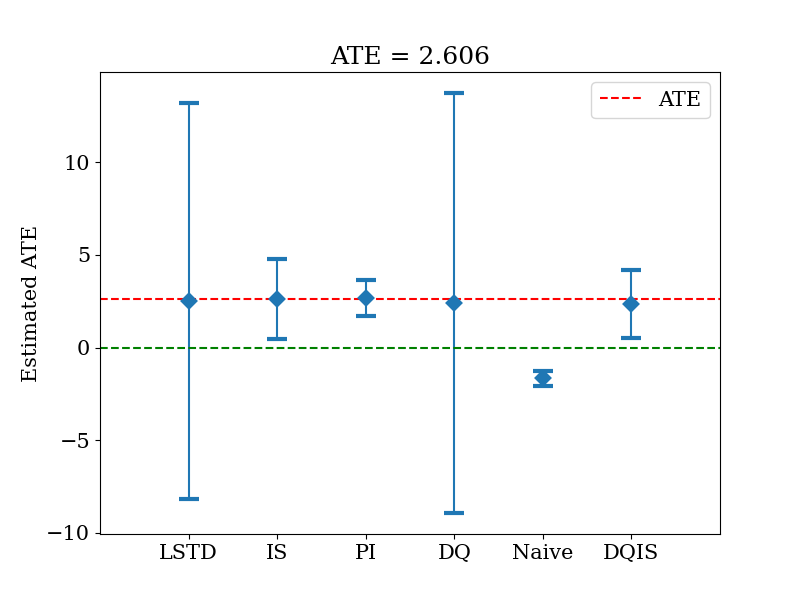}
    \caption{Synthetic simulations for Example~\ref{exp: CLV} with $\textrm{ATE}=2.606$. The result for SST policy with $n=2$ test arms and experiment horizon $T=350$. The empirical means and standard errors are obtained through $1000$ independent trajectory rollouts.}
    \label{fig: DQ}
\end{figure}

We then further implement Naive, DQ and DQ-Is in the same simulation environment of Example~\ref{exp: CLV} and the results are concluded in Figure~\ref{fig: DQ}.

% As outlined in the table, both naive and DQ estimators perform poorly in terms of bias and variance compared to unbiased estimators specifically designed for the discounted reward setting in this paper. The simulation reveals that DQ actually amplifies the bias relative to the naive estimator, likely due to the large discount factor $\gamma=0.83$ in this scenario. Regarding variance, the results align with intuitions from the average reward setting: DQ exhibits higher variance than the naive estimator. However, even the naive estimator shows substantially higher variance compared to A/B testing in the discounted reward setting. In conclusion, we believe designing estimators that achieve optimal bias-variance trade-offs in the discounted reward setting remains an open question.
\noindent\textbf{DQ trades-off variance for bias compared to Naive. } The simulation results first verify the DQ theory of bias. In particular DQ, the first order expansion, exhibits smaller variance than Naive, the zeroth order expansion.
\newline\noindent\textbf{IS is compatible with DQ.} The simulation results demonstrate that both DQ and DQ-IS exhibit negligible bias. However, DQ-IS shows a significantly reduced variance compared to the original DQ method. These findings validate the orthogonal application of the information sharing idea. 
\newline\noindent\textbf{DQ exhibits minimal bias-variance trade-off with SST.} A more surprising result is that both DQ and DQ-IS exhibit negligible bias, while their variances are comparable to that of two unbiased estimator, A/B testing and IS. This suggests that the structure of treatments may potentially impact the performance of bias-variance trade-off estimators . We believe this phenomenon warrants further study, as it may be of independent interest to DQ theory and the broader field of bias-variance trade-offs in RL.

	%%%%%%%%%%%%%%%%%
\end{document}